\documentclass[fleqn,usenatbib]{mnras}

\usepackage[T1]{fontenc}

\DeclareRobustCommand{\VAN}[3]{#2}
\let\VANthebibliography\thebibliography
\def\thebibliography{\DeclareRobustCommand{\VAN}[3]{##3}\VANthebibliography}

\usepackage{mathrsfs}
\usepackage{lmodern}
\usepackage{lineno}
\usepackage{array}
\usepackage{rotating}
\usepackage{tabularray}
\usepackage{amsmath} 
\usepackage{adjustbox}
\usepackage{booktabs}
\usepackage{longtable}
\usepackage{graphicx,natbib,amssymb}
\usepackage{epstopdf}
\usepackage{xcolor}
\usepackage{hyperref}
\usepackage{adjustbox}
\usepackage{multirow}
\usepackage{multicol}
\usepackage{appendix}
\usepackage{caption}
\usepackage{enumitem}
\usepackage{subcaption}
\usepackage{textcomp}
\usepackage[utf8]{inputenc}
\usepackage{xurl}
\usepackage{float}
\usepackage{booktabs, caption, makecell}

\DeclareGraphicsExtensions{.eps,.cps,.ps,.eps.gz,.ps.gz,.eps.Z}
\def \apj{ApJ}
\def \apjl{ApJL}

\def \aap{A\&A}

\def \mnras{MNRAS}
\def \nat{NATURE}
\def \pasp{PASP}
\def \pasj{PASJ}

\usepackage{pdflscape}
\usepackage{dblfloatfix}
\usepackage{rotating}
\usepackage{longtable}
\usepackage[flushleft]{threeparttable}
\usepackage{tablefootnote}
\usepackage{lscape}

\newcommand\Msol{\(\textup{M}_\odot\)}
\newcommand\Rsol{\(\textup{R}_\odot\)}
\newcommand\Trise{$t_{25\_75}$}
\newcommand\Tplat{$t_{\text{plat}}$}
\newcommand\Ttwosix{$t_{20\_60}$}
\newcommand\Tsixnine{$t_{60\_90}$}
\newcommand\Ttwofive{$t_{20\_50}$}
\newcommand\Tfiveeight{$t_{50\_80}$}
\newcommand\ColPeak{$(g-r)_{g,max}$}
\newcommand\Mten{$M_{g,10 d}$}
\newcommand\Mfive{$M_{g,5 d}$}
\newcommand\Mabs{$M_{g,peak}$}
\newcommand\Mcsm{$M_{CSM}$}
\newcommand\Rcsm{$R_{CSM}$}
\newcommand\MLR{$\dot M$}
\newcommand\Mfe{$M_{Fe,Core}$}
\usepackage{cleveref}
\usepackage{orcidlink}

\crefformat{section}{\S#2#1#3} % see manual of cleveref, Section~8.2.1
\crefformat{subsection}{\S#2#1#3}
\crefformat{subsubsection}{\S#2#1#3}

\title[Hinds et al.]{Inferring CSM Properties of Type II SNe Using a Magnitude-Limited ZTF Sample}

\newcommand{\ljmu}{1}
\newcommand{\stockholm}{2}
\newcommand{\northwestern}{3}
\newcommand{\skai}{4}
\newcommand{\caltechoptical}{5}
\newcommand{\caltechphysics}{6}
\newcommand{\naojapan}{7}
\newcommand{\sokendai}{8}
\newcommand{\monash}{9}
\newcommand{\caltechcahill}{10}
\newcommand{\dirac}{11}
\newcommand{\ipac}{12}
\newcommand{\minnesota}{13}
\newcommand{\hubble}{14}
\newcommand{\datadriven}{15}
\newcommand{\drexel}{16}

\author[Hinds et al.]{K-Ryan Hinds$^{\ljmu}$\orcidlink{0000-0002-0129-806X}, Daniel A. Perley$^{\ljmu}$\orcidlink{0000-0001-8472-1996}, Jesper Sollerman$^{\stockholm}$\orcidlink{0000-0003-1546-6615}, Adam A. Miller$^{\northwestern,\skai}$\orcidlink{0000-0001-9515-478X}, Christoffer Fremling$^{\caltechoptical,\caltechphysics}$\orcidlink{0000-0002-4223-103X}, \newauthor Takashi J. Moriya$^{\naojapan,\sokendai,\monash}$\orcidlink{0000-0003-1169-1954}, Kaustav K. Das$^{\caltechcahill}$\orcidlink{0000-0001-8372-997X}, Yu-Jing Qin$^{\caltechcahill}$\orcidlink{0000-0003-3658-6026}, Eric C. Bellm$^{\dirac}$\orcidlink{0000-0001-8018-5348}, Tracy X. Chen$^{\ipac}$\orcidlink{0000-0001-9152-6224}, \newauthor Michael Coughlin$^{\minnesota}$\orcidlink{0000-0002-8262-2924}, Wynn~V.~Jacobson-Gal\'{a}n$^{\caltechcahill,\hubble}$\orcidlink{0000-0002-3934-2644}, Mansi Kasliwal$^{\caltechcahill}$\orcidlink{0000-0002-5619-4938}, Shri Kulkarni$^{\caltechcahill}$\orcidlink{0000-0001-5390-8563}, \newauthor Frank J. Masci$^{\ipac}$\orcidlink{0000-0002-8532-9395},  Ashish~A.~Mahabal$^{\caltechphysics,\datadriven}$\orcidlink{0000-0003-2242-0244}, Priscila~J.~Pessi$^{\stockholm}$\orcidlink{0000-0002-8041-8559}, Josiah Purdum$^{\caltechoptical}$\orcidlink{0000-0003-1227-3738}, Reed Riddle$^{\caltechoptical}$\orcidlink{0000-0002-0387-370X}, \newauthor Avinash Singh$^{\stockholm}$\orcidlink{0000-0003-2091-622X}, Roger Smith$^{\caltechoptical}$\orcidlink{0000-0001-7062-9726}, Niharika Sravan$^{\drexel}$ \\
$^{\ljmu.}$ Astrophysics Research Institute, Liverpool John Moores University, 146 Brownlow Hill, Liverpool L3 5RF, UK \\
$^{\stockholm.}$ Department of Astronomy, The Oskar Klein Center, Stockholm University, AlbaNova University Center, SE 106 91 Stockholm, Sweden \\
$^{\northwestern.}$ Center for Interdisciplinary Exploration and Research in Astrophysics and Department of Physics and Astronomy, Northwestern University, \\1800 Sherman Ave, Evanston, IL 60201, USA \\
$^{\skai.}$ NSF-Simons AI Institute for the Sky (SkAI), 172 E. Chestnut St., Chicago, IL 60611, USA \\
$^{\caltechoptical.}$ Caltech Optical Observatories, California Institute of Technology, Pasadena, CA 91125, USA \\
$^{\caltechphysics.}$ Division of Physics, Mathematics and Astronomy, California Institute of Technology, Pasadena, CA 91125, USA \\
$^{\naojapan.}$ National Astronomical Observatory of Japan, National Institutes of Natural Sciences, 2-21-1 Osawa, Mitaka, Tokyo 181-8588, Japan \\
$^{\sokendai.}$ Graduate Institute for Advanced Studies, SOKENDAI, 2-21-1 Osawa, Mitaka, Tokyo 181-8588, Japan \\
$^{\monash.}$ School of Physics and Astronomy, Monash University, Clayton, VIC 3800, Australia \\
$^{\caltechcahill.}$ Department of Astronomy and Astrophysics, Cahill Center for Astrophysics, California Institute of Technology, MC 249-17,  \\1200 E California Boulevard, Pasadena, CA, 91125, USA \\
$^{\dirac.}$ DIRAC Institute, Department of Astronomy, University of Washington, 3910 15th Avenue NE, Seattle, WA 98195, USA \\
$^{\ipac.}$ IPAC, California Institute of Technology, 1200 E. California Blvd, Pasadena, CA 91125, USA \\
$^{\minnesota.}$ School of Physics and Astronomy, University of Minnesota, Minneapolis, MN 55455, USA \\
$^{\hubble.}$ NASA Hubble Fellow \\ 
$^{\datadriven.}$ Center for Data Driven Discovery, California Institute of Technology, Pasadena, CA 91125, USA \\
$^{\drexel.}$ Department of Physics, Drexel University, Philadelphia, PA 19104, USA}

\begin{document}

\maketitle

\begin{abstract} 
Although all Type II supernovae (SNe) originate from massive stars possessing a hydrogen-rich envelope, their light curve morphology is diverse, reflecting poorly characterised heterogeneity in the physical properties of their progenitor systems. Here, we present a detailed light curve analysis of a magnitude-limited sample of 639 Type II SNe from the Zwicky Transient Facility Bright Transient Survey. Using Gaussian processes, we systematically measure empirical light curve features (e.g., rise times, peak colours and luminosities) in a robust sampling-independent manner. We focus on rise times as they are highly sensitive to pre-explosion progenitor properties, especially the presence of a dense circumstellar medium (CSM) shed by the progenitor in the years immediately pre-explosion. By correlating our feature measurements with physical parameters from an extensive grid of STELLA hydrodynamical models with varying progenitor properties (CSM structure, $\dot M$, $R_{CSM}$ and $M_{ZAMS}$), we quantify the proportion of events with sufficient pre-explosion mass-loss to significantly alter the initial light curve (roughly $M_{CSM} \geq 10^{-2.5} M_{\odot}$) in a highly complete sample of 377 spectroscopically classified Type II SNe. We find that 67 $\pm$ 6\% of observed SNe in our magnitude-limited sample show evidence for substantial CSM ($M_{CSM} \geq 10^{-2.5} M_{\odot}$) close to the progenitor ($R_{CSM} <10^{15}$~cm) at the time of explosion. After applying a volumetric-correction, we find 36$^{+5}_{-7}$\% of all Type II SN progenitors possess substantial CSM within $10^{15}$~cm at the time of explosion. This high fraction of progenitors with dense CSM, supported by photometric and spectroscopic evidence of previous SNe, reveals mass-loss rates significantly exceeding those measured in local group red supergiants or predicted by current theoretical models.

\end{abstract}


%\linenumbers

\section{Introduction}
\label{sec:Intro}

Light curves of core-collapse supernovae (CCSNe), Type II SNe in particular, exhibit a large amount of diversity, varying across orders of magnitude in rise times, luminosities, and durations. The simple progenitor scenario, in which the initial mass is the only factor affecting the SN type or its light curve, cannot adequately explain the extensive observational diversity we see in photometry and spectroscopy -- particularly with the acknowledgement of the role binarity plays in stellar evolution \citep[e.g.,][]{Eldridge_2008, Sana_2012, Eldridge_2018, Zapartas_2019, Zapartas_2021} via binary induced mass transfer and mergers.

An area being explored in greater detail is the degree to which diversity arises from stars with similar initial masses and evolutionary histories that, nonetheless, produce distinct observational signatures at the time of explosion; e.g., varying mass of H envelopes, progenitor radii and H-richness of the outer envelope \citep[e.g.,][]{Popov_1993, Chieffi_2003, Young_2004, Reynolds_2020, Humphreys_2020, Hiramatsu_2021, Moriya_2023, Dessart_2023}. Type II SNe result from the core-collapse of stars with initial masses between 8--20~\Msol\ \citep[e.g.,][]{Eldridge_2004, Smartt_2009, Smartt_2015, Vandyk_2017, Beasor_2020}. The most common subtype, Type IIP, originate from red supergiants (RSGs) -- a connection confirmed through pre-explosion HST imaging \citep[see][]{Smartt_2009, Smartt_2015}. Their light curves exhibit H-recombination powered $\sim$100~d plateaus following steep rises to peak brightness, typically occurring within days to a week \citep[e.g.,][]{Langer_2012, Anderson_2014, Gonz_2015, Rubin_2016, Valenti_2016}.

Less common hydrogen-rich subtypes include: Type IIb SNe showing H-to-He spectral evolution from thin H envelopes \citep{Podsiadlowski_1993, Benson_1994, Woosley_1994, Jerkstrand_2014}; Type IIn SNe with slower rises and narrow emission lines from circumstellar material (CSM) interactions \citep{Schlegel_1990, Mauron_2011, Smith_2014, Arcavi_2017}; and SN 1987A-like events from blue supergiants with extended $>$30~d nickel-powered rises \citep{Schaeffer_1987, Suntzeff_1990, Arnett_1989A, Woosley_1988, Schlegel_1990, Arcavi_2017, Singh_2019b, Sit_2023}.

From the emergence of narrow emission lines in early spectra of young SNe \citep[flash ionisation;][]{Gal-Yam_2014, Groh_2014A, Yaron_2017, Gal-Yam_2017, Bruch_2022}, strong evidence has been presented for the presence of a substantial mass of dense material close to the progenitor at the time of core-collapse. Narrow lines are likely the result of shock breakout (SBO) shock-heating and ionising a slow-moving, dense material \citep[e.g.,][]{Yaron_2017, Irani_2023, Jacobson_2024b}. As the narrow lines typically persist for only a $\sim$ few days post-explosion, it is assumed that the CSM required is the result of mass-loss from the star in the years immediately preceding core-collapse \citep[e.g.,][]{Das_2017, Davies_2022, Tinyanont_2022, Pearson_2022}.

Measurements of the CSM properties from the flash ionisation allow for constraints on the mass-loss rate, \MLR, and late-stage RSG instabilities experienced in the centuries-decades-years immediately before core-collapse \citep[e.g.,][]{Mauron_2011, Yaron_2017, Morozova_2018, Stroh_2021, Pearson_2022, Tinyanont_2022, Bruch_2021, Bruch_2022, Moriya_2023}. These analyses typically assume that the CSM is an unbound, spherically symmetric material escaping with velocities of order $\sim$10 -- 100 km s$^{-1}$ \citep{Smith_2014, Morozova_2017}, following a density profile that decreases with radius \citep[$\rho \propto r^{-2}$ for steady-state mass loss][]{Morozova_2018, Moriya_2018, Davies_2022, Moriya_2023}. \citet{Bruch_2021, Bruch_2022} find that $\sim$~60\% of Type II SNe show evidence for significant amounts of dense CSM confined to a region around the progenitor at the time of explosion -- although, this figure is not corrected for observational biases and not volume limited. Potential precursor events \citep[e.g.,][]{Fraser_2013, Jacobson_2022, Dong_2024, Warwick_2025} provide further evidence of eruptions close to the `classical' core-collapse event. 

The notion that many Type IIP SNe progenitors are surrounded by dense CSM at the time of explosion is further supported by detailed studies of nearby events: SN~2021yja \citep[$\approx$23 Mpc;][]{Kozyreva_2022, Hossein_2022}, SN~2023ixf \citep[$\approx$7 Mpc;][]{Bostroem_2023, Hosseinzadeh_2023, Hiramatsu_2023, Jencson_2023, Jacobson_2023, Li_2023, Singh_2024, Zimmerman_2024} and SN~2024ggi \citep[$\approx$7 Mpc;][]{Chen_2024, Chen_2024b, Xiang_2024, TPessi_2024, Jacobson_2024a, Shrestha_2024}, which, combined with early photometric and spectroscopic data, confirm CSM around their RSG progenitors. In these cases, dense, optically thick CSM causes the SBO to occur within the CSM rather than at the stellar surface \citep{Forster_2018, Tinyanont_2022, Pearson_2022}, producing rapid rises and enhanced peak luminosities \citep[e.g.,][]{Moriya_2011, Das_2017, Morozova_2017, Morozova_2018, Bruch_2021, Bruch_2022, Tinyanont_2022, Pearson_2022, Moriya_2023, Li_2023}.

\MLR\ for RSGs have been measured through multiple techniques: mid-IR observations of circumstellar dust in clusters show \MLR~$\sim$10$^{-6}$~--~10$^{-5}$~\Msol~yr$^{-1}$ \citep[e.g.,][]{Beasor_2018, Beasor_2020}, consistent with rates derived from molecular line and radio measurements \citep[e.g.,][]{Mauron_2011} and comparing pre-explosion progenitor properties to theoretical stellar evolution models \citep[e.g.,][]{Smartt_2009}. Type IIn progenitors exhibit much higher rates of 10$^{-3}$~--~1~\Msol~yr$^{-1}$, derived from multi-wavelength observations \citep[e.g.,][]{Kiewe_2012, Taddia_2013, Fransson_2014}, and combined X-ray, radio, and spectroscopic signatures \citep[e.g.,][]{Smith_2017b,Smith_2017}.

However, \MLR\ inferred from RSG observations alone are insufficient to produce the measured \Mcsm\ and \Rcsm\ on the timescale of decades to months pre-explosion \citep[e.g.,][]{Bruch_2021, Bruch_2022}. Popular mechanisms for end-of-life mass-loss include: wave-driven energy heating into the stellar envelope \citep[e.g.,][]{Fuller_2017a, Morozova_2020, Wu_2021},  radiation-driven mass-loss \citep[e.g.,][]{Vink_2008, Vink_2023}, instabilities caused by explosive shell burning \citep[e.g.,][]{Arnett_Meakin_2011, Smith_2014b}, common envelope interactions caused by binary interactions \citep[e.g.,][]{Chevalier_2012, Sana_2012} and RSG `superwinds' \citep[e.g.,][]{Davies_2022}. 

Previous studies have been successful in characterising smaller subsets of Type II SN properties such as luminosities and rise times \citep[e.g.,][]{Taddia_2013, Anderson_2014, Sanders_2015, Gall_2015, Rubin_2016, Valenti_2016, Graur_2017b, Davis_2019} but are limited to small numbers or incomplete samples, made up of well-observed SNe detected in heterogeneous galaxy-targeted surveys. Infrequent and inconsistent survey cadence lead to inadequate coverage on the rise, limiting the amount of information one can infer from the rising light curve. The wide area, high cadences and untargeted nature of modern surveys allow for larger, more complete samples to be curated -- lending to more detailed statistical analysis of population characteristics and their frequencies. This work builds upon previous studies by making use of robust statistical methods and large, highly complete surveys.

To address these questions, we present detailed light curve analysis of spectroscopically classified Type II SNe from the Zwicky Transient Facility \citep[ZTF;][]{Bellm_2019a, Bellm_2019b,Graham_2019, Masci_2019, Dekany_2020}. In Section~\cref{sec:Method I} we introduce our sample, present the forced photometry light curves and the Gaussian process methodology and data analysis processes used in this work. We present the sample in Section~\cref{sec:Population Properties} and explore the diversity of Type II SNe. In Section~\cref{sec:Methods II}, we empirically infer progenitor properties using a correlation-based analysis based on previous studies. In Section~\cref{sec:CSM Analysis}, we present the volume corrected  \citep[$V_{max}$ method;][]{Schmidt_1968} distributions for \Mcsm, \Rcsm\ and \MLR. We then analyse and discuss the implications in Section~\cref{sec:Discussion}.

Throughout the paper, we correct for Galactic extinction using the NASA Extragalactic Database (NED) extinction tool (using the dust map from \citealt{Schlafly_2011}). We assume a cosmological model with $\Omega_M = 0.3$, $\Omega_\Lambda = 0.7$ and $h = 0.7$.

\section{Methods I -- Sample, Forced Photometry and Light Curve Modelling}
\label{sec:Method I}

\subsection{The Zwicky Transient Facility and The Bright Transient Survey}
\label{sec:Survey}

Of the total observing time available to ZTF, a major fraction has been devoted to public surveys -- 40\% in the initial 2.5 years, and 50\% in subsequent phases. Most of this public observing time is used for a Northern Sky Survey (NSS) of fields above declination $-30^\circ$ in ZTF $g$ and $r$ bands \citep{Bellm_2019b}. The NSS began as a 3~d cadence survey and now runs at 2~d cadence. The public surveys generate alerts which are distributed to various community alert brokers \citep{Patterson_2019}. ZTF $i$ band observations are available for some fields which overlap partnership surveys.

The Bright Transient Survey (BTS), described in \citet{Fremling_2020}, \citet{Perley_2020a} and Qin et al. 2025 (in prep.), is a magnitude-limited survey aiming to spectroscopically classify all extragalactic transients in the northern hemisphere, satisfying a few basic conditions: a peak apparent magnitude, $m_{peak}$, $\leq$ 18.5~mag, visibility from Palomar, and a location out outside of the Galactic Plane. As of December 31st 2024, the BTS catalogue includes $>$10,000 classified SNe brighter than 19~mag; spectroscopic classification is 95.5\% complete down to 18.5~mag for events passing visibility and cadence criteria \citep[see][for a review]{Perley_2020a}\footnote{These statistics are available on the ZTF Bright Transient Survey homepage: \url{https://sites.astro.caltech.edu/ztf/bts/bts.php}.}. The 2 -- 3~d or less cadence and sensitive nature of the survey are required to adequately sample enough of the rise to constrain it with some certainty, and secure detections during the early phase of the light curve, close to the explosion time. 

Final classifications (used here), volumetric rates and luminosity functions from the BTS sample will be presented in the upcoming paper Qin et al. 2025 (in prep.) which covers the period starting 2018 to the end of 2024. Both Qin et al. 2025 (in prep.) and this work have made use of ZTF observing time, instruments and software: Spectral Energy Distribution Machine \citep[SEDM;][]{Blagorodnova_2018, Rigault_2019, Kim_2022}, the DouBle Spectrograph \citep[DBSP;][]{Oke_1982}, Global Relay of Observatories Watching Transients Happen Marshal \citep[GROWTH;][]{Kasliwal_2019} and the Fritz SkyPortal Marshal  \citep[][]{Duev_2019, SkyPortal_2019, Duev_2021, Coughlin_2023}.

This study analyses the spectroscopically classified SNe from the BTS database\footnote{Finalised in Qin et al. 2025 (in prep.)}, incorporating both BTS classifications and TNS reports archived in the BTS from May 1st 2018 to December 31st 2023, retrieved via the BTS internal Sample Explorer\footnote{A public version is available at: \url{https://sites.astro.caltech.edu/ztf/bts/explorer.php}}. Beyond the apparent magnitude threshold, Palomar visibility constraints, and Galactic plane exclusion previously discussed, the BTS requires: sufficient temporal coverage spanning 7.5 -- 16.5~d pre-peak to 16.5 -- 28.5 d post-peak, with multiple observations near peak brightness; spectroscopic accessibility up to 30~d post-peak; the transient must be absent in the reference image; and alerts to pass the BTS alert stream filtering criteria detailed in \citet{Perley_2020a}.

The quality cuts ensure light curves are sampled during the rise to peak and well after peak, and are generally independent of light curve properties. Key values drawn from the sample that are used to comment on demographics have the additional criterion of light curves peaking brighter than 18.5~mag. From hereon, Type II SNe refer to SNe spectroscopically classified as Type II or Type IIP and do not include Type IIn or Type IIb, which are referred to as such.

\subsection{Forced Photometry Light Curve Analysis}
\label{sec:Forced Phot}

\begin{figure*}
    \centering
     \includegraphics[width=1.05\textwidth]{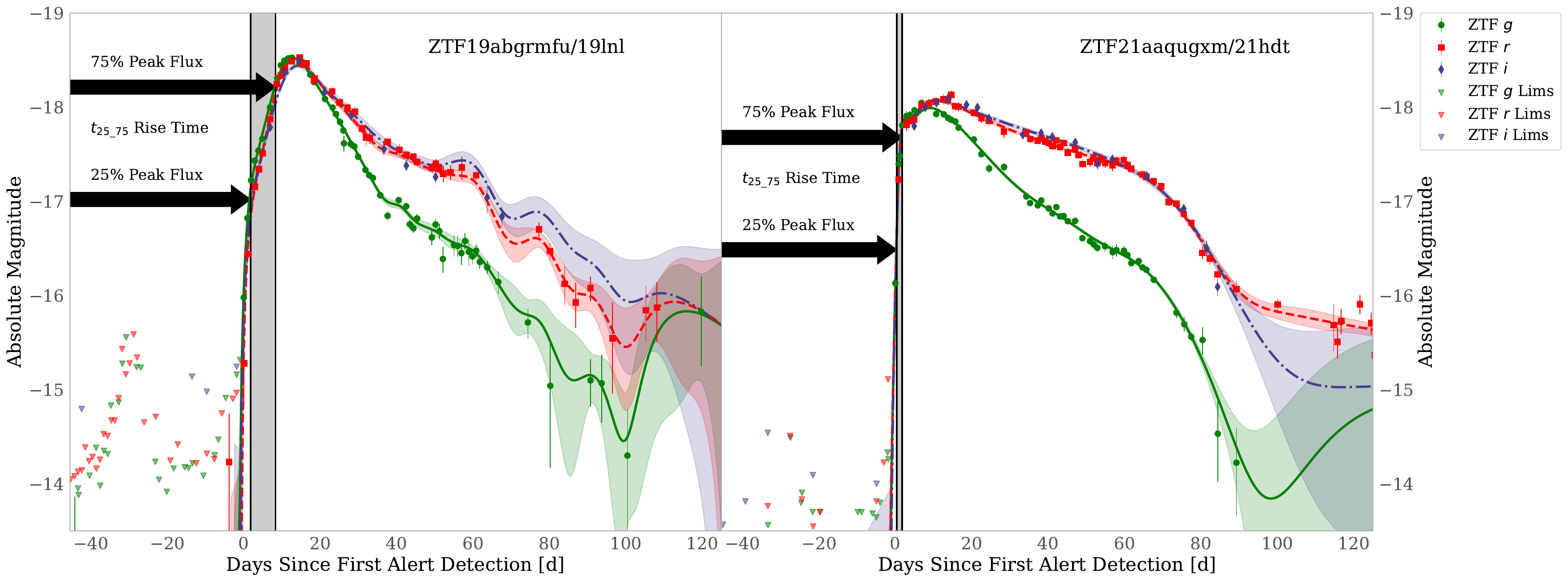}
    \caption{Type II SNe ZTF19abgrmfu/SN~2019lnl at z = 0.035 (\textit{left}) and ZTF21aaqugxm/SN~2021hdt at z = 0.019 (\textit{right}). ZTF~$gri$ forced photometry light curve modelled with 2D Gaussian process regression. We have annotated how a rise time metric (time to rise from 25~--~75\% of the peak flux \Trise) is measured. The inverted triangles represent the upper limits generated by the fps pipeline -- where the limit is determined to be the maximum of [flux + 2$\times\sigma_{\textrm{flux}}$,3$\times\sigma_{\textrm{flux}}$]. Green circles and solid lines represent ZTF~$g$, red squares and dotted line represent ZTF~$r$ and dark blue diamonds and dark blue dash-dotted line represent ZTF~$i$. The shaded regions represent the 68\% CI.}
    \label{fig:forced_lc}
\end{figure*}

The ZTF real-time data stream operates by producing alert packets, where an alert is generated based on real-time and historical contextual information \citep[][]{Masci_2019}. Point source function (PSF) photometry and difference imaging using ZTF archives generate upwards of 100,000 alerts nightly. Photometric measurements are performed based on image-subtracted photometry (\texttt{ZOGY}; \citealt{Zackay_2016}). The distributed alert packets do not allow for measurements below the detection threshold and do not fix the position, creating room to miss detections if the software does not recover an alert. 

Photometry for this study is produced using the ZTF forced photometry service \citep[fps;][]{Masci_2023}, with post-processing conducted following the procedures in Miller et al.\ (2025, in prep.). Briefly, the fps estimates the PSF flux at a user-specified location in all ZTF difference images with coverage of the specified position. The flux measurement uses the same PSF model defined by the \texttt{ZOGY} algorithm that is used to perform image subtraction in the production of ZTF real-time alerts. Observations in which the fps pipeline processing produces a flag, typically because the photometric calibration is excessively noisy or the initial image subtraction failed, are excluded from the analysis. The fps flux measurements require a systematic baseline correction, i.e., there is a small constant offset that needs to be removed to make the pre-SN flux measurements consistent with zero flux \citep[see][]{Masci_2023}. The baseline is estimated using observations that were obtained $> 100$~d before maximum and several hundred days after maximum, where the duration after the peak is determined by conservatively assuming the transient is purely powered by radioactive $^{56}$Co decay. Following the baseline correction, the uncertainties for the individual flux measurements are adjusted to account for a systematic trend whereby brighter sources have underestimated uncertainties (see Miller et al., 2025, in prep.\ for further details). As a final output, this post-processing produces a measurement of the transient flux and its uncertainty in units of $\mu$Jy, including in images where there is no flux detected from the transient. For this study, Public + Partnership + Caltech ZTF data was used.

\subsection{Gaussian Process Regression}
\label{sec:GPR}

CCSN light curves are difficult to model due to the extensive variety in their photometric behaviour. There exist analytical attempts to address the problem using parametric fitting functions \citep[e.g.,][]{Villar_2017, Villar_2019} and there are advancements in theoretical models to produce synthetic light curves with more likeness to observed light curves \citep[][and references therein]{Morozova_2017, Das_2017, Moriya_2023}. Whilst certain parameterisation and generalised empirical models have proven to be adequate in some scenarios, few of these models can fully characterise the diversity of parameters and properties present in transients being uncovered by large surveys. Thus, we are motivated to use a non-parametric technique such as Gaussian process regression (GPR; \citealt{Rasmussen_2004}).% which reduces the need for fine-tuning a model and its parameters for individual light curves or SN types.

GPR is a non-parametric, Bayesian machine learning method for modelling data with functions of an unknown form \citep[see][for a review]{araa_2022}. For single-band SN light curve interpolation, the unknown function a 1-dimensional Gaussian process (GP) approximates is flux as a function of time. We include the effective wavelength, $\lambda_{\rm{eff}}$, of each filter band, and train in 2-dimensions -- e.g., flux as a function of time and effective wavelength -- which is often expressed as probability by Eq. \ref{eqn:gpr_lc_prob}, similar to methodology used in \citet{Thornton_2024}.

\begin{equation}
    P(f|t,\lambda_{\rm{eff}}) = \mathcal{N}(\mu(t,\,\lambda_{\rm{eff}}),\textbf{K})
    \label{eqn:gpr_lc_prob}
\end{equation}

where $f$ is flux and $\lambda_{\rm{eff}}$ for ZTF~$g$, $r$ and $i$ is 4753.15~\AA, 6369.99~\AA\ and 7915.49~\AA\ respectively \citep[][]{Rodrigo_2020,Rodrigo_2024}. The data is input in the observer frame and for plotting purposes, we plot in the observer frame. For parameter measurements (luminosities, timescales and colours), we standardise to rest frame ZTF~$g$ by predicting at $\lambda_{\rm{eff},g}$$\times \left(1 + z\right)$.  This approach allows us to consistently compare physical parameters across our sample while preserving the original photometric information.

For modelling diverse SN light curves, we use a Matèrn-5/2 covariance function (\textbf{K}) that captures both smooth evolution in addition to the sharp transitions characteristic of SNe -- the kernel includes an additive white noise term to account for photometric uncertainties . We implement the GPR using the \texttt{python} package \texttt{george} \citep{georgeGPR_2014}, combining the Matèrn-5/2 kernel with the function from \citet{Villar_2019} -- modified in \citet{Sanchez-Saez_2021}, see their Eq. A5 -- to constrain the behaviour in coverage gaps. 
%\footnote{The main difference being to include a smooth transition function between the two components to ensure continuity.}

GPR allows robust parameter extraction despite heterogeneous sampling and measurement uncertainties, facilitating empirical correlation analysis without detailed individual modelling. By fitting in flux space, we incorporate non-detections to better constrain early light curve evolution.

Fig.~\ref{fig:forced_lc} demonstrates our 2D GPR modelling of ZTF~$gri$ light curves for ZTF19abgrmfu/SN~2019lnl and ZTF21aaqugxm/SN~2021hdt. This approach leverages cross-filter correlations to simultaneously predict temporal and spectral evolution -- particularly valuable when sampling is irregular across bands. While ZTF~$g$ and $r$ observations maintain a regular 2 -- 3~d cadence, ZTF~$i$ band data is often sparse due to partnership-specific scheduling. Our multi-band GPR uses well-sampled bands to constrain the less frequently observed ones, enabling more consistent and precise measurements of colour, rise times, and peak magnitudes across all bands.

Our 2D GPR approach uses a single length scale parameter to handle heterogeneous sampling by controlling correlation strength between observations. While this flexibility accommodates diverse light curve shapes, it presents challenges: the kernel must balance modelling rapid early evolution with slower late-time decline, potentially overfitting lower signal-to-noise data during the radioactive decay phase. Even with the \citet{Villar_2019} function providing a smooth mean function, the GP's flexibility can introduce unphysical variations at late times. The method also assumes consistent colour evolution, potentially misrepresenting rapid colour changes.

However, these limitations primarily affect the faint and late phases, having minimal impact on our scientific conclusions since we focus on bright events ($m_{peak}$~$\leq$~18.5~mag) and measure parameters during well-sampled phases near peak brightness. %Our approach thus optimises photometric characterisation during the most physically informative epochs while maintaining the advantages of a non-parametric fitting technique. 

\subsubsection{Feature Extraction}
\label{sec:GPR_LC}

Combining the fps pipeline and GPR developed for this work, we have used 2D GPR to interpolate across all available filters for each SN in the BTS sample, including Type Ia SNe. Using the interpolated light curves and $\lambda_{\rm{eff}}$ information, we have empirically measured $>$ 20 metrics (see Table~\ref{tab:parameter_descriptions}) for each band where the coverage allowed for measurements to be reliably taken -- for example, coverage constraints are placed on the rise to ensure reliability, we explore this in more detail in a proceeding section. Metrics relevant for this paper rely first on the peak flux (flux at maximum light) and include: rise times from 10\%, 25\%, 50\% of the peak flux to the peak flux or 50\%, 60\%, 75\%, 80\%, 90\% of the peak flux; the peak flux; magnitude; luminosity; time of the peak relative to the first alert detection; colour at several times before, at and after the peak; plateau length and plateau magnitude; and other metrics for analysis in future works. As mentioned, we standardise these measurements by predicting the GP model for each filter in the rest-frame ZTF band. %We then perform a standard k-correction, $1/\left(1+ z\right)$, for luminosities.

\begin{table*}
\begin{threeparttable}

    \centering
    \begin{tabular}{cccc}
    \hline
    Parameter & Symbol & Unit & Definition\\
    \hline \hline
    \multicolumn{4}{c}{P1 - Parameters measured empirically from the GPR model used here} \\ \hline
        Peak Apparent Magnitude & $m_{peak}$ & mag & Apparent magnitude at peak \tnote{1} \\
        Time of Peak & $t_{peak}$ & d & Time at which the peak occurs \tnote{2} \\
        X -- Y\% Rise Time & $t_{\texttt{X}\_\texttt{Y}}$ & d & Time between X\% and Y\% of the peak (e.g., 25~--~75\%) \tnote{3}\\ 
        Fade Time & $t_{Peak\_50}$ & d & e.g., time between peak and 50\% peak flux  \\
        Duration Time & $t_{50}$ & d & e.g., time spent above 50\% peak flux \\
        Apparent Magnitude at X d & $m_{g,X d}$ & mag & Apparent ZTF~$g$ magnitude at 5, 10 \& 50~d post ZTF~$g$ peak \\
        Plateau Duration & \Tplat & d & Measured using the gradient of the GP interpolated light curve  \\
        Colour & $(g-r)_{r,- X d}$ & mag & e.g., ZTF~$g-r$ calculated X d before r-band peak \\
        \hline
        \multicolumn{4}{c}{P2 - Additional parameters measured empirically from the GPR model not used here} \\ \hline
        Plateau Colour & $(g-r)_{plat}$ & mag & Colour at the end of the plateau \\
        MJD Explosion Time & $T_{exp}$ & d & Explosion time \tnote{2} \\
        Colour Evolution & $g-r$ & mag & ZTF~$g-r$ colour relative to ZTF~$r$ band peak\\
        Peak Luminosity & $L_{peak}$ & erg s$^{-1}$ & Luminosity at peak\\
        Plateau Slope & $\nabla_{plat}$ &~mag d$^{-1}$ & Gradient of the plateau \\
        No. Bumps & $N_{bumps}$ & - & The number of peaks in the light curves \\
        Decline Rates & $\Delta m_{X d}$ &~mag d$^{-1}$ & Difference in magnitude between peak and peak + X d\\
        Optical Energy & $E_{opt}$ &erg & Total integrated optical energy in each band ($\int\nu f_{\nu} d\nu$) \\ \hline
        \multicolumn{4}{c}{P3 - Parameters indirectly measured from relationships involving P1 parameters} \\ \hline
        CSM Mass & \Mcsm & \Msol & Mass of circumstellar material\\
        Mass-Loss Rate & $\dot M$ & \Msol~yr$^{-1}$ & Rate of mass-loss of the progenitor\\
        CSM Radial Extent & \Rcsm &~cm & Radial extent of circumstellar material\\
        Iron Core Mass & \Mfe\ & \Msol & Iron core mass of progenitor prior to explosion\\
        \hline
        
    \hline
    \end{tabular}

\tnote{1}~Defined as being the peak with the highest flux and it must have previous detections or non-detections on the rise to confirm it is the peak. In the case of multiple peaks, all peaks are identified and the peaks are ranked in order of strength and time (earliest first). All relative time intervals are reported in the rest frame; \tnote{2}~Relative to first alert detection; \tnote{3}~We also measure 20~--~60\%, 60~--~90\%, 20~--~50\% \& 50~--~80\% rise times.
\end{threeparttable}
\caption{Description of parameters empirically measured directly using GPR interpolations (P1), or calculated using measured parameters and established relationships -- see Section~\cref{sec:Measuring CSM Mass and Radius} for \Mcsm, \MLR\ and \Rcsm. We standardise all measurements by predicting the GP model at $\lambda_{eff}$$\times~(1+z)$ such that all measurements are in the rest-frame ZTF band or relative to the rest-frame measurements. Most parameters are filter-dependent and have been measured for each filter. The table is split into sections: P1 are the empirical GP measured parameters used specifically in this work, P2 are parameters also empirically measured using the GP interpolation but not used in this work and P3 which shows parameters estimated from relationships or methods involving parameters in P1. \\}
\label{tab:parameter_descriptions}
\end{table*}

We estimate parameter uncertainties by drawing 1,000 samples from the GPR posterior distribution, leveraging the probabilistic nature of the GP. The 1~$\sigma$ uncertainties are derived from the resulting distribution of measured values, capturing both photometric uncertainties and the range of light curve behaviours consistent with our data.

\subsection{Galactic and Host Extinction Corrections}
\label{sec:Extinction}
We correct for line of sight Galactic extinction using NED extinction tools (based on the dust map from \citealt{Schlafly_2011}). For \Mabs\ used in this work, we calculate the peak of the GP model, correct for Galactic extinction at the interpolated central wavelength and apply a uniform K-correction of $2.5\log_{10}\left(1+z\right)$, which is typically small.

\begin{figure*}
    \centering
    % \hspace*{-1.45cm} 
    \includegraphics[width=1.0\textwidth]{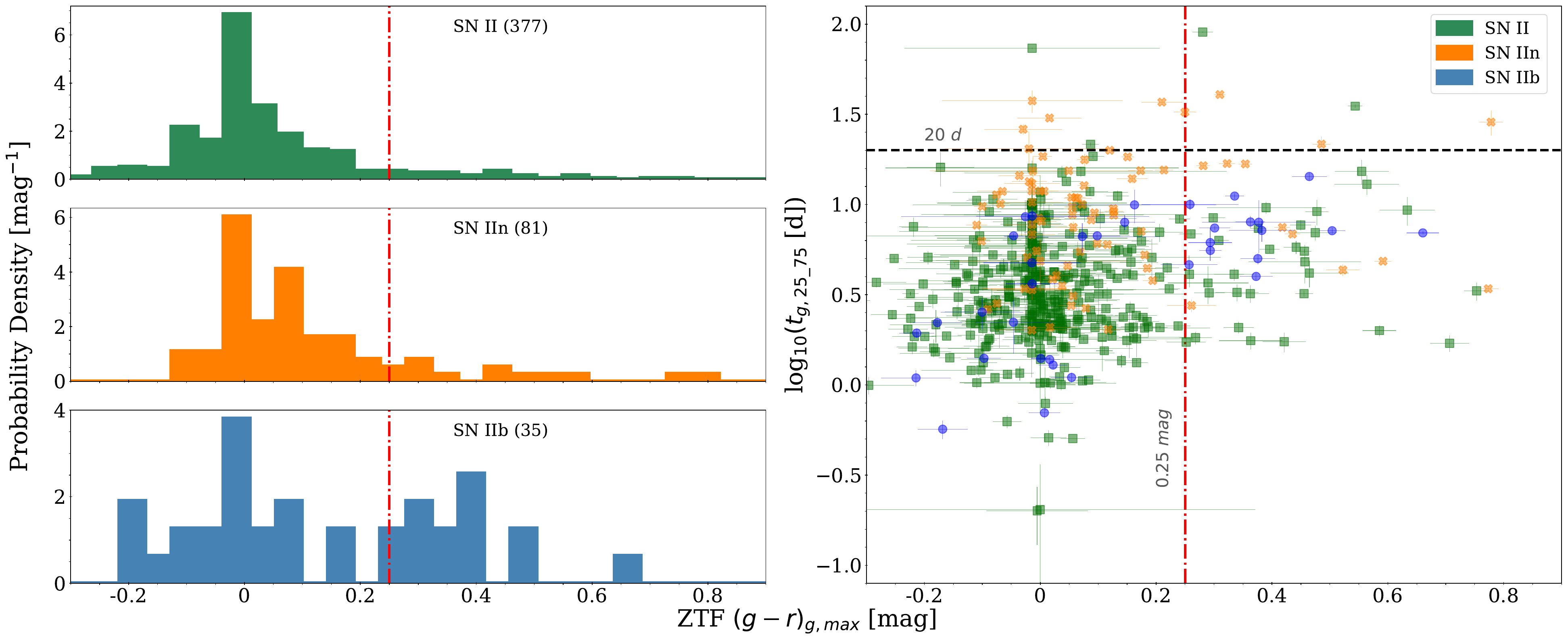}
    \caption{\ColPeak\ histograms (\textit{left}) and ZTF~$g$ time to rise from 25~--~75\% peak flux vs \ColPeak\ (\textit{right}) for Type II, Type IIn and Type IIb SNe (\textit{top left} to \textit{bottom left}) for those with a $m_{peak} \leq 18.5$~mag in any filter. The histograms for Type II and Type IIn are sharply peaked and have a tail in the red direction indicating a standard colour distribution and the presence of some `dusty', \ColPeak~$\geq 0.25$~mag, SNe. The horizontal line shows 20~d and the vertical line at 0.25~mag is the limit we define, beyond which we identify objects as `dusty'. This plot is used to distinguish the likely heavily host dust-extinguished SNe (rise~$<$~20~d, and red, $g - r >$~0.25~mag) from those likely intrinsically red (rise~$\geq$~20~d and red, $g - r >$~0.25~mag).}
    \label{fig:AllII_pcol_hist}
\end{figure*}

As ZTF is a survey in $gri$, a comprehensive host extinction correction is not feasible with the survey photometry alone. One can approximately find the host extinction using the ZTF~$g-r$ colour at the ZTF~$g$ peak, \ColPeak, and apply a correction based on this colour. For Type II and Type IIn SNe, we find the unweighted histogram of the peak colour \ColPeak, Fig.~\ref{fig:AllII_pcol_hist}, is sharply peaked around 0~mag, which is characteristic of a well-defined intrinsic population and suggests standard colour distribution, consistent with the findings of \citet{deJaeger_2018a}.

The Type II SN colour distribution, shown in Fig.~\ref{fig:AllII_pcol_hist}, has an asymmetric tail toward redder colours. This asymmetry is particularly informative -- if there existed a significant population of intrinsically red SNe, we would expect a more symmetric distribution or a secondary peak, rather than the observed sharp core with a red tail. Based on the distribution in Fig.\ref{fig:AllII_pcol_hist}, we find the 90th and 95th percentiles for Type II SN \ColPeak\ to be $\approx$~0.2 and $\approx$~0.39~mag, respectively. We establish \ColPeak~$\geq$~0.25~mag as the threshold for identifying 'dusty' events that necessitate host-extinction correction.  Using an Empirical Cumulative Distribution Function (ECDF), we find that $9^{+12}_{-6}$\% of the observed Type II population have \ColPeak~$\geq$~0.25~mag, confirming that heavily dust-extinguished events constitute a minority of the sample. For Type IIn and Type IIb SNe, we find $14^{+14}_{-8}$\% and 40~$\pm$~20\%, respectively, have \ColPeak~$\geq~0.25$~mag -- uncertainties reported here are the binomial CI.

Using a rise time definition we present in Section~\cref{sec:DA - Rise Distribution}, we investigate the rise time vs colour parameter space to understand how this correction is applied to SNe across a variety of rise times -- see Fig.~\ref{fig:AllII_pcol_hist}. The majority of SNe with rise times~$<$~20~d exhibit peak $g-r$ colours blueward of 0.25~mag. Objects redward of this threshold likely suffer significant dust-extinction, forming a skewed tail extending from an otherwise approximately normal colour distribution. Those with \ColPeak~$>$~0.25 and rise $>$~20~d are likely intrinsically red, owing to photons emitted from the core being trapped for longer which increases diffusion time and rise time to maximum light. We perform this exercise to avoid applying an incorrect host-correction to those SNe that are likely intrinsically red, therefore artificially boosting their luminosity. As we have identified this population in the top right quadrant (rise~$>$~20~d and \ColPeak~$>$~0.25~mag) as intrinsically red, we correct the subset in the bottom right quadrant (rise~$<$~20~d and \ColPeak~$>$~0.25~mag). 

To ensure that our method is accurately identifying events affected by dust extinction (requiring host-extinction corrections) we inspected the host galaxy environments of the `dusty' SNe (22 Type II, 8 Type IIn and 11 Type IIb after 18.5~mag cut for completeness, see Table~\ref{tab:dust_ext_sample} for a summary of properties). We examined these environments for indicators that could explain the reddening, such as substantial dust content, location within dense spiral arms, the galactic bulge or edge-on or highly inclined host galaxy orientations. We see that these events are predominantly located in regions associated with significant dust content -- specifically within their host galactic plane or star-forming regions. These SNe also show a persistent red colour throughout their rise phase, consistent with dust extinction rather than intrinsic colour variation. This environmental association, combined with their photometric evolution, suggests that their red appearance stems from host galaxy extinction rather than intrinsic properties. We perform our later analysis both with and without host extinction corrections, finding no significant differences in our primary results.

One event in particular, ZTF18acbwaxk/SN~2018hna \citep[][]{Singh_2019b, Thevenot_2020,Tinyanont_2021, Maund_2021, Sit_2023, Xiang_2023}, is both red and rises slowly with similar light curve morphology and timescales to SN~1987A. This SN appears in a face-on dwarf galaxy, the minimal expected host extinction suggests its red colour is intrinsic -- for comparison, SN~1987A was $g-r \gtrsim$~0.4~mag at peak. We interpret this as an intrinsically red event and do not apply a host-correction.

We perform a correction for host extinction with the colour at peak, \ColPeak, for this subset of SNe described above. Relative to V-band ($A_V$ = 1~mag) and based on the assumption of a standard Milky Way $R_V$ = 3.1~mag \citep[e.g.,][]{MW_RV_Cardelli_1989} reddening law (as implemented in \texttt{python extinction} package). We assume an extinction of $A_g/A_V$~=~1.19~mag, $A_r/A_V$~=~0.84~mag and $A_i/A_V$~=~0.61~mag -- we ignore the effect of redshift here. For a ZTF~$g$ extinction relative to \ColPeak\ reddening of 1~mag, we find an extinction of $A_g = 3.37$~\ColPeak~mag. For our correction to ZTF~$g$ luminosities, we multiply the \ColPeak\ by 3.37 and apply this to ZTF~$g$ magnitudes of all Types. From this point on, we carry through this extinction correction for light curves where \ColPeak~$>$~0.25 and \Trise~$\leq$~20~d. However, the volumetric-correction weighting applied in Section~\cref{sec:LC VCD} is based on the uncorrected peak magnitude. 

To minimise over-weighting low-luminosity, nearby events in our magnitude-limited analysis, we adopt more precise luminosity distances from recent literature for events within $d_{l,max} \leq 50$~Mpc (Table~\ref{tab:close_events}).

\begin{table}
\begin{threeparttable}
\centering
    \begin{tabular}{cccc}
    \hline
     ZTF & TNS ID & $d_l$  [Mpc] & Reference \\
    \hline \hline
    ZTF18acbwaxk & 2018hna  & 12.82 $\pm$ 2.02 & \tnote{1}    \\
    ZTF18abwkrbl & 2018gjx  & 35.00 $\pm$ 5.00 & \tnote{2}  \\
    ZTF19abwztsb & 2019pjs  & 40.10            & \tnote{3,4} \\
    ZTF19acfejbj & 2019sox  & 48.78            &  \tnote{5}           \\ 
    ZTF20acrzwvx & 2020aatb & 40.50 $\pm$ 5.11 & \tnote{4,6} \\
    ZTF20acwqqjs & 2020acat & 35.30 $\pm$ 4.40 & \tnote{7}  \\
    ZTF20aapchqy & 2020cxd  & 22.00 $\pm$ 3.00 & \tnote{8}   \\
    ZTF20aatzhhl & 2020fqv  & 17.30 $\pm$ 3.60 & \tnote{9,10} \\
    ZTF20abeohfn & 2020mjm  & 28.30 $\pm$ 2.00 & \tnote{3,4} \\
    ZTF21aadoizf & 2021aai  & 20.90 $\pm$ 1.90 & \tnote{11}   \\
    ZTF21aaqgmjt & 2021gmj  & 13.10 $\pm$ 2.00 & \tnote{12}   \\
    ZTF21abvcxel & 2021wvw  & 44.12            &  \tnote{5}            \\
    ZTF22abtjefa & 2022aaad & 11.10            & \tnote{5}           \\
    ZTF22abtspsw & 2022aagp & 21.83 $\pm$ 3.00 & \tnote{4}   \\
    ZTF22aaotgrc & 2022ngb  & 43.07            & \tnote{4}           \\
    ZTF22aauurbv & 2022pgf  & 36.63 $\pm$ 2.60 & \tnote{4}   \\
    ZTF22abfzdkz & 2022uop  & 44.65            &  \tnote{5}            \\
    ZTF22abnejmu & 2022ycs  & 44.65            &  \tnote{5}            \\ \hline
    \end{tabular}
%\begin{tablenotes}\footnotesize 
\tnote{1}~\citet{Singh_2019b};  \tnote{2}~\citet{Prentice_2020};  \tnote{3}~\citet{Strotjohann_2021};  \tnote{4}~\citet{NED_1991};  \tnote{5}~NASA/IPAC Extragalactic Database~\citet{NED_1991};  \tnote{6}~\citet{Theureau_2005}; \tnote{7}~\citet{Medler_2022}; \tnote{8}~\citet{Yang_2021};  \tnote{9}~\citet{Tinyanont_2022};  \tnote{10}~\citet{Theureau_2007};  \tnote{11}~\citet{Valerin_2022};  \tnote{12}~\citet{Zimmerman_2021z}

%\end{tablenotes}
\end{threeparttable}
\caption{Sources of improved luminosity distance for events closer than 50~Mpc to improve the luminosity weighted volumetric-corrections of close events, particularly those at the extremes -- close and faint.}
    \label{tab:close_events}
\end{table}

\subsection{Rise Time}
\label{sec:Rise Times}
We focus on the rise time and how it relates to the CSM as recent studies \citep[e.g.,][]{Morozova_2016, Yaron_2017, Morozova_2018, Pearson_2022, Tinyanont_2022, Bruch_2021, Bruch_2022, Jacobson_2023, Hosseinzadeh_2018, Hosseinzadeh_2023, Irani_2023} have shown substantial evidence of the sensitive nature of the rise time to the progenitor properties and pre-explosion conditions (e.g., \Mcsm, \Rcsm, CSM density and progenitor radius). 

Traditional rise time measurements -- from explosion to peak -- require well-constrained explosion epochs through deep non-detections immediately before explosion and good sampling of the early rise. These observational constraints significantly limit the number of events for which explosion epochs can be reliably determined, though this is partially alleviated by fitting the early light curve with power-law or polynomial functions to approximate the explosion time \citep{Gall_2015, Gonz_2015, Bruch_2021, Bruch_2022}. To analyse our heterogeneously sampled dataset, we instead adopt a mathematically defined rise time that does not depend on explosion epoch constraints.

\subsubsection{Rise Distribution}
\label{sec:DA - Rise Distribution}

We define rise time (\Trise) as the interval from 25\% to 75\% of peak flux in ZTF~$g$ band. This definition offers several advantages: it's robust against sampling errors and low S/N measurements; captures the epoch where CSM signatures are strongest; and avoids plateau phases where other physical processes dominate. We select ZTF~$g$ band for its sensitivity to CSM interaction signatures in the blue optical \citep[e.g.,][]{Groh_2014A, Gal-Yam_2014, Yaron_2017, Kulkarni_2023} -- a choice validated by recent modelling of SN~2023ixf, where $g$ band provided optimal fits (minimum $\chi^2$ per progenitor model) across all bands \citep{Moriya_2024}. 

Events peaking at 18.5~mag have 25\% peak flux at 20th~mag (e.g., $+$~1.5~mag), which does not typically exceed the ZTF detection limit under favourable conditions but may fall below the threshold during sub-optimal conditions (e.g., bright time). Thus, this ensures we capture a substantial portion of the rise whilst remaining sensitive to fainter objects. Fig.~\ref{fig:forced_lc} shows our measurement methodology.

For consistent measurements on the rise, we place additional constraints on the rising light curve coverage to minimise the impact large gaps in coverage have on the GP modelling. For the measurement, we required at least one observation (detection or non-detection) in each of the following regions:

\begin{enumerate}[label=\Alph*.]
\centering
    
    \item $T_{g,75} - 4\ \ \leq T \leq T_{g,75} + 4\ $ [d]
    \item $T_{g,75} - 12 \leq T \leq T_{g,75} - 4\ $ [d]
    \item $T_{g,75} - 20 \leq T \leq T_{g,75} - 12$ [d]

\end{enumerate}

\begin{figure*}   
\centering
\includegraphics[width=1\textwidth]{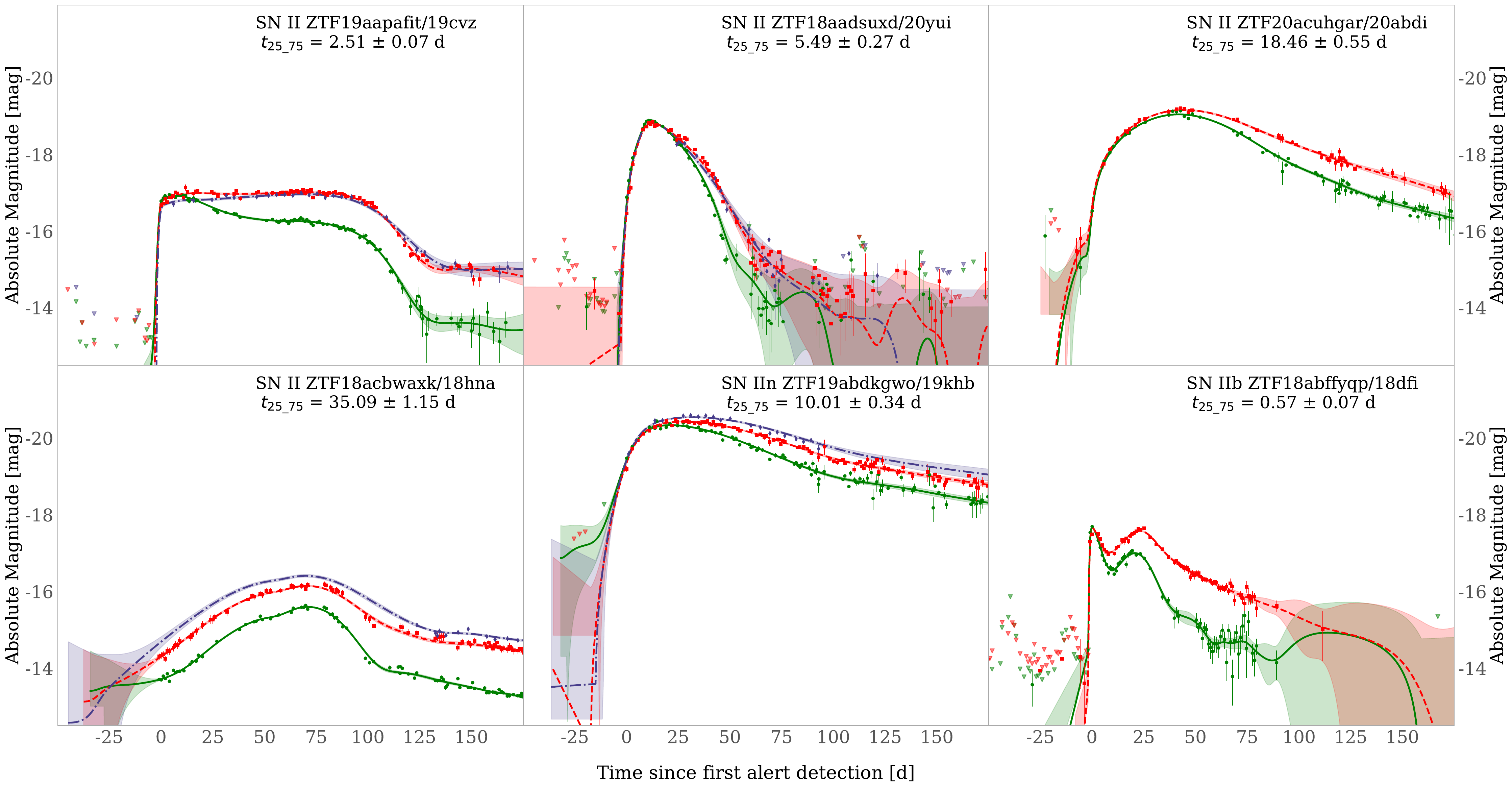}
\caption{ZTF~$gri$ forced photometry GPR light curve panel showing the diversity of Type II SNe and the ZTF~$g$ \Trise\ rise times.}
\label{fig:LC_Diversity}
\end{figure*}

\noindent Here, $T_{g,75}$ marks when the flux reaches 75\% of its peak during the rising phase in ZTF~$g$. To measure rise times in the ZTF~$g$ band, we begin by examining only the $g$ band data. When ZTF~$g$ band coverage is incomplete (missing regions B and/or C), we expand our analysis to include ZTF~$r$ and $i$ band data, evaluating them relative to $T_{g,75}$. In cases where ZTF~$g$ band data only covers region A, we classify the measurement as an upper limit if the ZTF~$r$ and $i$ bands cover either regions A and B together or regions A and C together. For early rise measurements ($T_{b,25}$, time at 25\% peak flux relative to ZTF~$g$), we require coverage of only region A. This requirement is typically met automatically through our $T_{g,75}$ criteria since most objects complete their rise within 20~d. The diversity of light curve morphologies is shown in Fig.~\ref{fig:LC_Diversity}. 

From our sample of SNe with forced photometry (as of December 31st 2023), we identified 1323 CCSNe that meet the quality criteria established by \citet{Perley_2020a}. Of these, 981 are hydrogen-rich CCSNe (including regular Type II/IIP, IIn, IIb, and H-rich superluminous SNe; SLSNe), while the remaining 342 are classified as stripped-envelope SNe (SESNe). The other SNe in the quality sample, 4009, are Type Ia SNe (see Table~\ref{tab:Sample_Numbers}). Our Type IIb sample is relatively small due to classification challenges inherent to this subtype. The limited spectral resolution of the SEDM makes it difficult to identify the characteristic evolution -- specifically, the disappearance of H features and the emergence of often weak He lines in later spectra. Comprehensive classification typically requires multiple spectra obtained at different epochs, which is not always feasible. While this likely results in some incompleteness in our Type IIb sample, the impact on our overall study conclusions is minimal, as these events represent a small fraction of the H-rich SN population. We discuss and quantify this systematic impact of this in \cref{app:Systematic Misclassification}.

\begin{table}
    \centering
    \begin{tabular}{ccccc}
    \hline
        Type & Total & BTS Cut & Rise Cut & $m_{peak} \leq$~18.5~mag\\ \hline \hline
        \multicolumn{5}{c}{\multirow{2}{*}{\textit{H-rich CCSNe}}}  \\ 
        &&&& \\ \hline \hline
        II/IIP & 1387 & 716 & 479 & 377\\
        IIn    & 241 & 145 & 94  & 71\\
        IIb    & 136 & 97  & 50  & 35 \\
        SL II   & 38 & 23  & 16  & 10 \\ \hline
        Tot. & 1802 & 981 & 639 & 493 \\\hline \hline
        \multicolumn{5}{c}{\multirow{2}{*}{\textit{H-poor CCSNe}}}  \\ 
        &&&& \\ \hline\hline
        Ib/c &  363 & 223 & - &-\\
        Ic-BL    & 63 & 45  & -  &-\\
        Ibn   & 36 & 27 & -  &-\\
        Icn & 1 & 1 &- &-\\
        SL I & 75 & 46 & -&-\\ \hline
        Tot. & 538 & 342 &- &-\\ \hline \hline
        \multicolumn{5}{c}{\multirow{2}{*}{\textit{Type Ia SNe}}}  \\ 
        &&&& \\ \hline \hline
        SN Ia & 6329 & 4009 &- &-\\
        \hline
    \end{tabular}
    \caption{Figures showing the number of SNe in the Bright Transient Survey quality sample \citep[see][]{Perley_2020a} between the 1st of May 2018 and the 31st of December 2023, in addition to the number of SNe that make up the final sample of this work after applying rise time constraints to ensure an adequate sampling of the rise and a $m_{peak}$ cut for volumetric weighting -- $m_{peak}$ can be in any ZTF band.}
    \label{tab:Sample_Numbers}
\end{table}

\begin{figure*}
    \centering
    % \hspace*{-1.45cm} 
    \includegraphics[width=1.0\textwidth]{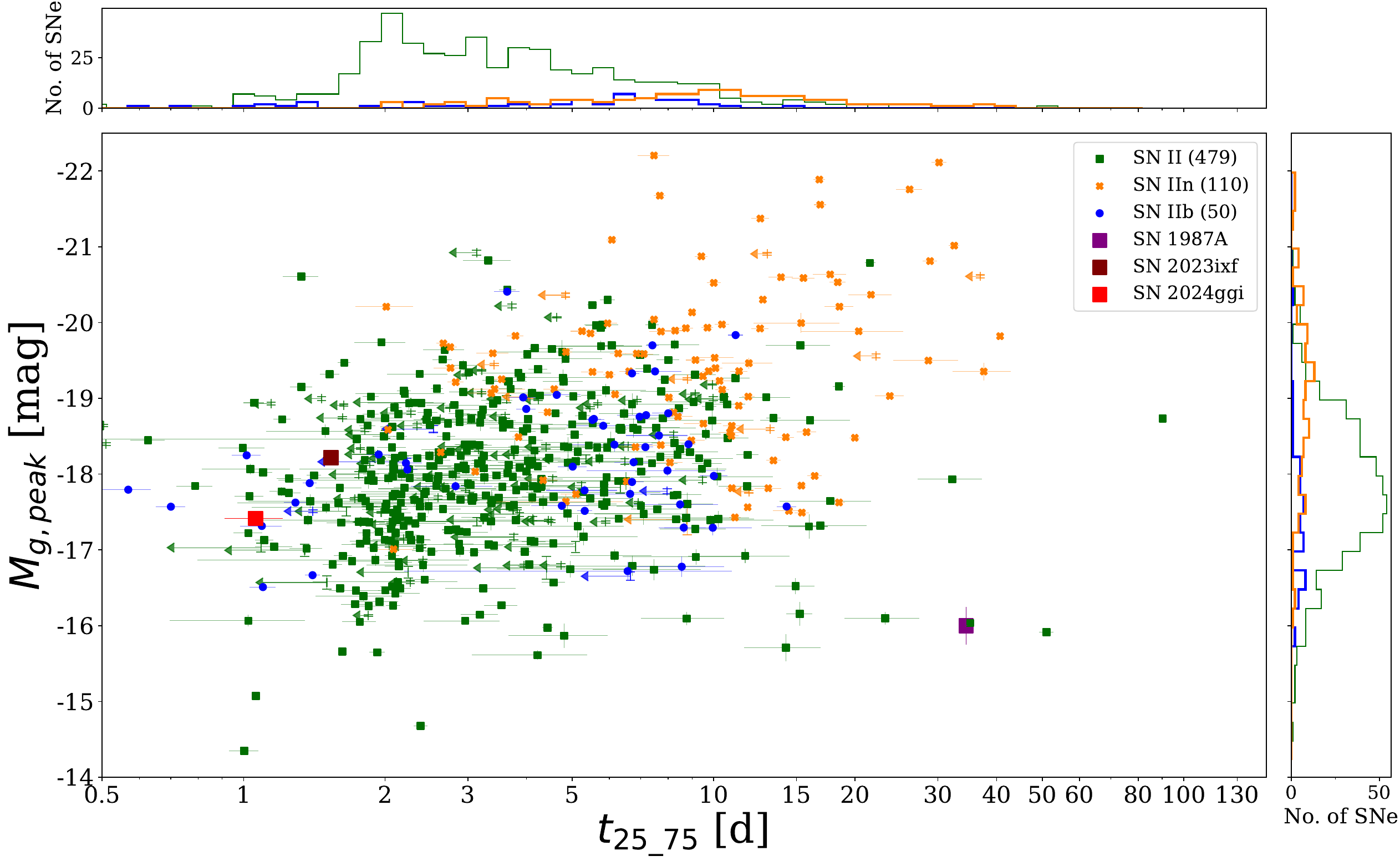}
    \caption{Rise time distribution in ZTF~$g$ band for spectroscopically classified SNe II, SNe IIn  and SNe IIb. The purple square is SN~1987A, the brown square is SN~2023ixf and the red square is SN~2024ggi. Arrows represent upper limits on the rise times as they meet only two of the criteria from Section~\cref{sec:DA - Rise Distribution}. These peak magnitudes are corrected for Galactic and host extinction, as described in Section~\cref{sec:Extinction}.}
    \label{fig:rise distribution}
\end{figure*}

Given the coverage constraints and the availability of forced photometry at the time of writing, 639 H-rich SNe make it through our quality cuts, allowing for constraining measurements of the \Trise\ metric to be made -- see Table~\ref{tab:Sample_Numbers} for a breakdown. For this paper, we consider SLSNe II as Type IIn based on the ambiguous boundary between the classes. Additionally, for the 23 SLSNe passing the quality cuts from the BTS and outlined here, we checked their spectra and found obvious narrow lines indicative of Type IIn SNe in all except ZTF19ackzvdp/SN~2019uba which showed slightly broader emission lines \citep[see also][]{Nyholm_2020, Kangas_2022, PPessi_2023}, prompting us to exclude this from the sample completely. Fig.~\ref{fig:rise distribution} is the rise time distribution using \Trise\ rise including comparison events from the literature.

In Fig.~\ref{fig:rise distribution}, we have included the well-studied Type II SNe SN~2023ixf, SN~2024ggi and SN~1987A to determine where amongst the larger population these events lie. For SN~2023ixf\footnote{Data gathered from \url{https://www.wiserep.org/object/23278}} and SN~2024ggi\footnote{Data gathered from \citet{Shrestha_2024}}, we use publicly available data and model the multi-band light curve using the same process as described in Section~\cref{sec:GPR}, and for SN~1987A we use figures from \citet{Schaeffer_1987, Catchpole_1989_87A, Suntzeff_1992_87A, Suntzeff_1997_87A, Fransson_2007_87A} to extract the data using a data extractor\footnote{WebPlotDigitizer: \url{https://apps.automeris.io/wpd/}}. As shown in Fig.~\ref{fig:rise distribution}, Type II SNe occupy a large range in this parameter space, highlighting the large diversity present in H-rich CCSNe. The Type IIn SNe are preferentially more luminous and generally longer rising, attributed to their larger and more dense CSM components driving a prolonged CSM interaction. 

\subsubsection{Rise Time Limitations}
\label{sec:RiseLimits}

A possible bias emerges when measuring rise times similar to or less than the survey cadence of 2~--~3~d. The discrete sampling of the light curve means rise times on these timescales are typically less well-constrained than longer-rising events, where multiple observations sample the rising phase. With this in mind, we tested the predictive power of the GPR method by taking well-sampled light curves (with cadences $\sim$ 2~d or less between peak $-$ 50 and peak $+$ 200~d where the origin is the first alert detection) and resampled the light curves based on the sampling function of 20 events with much worse and more irregular cadences\footnote{Chosen by finding events with an average cadence from peak $- 50$~--~peak + 200~d $\geq$ 5~d.}. Using the actual light curve of the well-observed SN as the `ground truth', we shift it according to the sampling function of another light curve to emulate the `ground truth' light curve being sampled differently.

Applying our standard rise time definition and constraints to these resampled light curves, we confirm that events maintain their classification as fast (\Trise~$\leq$~5~d) or slow (\Trise~$>$~5~d) risers regardless of sampling pattern. This was done to explore the range of a measured rise time based on the sampling function applied to a light curve. We find the range in rise times is increased for shorter rise times compared to longer risers, particularly at \Trise~$\leq 5$~d -- see Fig.~\ref{fig:RiseResample}. For \Trise\ between 1~--~2~d we see a range of $\approx$~0.7~dex, for \Trise\ between 3~--~5~d we see a range of $\approx$~0.4~dex and for \Trise~$> 5$~d there is a range $\approx$~0.2~dex.

Our \Trise\ metric requires consideration for Type IIb SNe, which often show double-peaked light curves due to the SC peak lasting hours to days, followed by a radioactively-powered peak \citep{Chevalier_1992, Richmond_1994, Chevalier_2008}. As we are only concerned with measuring the rise time and peak magnitude for Type IIb and Type IIn and not progenitor properties, this remains an adequate descriptive measurement to characterise these SNe.

\begin{figure*}
    \centering
    % \hspace*{-1.45cm} 
    \includegraphics[width=1\textwidth]{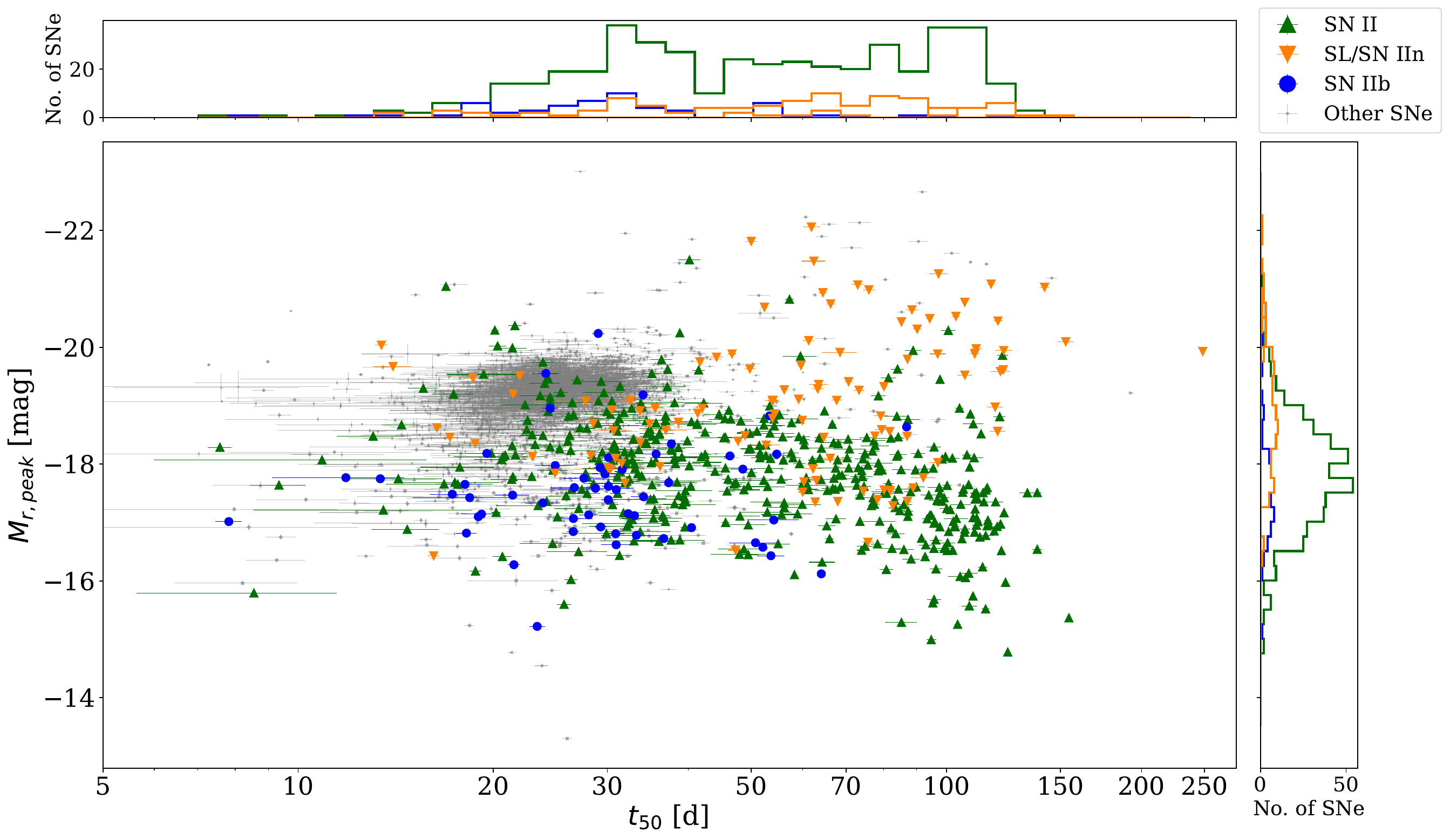}
    \caption{Phase-space diagram showing peak absolute magnitude in ZTF~$r$ ($M_{r,peak}$) vs. rest-frame duration above half-maximum brightness ($t_{50}$) for all SNe with $m_{peak} \leq 19$ mag classified in the BTS through December 31, 2023. Type II SNe are colour-coded by subtype, with other SN classes (e.g., Ia, Ib/c, SLSNe-I) shown in grey for comparison. $M_{r,peak}$ corrected for Galactic extinction only.}
    \label{fig:BTS_SampleDurMag}
\end{figure*}

\section{Population Properties}
\label{sec:Population Properties}

In Fig.~\ref{fig:BTS_SampleDurMag} we show the luminosity - duration phase-space distribution of all classified SNe (Type I and Type II) in the BTS sample (as of December 31st 2023). Fig.~\ref{fig:BTS_SampleDurMag} is included to both show the increased number of events as compared to \citet{Perley_2020a} (Fig.~7a from 1D interpolation of data points)\footnote{\label{BTSHome}Also available to see on the BTS Homepage: \url{https://sites.astro.caltech.edu/ztf/bts/bts.php}} and to demonstrate the reliability of GPR when measuring light curve parameters due to the likeness between the two figures. Classifications and redshifts used in this work are part of the upcoming BTS classification paper (Qin et al., in prep.) for events with a $m_{peak}$~$\leq$~18.5~mag. For events with a $m_{peak}$~$>$~18.5~mag, we used the current TNS classifications stored on an internal BTS catalogue\footref{BTSHome}. We do not expect the provisional nature of these classifications to significantly impact the study.

\subsection{Overall Distribution}
\label{sec:OverallDistributionAnalysis}

Studies of the relationship between rise time and peak luminosity in Type II SNe have yielded conflicting results. Significant correlations have been reported for Type II, IIb, and IIn SNe \citep[e.g.,][]{Gonz_2015, Pessi_2019, Hiramatsu_2024}, suggesting the rise and peak may be intrinsically coupled by their power source. However, other analyses find no significant correlation \citep[e.g.,][]{Gall_2015, Rubin_2016, Valenti_2016, Nyholm_2020}. These discrepant findings likely stem from small sample numbers, limiting their statistical power to comment of population characteristics.

Motivated by the possibility of an enhanced population with fast rises and luminous peaks, either from SBO occurring in the CSM or at the surface of the star, we tested the strength of any existing correlation between \Trise\ and \Mabs. For Type II  SNe, a Spearman rank test between \Trise\ in ZTF~$g$ and \Mabs\ finds a correlation, with a correlation coefficient $\rho = - 0.21$ and p-value $p < 10^{-5}$. The scatter present in the correlation is likely caused by the large diversity of Type II SNe and lack of clear division within Type II SNe (e.g., IIP vs IIL vs 87A-like) as from Fig.~\ref{fig:rise distribution}, it can be seen that Type II SNe occupy both bright and faint, fast and slow regions, see Fig.~\ref{fig:LC_Diversity}. 

In our sample of 110 Type IIn SNe (which includes 16 SNe classified as super-luminous), we find  $\rho = - 0.18$ and p-value $p \sim 0.05$, a correlation both weaker and less significant than the recent findings of \citet{Hiramatsu_2024}. We see great diversity in our Type IIn light curve morphology, which seems to suggest a range of progenitor pathways are possible, with a large range in \Trise\ of $\sim$~2~d to 40~d and \Mabs\ of $-22.20$ to $-17.01$~mag. As Type IIn SNe are well-understood to be CSM-driven \citep[e.g.,][]{Schlegel_1990, Fassia_2000, Smartt_2009, Smith_2010, Taddia_2013, Ransome_2021}, and under this scenario it is expected that both the rise time and luminosity increase with the amount of CSM present (continuing the interaction), this possible correlation is not surprising (Section~\cref{sec:Discussion}). For Type IIb SNe, we find no significant correlation as $p > 0.1$. 

\subsection{Volume Corrected Distributions}
\label{sec:LC VCD}

With a highly complete magnitude-limited survey, we can perform a volume correction such that we can offer a more accurate representation of the true distribution of properties for a given population of SNe. The corrections account for intrinsic observational biases that favour the detection of more luminous events since they can be observed to greater distances in a magnitude-limited survey \citep[Malquist bias;][]{Malmquist_1920}. 

The volumetric correction we apply is according to a $1/V_{max}$ weighting \citep[][]{Schmidt_1968}. Initially, we perform a magnitude cut at $m_{peak}$~=~18.5~mag given that the BTS is $\gtrsim$ 95\% complete at this level. The peak magnitude cut reduces the sample to 377 Type II SNe, 81 Type IIn SNe and 35 Type IIb SNe. $d_{l,max}$ is calculated assuming a limiting magnitude of 18.5~mag, an average Galactic extinction, $A_{g,Gal}$, of 0.19~mag (calculated using the $A_{g,Gal}$ of the sample) and the ZTF~$g$ peak absolute magnitude of the GP model, \Mabs\ -- which is standardised to the rest frame ZTF~$g$ band, corrected only for Milky Way extinction and not host extinction. We calculate $V_{max}$ to be $d_{l,max}^3/(1 + z)^3$ (based on the comoving distance), and use its reciprocal for weighting after normalising the weights to unity. %We show the redshift distributions for the whole sample and those with $m_{peak} \leq$~18.5~mag in Fig.~\ref{fig:RedshiftDist}.
\begin{figure}
    \begin{subfigure}[b]{0.47\textwidth}
         \includegraphics[width=1\textwidth]{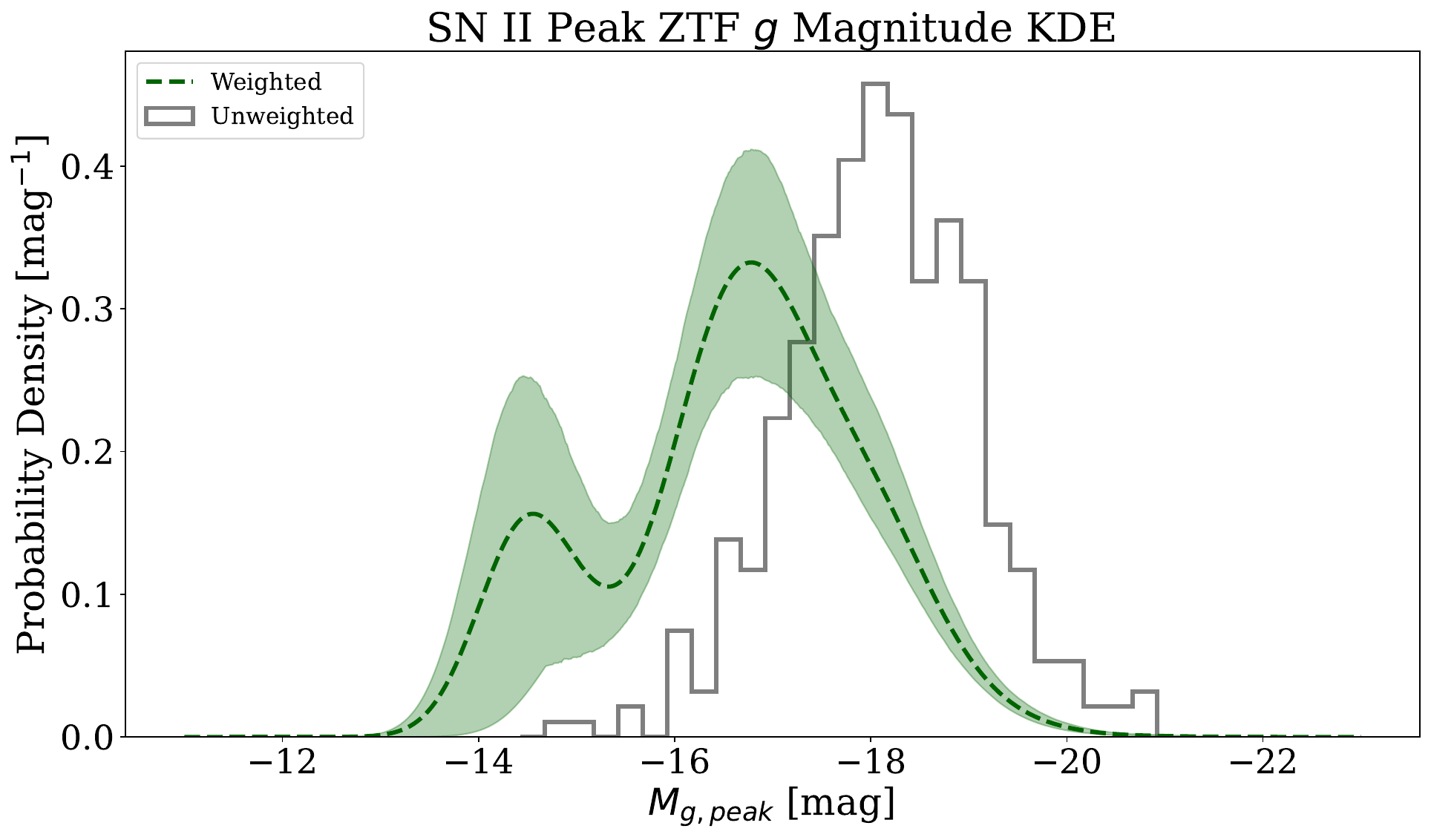}
         \caption{Type II \Mabs\ KDE distribution.}
         \label{fig:II_abs_mag}
\end{subfigure}
\begin{subfigure}[b]{0.47\textwidth}
         \includegraphics[width=1\textwidth]{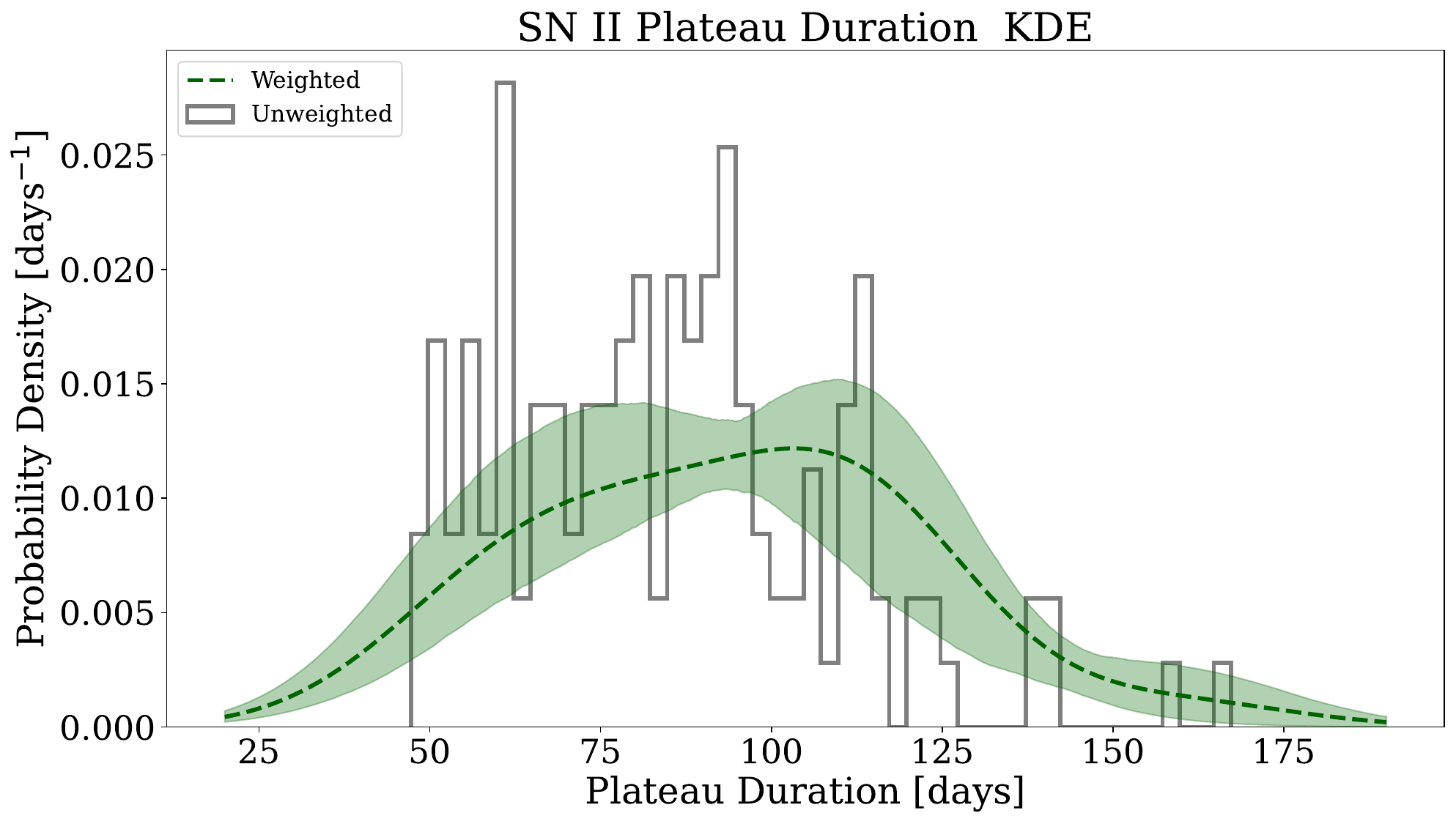}
         \caption{Type II \Tplat\ KDE distribution.}
         \label{fig:II_plat}
\end{subfigure}
\caption{\Mabs\ and \Tplat\ KDE distribution for spectroscopically classified Type II SNe. We plot the weighted KDE distributions in the darker colours (dashed line) and the unweighted histogram in solid black.} 
\end{figure}

\begin{figure}
    \begin{subfigure}[b]{0.47\textwidth}
         \includegraphics[width=1\textwidth]{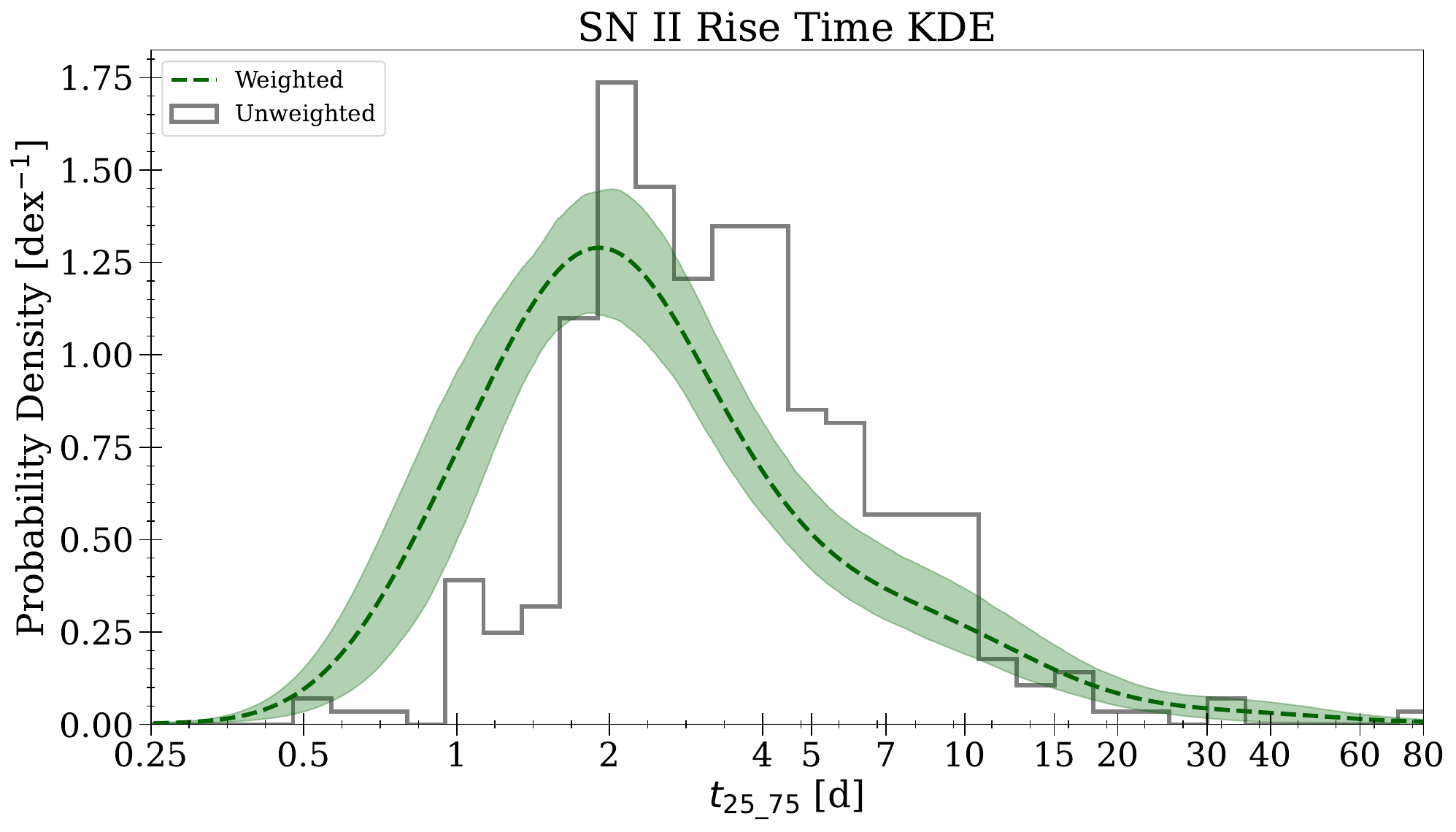}
         \caption{Type II \Trise\ KDE distribution.}
         \label{fig:II_rise}
\end{subfigure}
\begin{subfigure}[b]{0.47\textwidth}
         \includegraphics[width=1\textwidth]{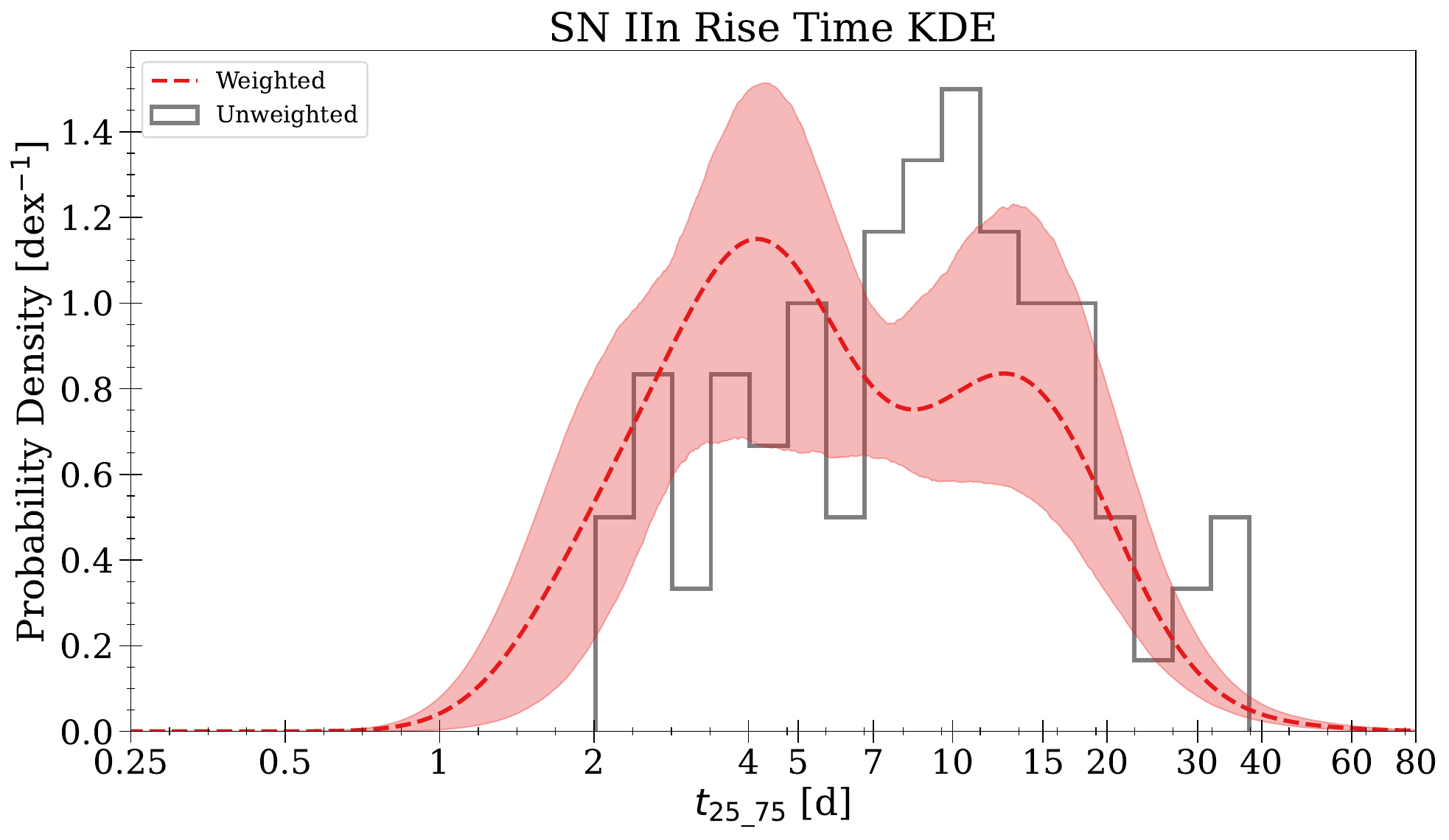}
         \caption{Type IIn \Trise\ KDE distribution.}
         \label{fig:IIn_rise}
\end{subfigure}
\begin{subfigure}[b]{0.47\textwidth}
         \includegraphics[width=1\textwidth]{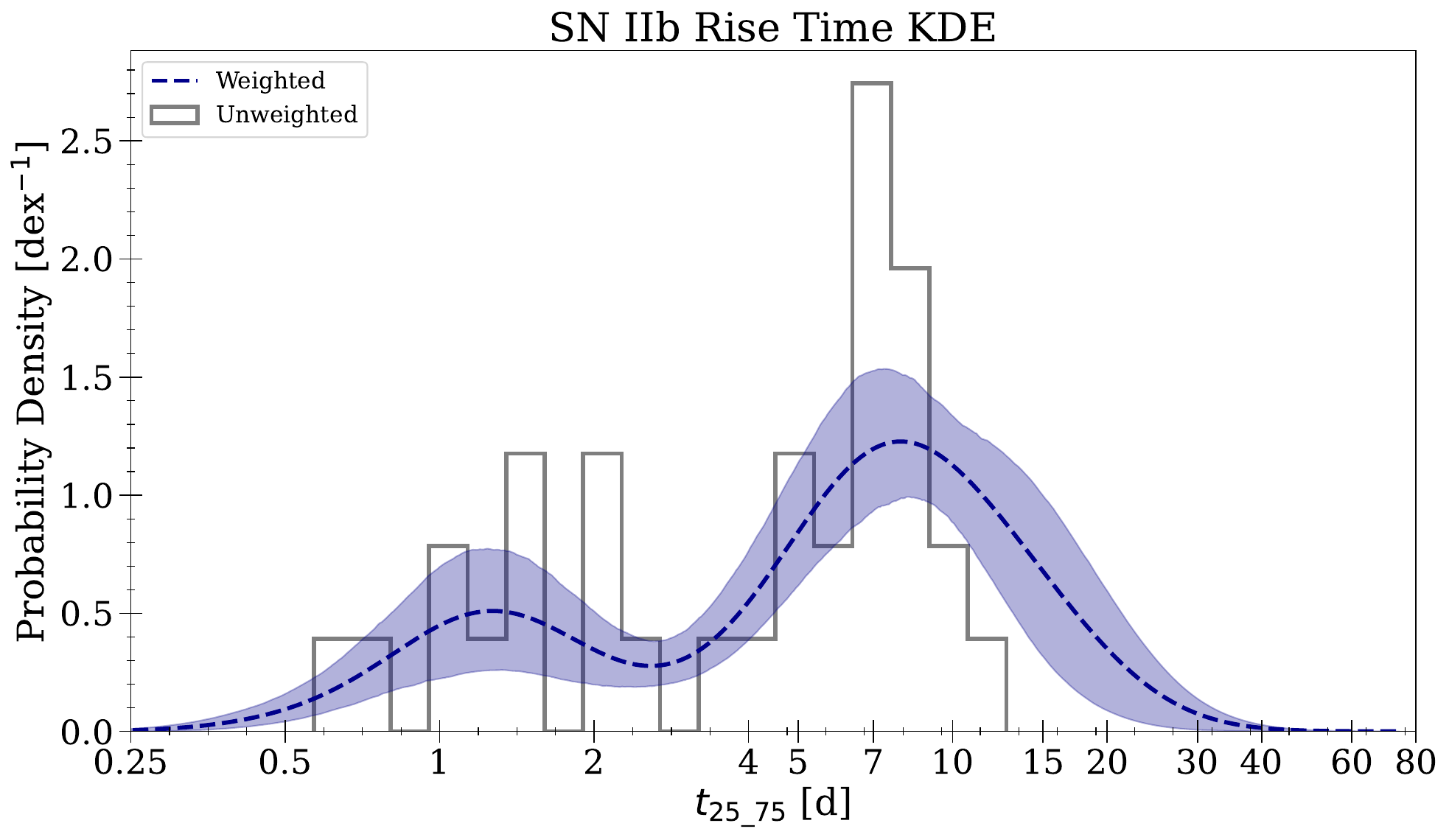}
         \caption{Type IIb \Trise\ KDE distribution.}
         \label{fig:IIb_rise}
\end{subfigure}
\caption{\Trise\ KDE distribution for spectroscopically classified Type II, Type IIn and Type IIb with their associated 80\% CIs. We plot the weighted KDE distributions in the darker colours (dashed line) and the unweighted histogram in solid black. }
\end{figure}

To represent the intrinsic distribution of parameters, we use Kernel Density Estimation (KDE). For each observed value, we generate a normalised Gaussian kernel centred on that real value and weight each Gaussian as $1/V_{max}$. The width (sigma) of each kernel is optimised using cross-validation \citep[e.g.,][]{Wu_1997}. This approach estimates the underlying probability density function of the parameter distribution. We also normalise the KDE by the sum of weights, allowing us to account for Malmquist bias.

We quantify uncertainty in the weighted KDE as a 80\% CI, calculated by bootstrapping our sample with replacement. Similarly, we compute the ECDF with 95\% CI for unweighted distributions and bootstrapped 80\% confidence intervals for weighted distributions.

\subsection{Data Exploration}

\begin{table*}
\centering
%\hspace{-1.7cm}
\begin{tabular}{ccllllcllll}
%\begin{tabular}{cccccccccccc}
\hline
\multirow{2}{*}{Type} & \multicolumn{4}{c}{Weighted} & \multicolumn{4}{c}{Unweighted} \\
%\multirow{2}{*}{Type} & \multicolumn{4}{c}{Weighted} & \multicolumn{4}{c}{Unweighted} & \multirow{2}{*}{Range} &\multirow{2}{*}{No.} \\
 & Mean & 25th\%ile & 50th\%ile  & 75th\%ile  &  Mean & 25th\%ile & 50th\%ile  & 75th\%ile    \\ \hline\hline

& \multicolumn{8}{c}{\multirow{2}{*}{\Trise\ [d]}}   \\ 
 &&&&&&&&&  \\ \hline
II (377)  & $2.48^{+0.30}_{-0.27}$ & $ 1.65^{+0.57}_{-0.11} $ & $2.18^{+0.25}_{-0.23} $ & $ 3.38^{+0.31}_{-0.74} $  & $3.39^{+0.12}_{-0.12}$ & $ 2.13^{+0.07}_{-0.08} $ & $3.21^{+0.11}_{-0.11} $ & $ 5.09^{+0.47}_{-0.64} $   \\ 
IIn (81) & $6.01^{+1.09}_{-0.92}$ & $ 3.41^{+0.66}_{-0.94} $ & $5.59^{+1.44}_{-1.14} $ & $ 11.83^{+2.10}_{-1.71} $  & $8.54^{+0.73}_{-0.67}$ & $ 4.85^{+1.35}_{-1.48} $ & $8.73^{+0.86}_{-0.78} $ & $ 14.32^{+3.06}_{-2.22} $   \\
IIb (35)& $4.97^{+1.27}_{-1.01}$ & $ 2.00^{+0.61}_{-2.68} $ & $6.32^{+1.72}_{-1.35} $ & $ 8.59^{+0.67}_{-1.69} $  & $3.99^{+0.61}_{-0.53}$ & $ 1.79^{+0.69}_{-1.53} $ & $4.87^{+0.93}_{-0.78} $ & $ 7.54^{+0.89}_{-1.03} $  \\ \hline

& \multicolumn{8}{c}{\multirow{2}{*}{\Mabs\ [mag]}}   \\ 
 &&&&&&&&& \\ \hline
 %\vspace{0.001cm} \\
II (377)  &$-16.59 \pm 0.29 $ & $ -16.03^{+0.28}_{-0.39} $ & $-16.71 \pm 0.25 $ & $ -17.44^{+0.22}_{-0.09} $ & $-18.10 \pm 0.05 $ & $ -17.54^{+0.10}_{-0.15} $ &$-18.11 \pm 0.05 $ & $ -18.74^{+0.16}_{-0.08} $ \\  
IIn (81) &$-18.19 \pm 0.19 $ & $ -17.68^{+0.14}_{-0.12} $ & $-18.02 \pm 0.21 $ & $ -18.82^{+0.56}_{-0.43} $ & $-19.35 \pm 0.13 $ & $ -18.49^{+0.56}_{-0.43} $ &$-19.32 \pm 0.13 $ & $ -19.98^{+0.55}_{-0.27} $ \\  
IIb (35) &$-17.79 \pm 0.17 $ & $ -17.45^{+0.12}_{-0.16} $ & $-17.79 \pm 0.19 $ & $ -18.37^{+0.35}_{-0.24} $ & $-18.15 \pm 0.14 $ & $ -17.58^{+0.31}_{-0.27} $ &$-18.09 \pm 0.14 $ & $ -18.76^{+0.46}_{-0.40} $    \\  \hline
%\vspace{0.001cm} \\

& \multicolumn{8}{c}{\multirow{2}{*}{\ColPeak\ [mag]} }  \\ 
 &&&&&&&&& \\ \hline
%\vspace{0.001cm} \\
II (377) & $0.17 \pm 0.05 $ & $ 0.01^{+0.01}_{-0.02} $ & $0.09 \pm 0.03 $ & $ 0.25^{+0.09}_{-0.14} $  & $0.04 \pm 0.01 $ & $ -0.04^{+0.02}_{-0.02} $ & $0.02 \pm 0.01 $ & $ 0.08^{+0.02}_{-0.03} $   \\ 
IIn (81) & $0.21 \pm 0.08 $ & $ 0.00^{+0.01}_{-0.02} $ & $0.10 \pm 0.07 $ & $ 0.49^{+0.32}_{-0.02} $  & $0.11 \pm 0.02 $ & $ -0.01^{+0.01}_{-0.02} $ & $0.08 \pm 0.01 $ & $ 0.16^{+0.06}_{-0.05} $  \\ 
IIb (35)& $0.25 \pm 0.06 $ & $ 0.00^{+0.02}_{-0.04} $ & $0.26 \pm 0.06 $ & $ 0.41^{+0.04}_{-0.04} $  & $0.13 \pm 0.04 $ & $ -0.03^{+0.11}_{-0.04} $ & $0.12 \pm 0.04 $ & $ 0.31^{+0.16}_{-0.07} $  \\ \hline
%\vspace{0.001cm} \\

& \multicolumn{8}{c}{\multirow{2}{*}{\Tplat\ [d]} } \\ 
 &&&&&&&&& \\ \hline
 %\vspace{0.001cm} \\
II (151) &  $93.86 \pm 6.39 $ & $ 71.89^{+10.63}_{-6.87} $ & $ 93.88 \pm 7.96 $ & $ 112.50^{+14.85}_{-1.25} $ & $83.98 \pm 2.11 $ & $ 61.25^{+2.51}_{-5.47} $ & $ 82.29 \pm 2.19 $ & $ 96.45^{+4.75}_{-9.81} $  \\ \hline
%\vspace{0.001cm} \\

\end{tabular}
\caption{Mean and median of the volume corrected KDE for \Trise, ZTF \Mabs, \ColPeak\ and \Tplat\ in the final sample, measured directly using the GPR described in Sections \cref{sec:GPR} and \cref{sec:GPR_LC}. Uncertainties reported here are the 1~$\sigma$ standard deviation on the bootstrapped values.}
\label{tab:risemagminmax}

\end{table*}

After weighting the distributions, we create weighted KDEs for direct light curve properties of Type II SNe \Mabs, \Tplat, Figs. \ref{fig:II_abs_mag}, \ref{fig:II_plat}, and \Trise\ for all classes, Figs. \ref{fig:II_rise}, \ref{fig:IIn_rise} and \ref{fig:IIb_rise}. We extract various statistical properties relating to \Trise, \Mabs, \ColPeak\ and \Tplat\ of each subclass from the KDE distributions and show these in Table~\ref{tab:risemagminmax}, with uncertainties based on the 1~$\sigma$ standard deviations for each quantity.

Our demographic analysis of Type II SNe reveals a median absolute magnitude of \Mabs~=~$-$16.71~$\pm$~0.25~mag and a median rise time of \Trise~=~2.18$_{-0.23}^{+0.25}$~d. From the volume-weighted ECDF, we find that $82^{+11}_{-12}$\% of the population has \Mabs~$\leq$~$-$15~mag, with the first and third quartiles at $-$16.03 and $-$17.44~mag, respectively. Most notably, 84~$\pm$~3\% of the weighted population exhibits remarkably brief rise times ($\leq$~5~d), with first and third quartiles at 1.65~d and 3.38~d. These distributions highlight the significant heterogeneity within the Type II SN population.

\section{Methods II -- Comparison to Simulated Light Curves}
\label{sec:Methods II}

\begin{figure*}
    % \begin{center}
    \centering
    % \hspace{-0.25cm}
    \includegraphics[width=1.0\textwidth]{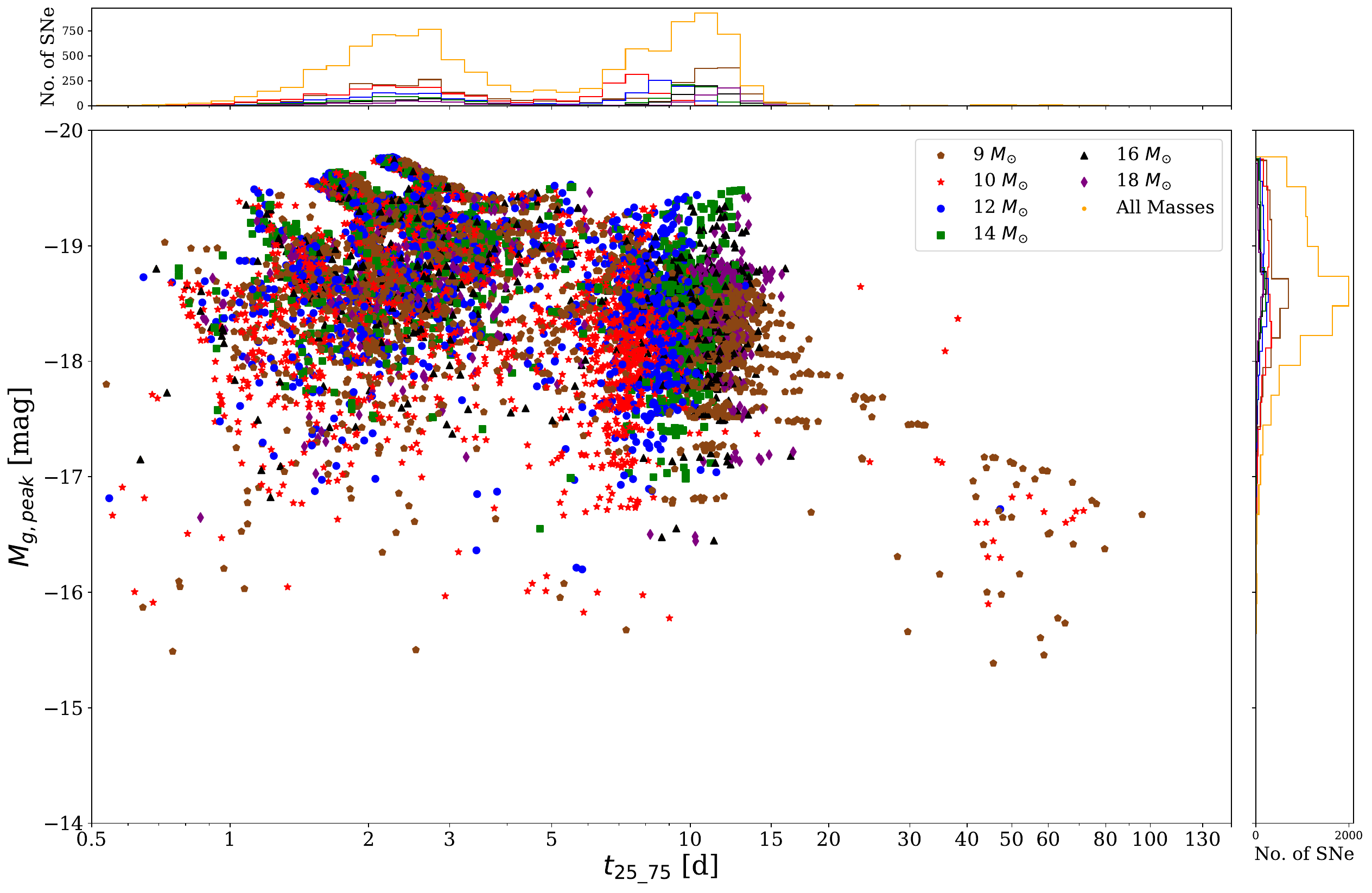}
    \caption{\Trise\ vs \Mabs\ distribution for the theoretical light curve grid from M23, with points drawn from a sample of 10,000 models colour-coded by progenitor mass. Models are weighted by $V_{max}\times M_{ZAMS}^{-2.35}$, combining volume-limited sampling (calculated using a magnitude limit of 18.5~mag) with the Salpeter initial mass function. This weighting scheme reproduces both the observational bias against fainter events and the natural frequency of different progenitor masses.}
    \label{fig:Moriya_absrise_massep}

\end{figure*}

Light curve models are increasingly incorporating CSM or extended stellar envelopes \citep[e.g.,][]{Hillier_2012, Dessart_2015, Dessart_2017, Das_2017, Morozova_2018, Pearson_2022, Tinyanont_2022, Moriya_2023, Morag_2023}. \citet{Morozova_2018} demonstrated that including CSM in \texttt{SNEC} \footnote{Supernova ExplosioN Code.} \citep{Morozova_2015} models significantly improved fits for 20 well-observed, multi-band light curves to estimate progenitor parameters, such as \Rcsm, CSM density, and \Mcsm\ \citep[see Figs. 2 and 3 and Table 2 in][]{Morozova_2018}. Similar conclusions emerge from studies by \citet{Das_2017}, \citet{Bruch_2022}, \citet{Moriya_2023}, \citet{Irani_2023}, and \citet{Jacobson_2024b}, which collectively find that substantial CSM masses near Type II SN progenitors are common and deposited shortly before core-collapse.

While fitting detailed physical models to the entire BTS Type II sample is possible, the computational demands and data heterogeneity make empirically derived relationships more practical for large-scale analysis. Our GP parameter catalogue enables efficient estimation of CSM properties through use of empirical relations. Additionally, we leverage this catalogue to investigate progenitor iron core masses (\Mfe), which significantly influence neutron star formation and properties \citep{Barker_2022a, Barker_2022b}. By applying simulation-based empirical relations to our light curve parameters, we provide constraints on \Mfe\ distributions (see \cref{app:Mfe}).

\subsection{Measuring Theoretical Light Curve Metrics}
\label{sec:Measuring LC and CSM}

To probe the physical origin of the rise time distribution shown in Fig.~\ref{fig:rise distribution}, we leverage our highly complete observational sample and the extensive grid of theoretical light curves from \citet{Moriya_2023} (hereafter M23) using \texttt{STELLA} \citep{Blinnikov_1998, Blinnikov_2000, Blinnikov_2006}. This comparison between observations and models enables us to systematically explore how progenitor and CSM properties shape the observed diversity. The models from M23 are some of the most comprehensive performed to date, as they sample several progenitor zero-age main sequence masses, ($M_{ZAMS}$; 9~--~18~\Msol), \MLR\ ($10^{-5}$~--~$10^{-1}$~\Msol~yr$^{-1}$), RSG wind structure parameter, ($\beta$\footnote{Determined by the efficiency of wind acceleration, for RSGs $\beta >$ 1 \citep{Moriya_2023}.}; 0.5~--~5) and \Rcsm\ ($10^{14}$~--~$10^{15}$~cm) among other progenitor properties (see Table 2 in M23 for more details). The published grid contains over 200,000 models sampled from these parameters and is a base for comparing observed light curves. 

To derive \Mcsm, we first calculate the wind velocity, $v_{wind}$, at \Rcsm\ using the velocity profile from Eq. 2 in \citet{Moriya_2023}, which depends on $\beta$ and the progenitor radius \Rsol. \Mcsm\ is calculated using \Mcsm~=~\MLR~$\times$~\Rcsm~/$v_{wind}$.

We measure various rise times -- including the same \Trise\ described in Section~\cref{sec:Rise Times} -- directly from the M23 ZTF bandpass light curves, in addition to absolute magnitudes (at peak and various $N$ days after peak), magnitude decline rates and colours, e.g., \ColPeak\ -- Fig.~\ref{fig:forced_lc} shows example measurements using ZTF SN light curves. 

With the empirically measured light curve parameters from M23, we create a similar luminosity-rise distribution plot\footnote{We also included unpublished 9 and 10~\Msol\ mass progenitor models with lower and higher explosion energies than in M23.} of \Trise\ vs \Mabs\ in ZTF~$g$ -- see Fig.~\ref{fig:rise distribution}. We apply a probabilistic weighting of $V_{max}\times M_{ZAMS}^{-2.35}$ to the M23 models to mimic the combined effects of the initial mass function \citep[IMF;][]{Salpeter_1955} and Malmquist bias on an observed sample, and draw 10,000 events to show in Fig.~\ref{fig:Moriya_absrise_massep}

In Fig.~\ref{fig:Moriya_absrise_massep}, we see a clear bi-modality, suggesting the dichotomy seen in observations is reflecting the transition between purely shock cooling (SC) dominated rises and rises dominated by the interaction heating from SBO shocking the CSM \citep[e.g.,][]{Irani_2023, Jacobson_2024b}. Correlation tests between progenitor-SN parameters confirm this dichotomy stems from significant relationships between CSM properties and light curve observables, as well as between different CSM parameters\footnote{Correlations between CSM parameters likely reflect physics pre-defined in the simulations.}. This is further evidenced by the colour gradient seen when we apply a colour map based on \MLR\  or \Rcsm\ -- see \cref{app:MoriyaCMAP} for further details. 

Two distinct populations emerge in the theoretical models: fast risers (\Trise~$\leq$~5~d) with moderately more luminous peaks, possessing higher \Mcsm\ and smaller \Rcsm\ (e.g., confined and dense); and slower risers (\Trise~$>$~5~d) with overall less luminous peaks, less massive \Mcsm\ and larger \Rcsm\ (e.g., less confined and less dense). Within the slower population, the most luminous events still require substantial \Mcsm, suggesting CSM mass remains a key driver of peak luminosity. This bi-modality emerges naturally from the underlying physics rather than parameter choices, hinting at fundamental differences in mass-loss mechanisms.

For slower-rising events, the correlation between \Trise\ and CSM parameters lessens, giving way to a stronger dependence on progenitor mass ($M_{ZAMS}$), which serves as a proxy for progenitor radius. This transition reflects both the increasing dominance of light travel time and diffusion processes in more massive, extended progenitors, as well as the slower cooling of the shocked envelope \citep[e.g.,][]{Nakar_2010, Ofek_2010, Tominaga_2011, Morozova_2016, Irani_2023}. Notably, for events with longer rise times ($\geq$~5~d), the progenitor radius influences rise time by affecting the SBO time and location \citep{Chevalier_2011, Moriya_2011, Svirski_2012, Gonz_2015, Morozova_2016}, though the light curve evolution is still predominantly shaped by CSM interaction rather than envelope cooling, depending critically on the CSM density profile \citep{Moriya_2023, Irani_2023}.

Notably, the contrast between the smooth distribution in our observational data (Fig.~\ref{fig:rise distribution}) and the distinct bi-modality in theoretical models (Fig.~\ref{fig:Moriya_absrise_massep}) implies that there are physical process that lead to restrictive prior distributions on progenitor properties, deviating notably from the exploratory, uniform parameter sampling employed in M23 for physical progenitor parameters. 

\begin{figure*}%\ContinuedFloat
\centering
    \includegraphics[width=1\textwidth]{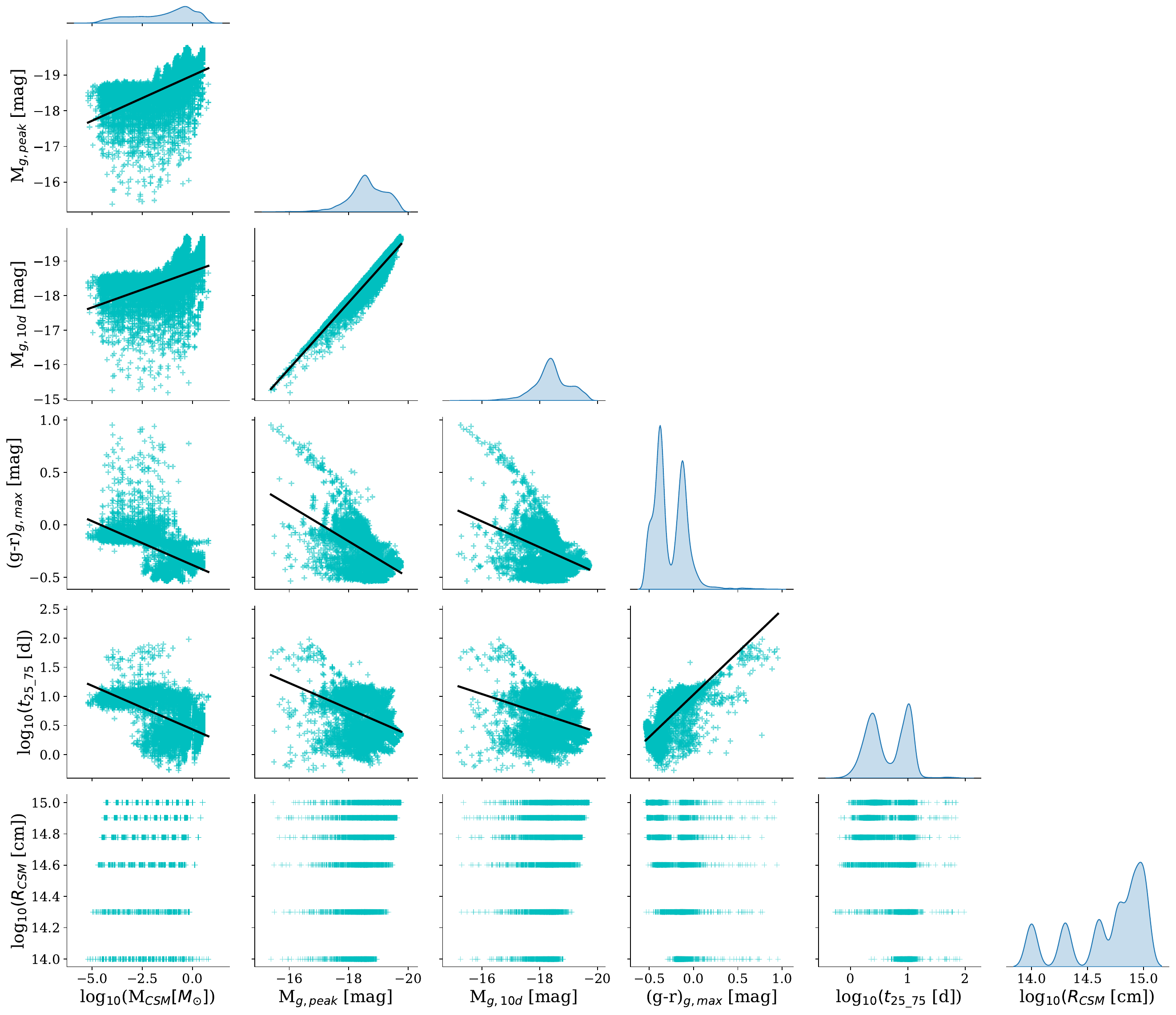}
     \caption{Corner plot showing the relationship between M23 theoretical light curve parameters measured in this work, \Trise, \Mabs, \ColPeak\ and \Mten\ to the \Mcsm\ and \Rcsm\ values returned in the modelling. The solid black line is a first-order polynomial fit to the data.}
     \label{fig:MoriyaCorner}
\end{figure*}

\begin{table*}
%\centering
%\hspace*{-1cm}
\begin{tabular}{cccccccccc}
\hline
\multirow{2}{*}{Data Regime}& \multirow{2}{*}{Formula} & \multirow{2}{*}{Order} & \multirow{2}{*}{D.o.F} & \multicolumn{3}{c}{\Mcsm} & \multicolumn{3}{c}{\Rcsm} \\ 
% \hline \hline
& & & & $R^2$ & BIC & RMS [dex] & $R^2$ & BIC & RMS [dex]\\ \hline \hline
All & Eq. \ref{eqn:polyorder1} & 1 &298980 & 0.47 &  839466 & 0.30 &  0.14&  175163 & 0.12  \\ 
All & Eq. \ref{eqn:polyorder2} & 2 &298970 & 0.57 &  777999 & 0.28 & 0.23&  141781 & 0.12  \\ 
All & Eq. \ref{eqn:polyorder3} & 3 &298950 & 0.62 &  736557 & 0.26 & 0.28&  121588 & 0.11  \\ \hline
$\leq 5$ d & Eq. \ref{eqn:polyorder1} & 1 &112263 & 0.45 &  210328 & 0.21   & 0.47 &  -89654 & 0.07  \\ 
$\leq 5$ d & Eq. \ref{eqn:polyorder2} & 2 &112253 & 0.52 &  193260 & 0.20   & 0.53 &  -103416 & 0.06  \\ 
$\leq 5$ d & Eq. \ref{eqn:polyorder3} & 3 &112233 & 0.58 &  180362 & 0.19   & 0.55 &  -109371 & 0.06  \\ \hline
$>$ 5 d & Eq. \ref{eqn:polyorder1} & 1 &186711 & 0.29 &  536827 & 0.31   & 0.04 &  146335 & 0.13  \\ 
$>$ 5 d & Eq. \ref{eqn:polyorder2} & 2 &186701 & 0.41 &  502875 & 0.29   & 0.07 &  138972 & 0.13  \\ 
$>$ 5 d & Eq. \ref{eqn:polyorder3} & 3 &186681 & 0.48 &  480037 & 0.27   & 0.12 &  129340 & 0.13  \\ \hline

\end{tabular}
\caption{Comparing the performance over different orders of polynomials used in the multiple regression to determine the relations between observed features and CSM properties, \Mcsm\ and \Rcsm, using models from M23. }
\label{tab:BIC_comparisons}
\end{table*}

\section{Physical Progenitor Property Inference Analysis}
\label{sec:CSM Analysis}

To quantify the percentage of RSGs that are surrounded by significant components of CSM at the time of core-collapse, we use the early light curve to infer the presence and properties of CSM. As the rise of Type II SNe is highly sensitive to the CSM parameters and progenitor radius \citep[e.g.,][]{Morozova_2016, Morozova_2018, Pearson_2022, Tinyanont_2022, Moriya_2023, Irani_2023}, we can estimate and place constraints on these properties from photometry alone. The high cadence and good coverage on the rise specifically, combined with the high completeness of the BTS lends itself well to such a detailed study.

\subsection{Defining Light Curve - CSM Relations}
\label{sec:CSMPolyReg}

We examined relationships between observational and progenitor parameters by performing systematic correlation tests across the M23 model grid. Fig. \ref{fig:MoriyaCorner} reveals significant pairwise correlations between progenitor properties (e.g., \Mcsm\ and \Rcsm) and the observational parameters measured in this work (\Mabs\ and \Trise), providing a statistical foundation for our subsequent parameter estimation.

When we analyse simple two-variable analysis, e.g., \Mcsm~=~$f\left(\text{\Mabs},\ \text{\Trise}\right)$ the relations show substantial scatter, indicating that these parameters alone cannot capture the complex CSM-ejecta interaction physics. To better characterise the evolution, we decompose the rise time into: 20~--~60\%, \Ttwosix, and 60~--~90\%, \Tsixnine, or 20~--~50\%, \Ttwofive, and 50~--~80\%, \Tfiveeight. This approach provides additional diagnostics through the shape of the rise. Additionally, we include ZTF~$g-r$ colour at ZTF~$g$ peak flux, \ColPeak, and the ZTF~$g$ absolute magnitude at 10~d post ZTF~$g$ peak, \Mten, to account for any contribution to the early plateau or post-peak behaviour by the CSM (or lack of). We perform multiple regression analysis, attempting to express \Mcsm\ empirically as \Mcsm~=~$f\left(\text{\Mabs},\ \text{\Ttwosix},\ \text{\Tsixnine},\ \text{\ColPeak},\ \text{\Mten}\right)$ and a similar expression for \Rcsm. %From Fig.~\ref{fig:MoriyaCorner}, it appears that \Mabs\ and \ColPeak\ are most strongly correlated to the \Mcsm\ and offer the strongest diagnostic for this parameter in addition to a correlation between \Trise\ and \Mcsm. %The predicted \Mcsm are all within 1-2$\sigma$ of the real mass and we can attribute some dispersion in the relations to the parameter range in M23.

\begin{figure*}%\ContinuedFloat
\centering
    \includegraphics[width=1\textwidth]{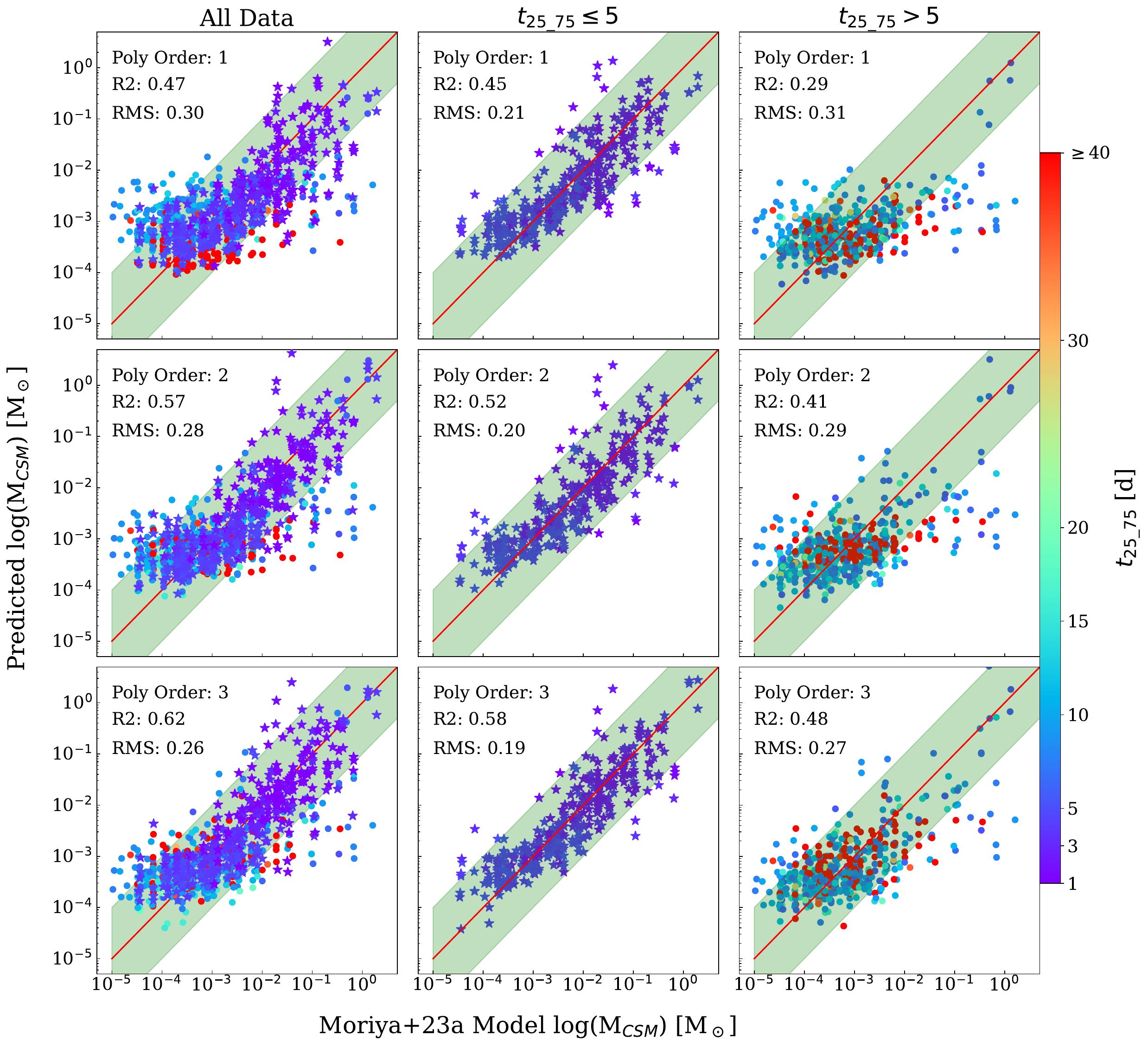}
     \caption{Multivariate analysis results of the predicted \Mcsm\ mass (y-axis) vs. the M23 \Mcsm\ mass (x-axis). The top, middle and bottom rows are polynomial orders 1, 2 and 3 respectively. The first column contains all the data and stars are those with \Trise~$\leq$~5~d with the 2nd and 3rd rows containing only data with \Trise~$\leq$~5~d and \Trise~$>$~5~d to show how the correlations predictive power decreases significantly with \Trise~$\geq$~5~d. The diagonal red line is the 1:1 line with the green shaded region showing 1 order of magnitude above and below. 1,000 models were used in the plot to avoid overcrowding.}
     \label{fig:MoriyaPolyGridMass}
\end{figure*}

We analyse the relationship between CSM parameters and observables using multivariate polynomial regression, implemented via the \texttt{statsmodels} \texttt{ols} package in \texttt{python}. For \Mcsm, we achieve $R^2 > 0.5$ across most polynomial orders, indicating robust correlations. The relationships with \Rcsm\ exhibit somewhat weaker but still significant correlations, as detailed in Table~\ref{tab:BIC_comparisons}. Our goal is to develop reliable predictive relationships that enable rapid estimation of physical parameters (\Mcsm\ and \Rcsm) from observable quantities such as \Trise\ and \Mabs.

The performance of each polynomial order is analysed using the Bayesian Information Criterion (BIC; Table~\ref{tab:BIC_comparisons}). We split the correlation testing into \Trise~$\leq$~5~d and $>$~5~d as the correlation appears strongest for the fast-rising models \Trise~$\leq$~5~d and under-predicts \Mcsm\ for several long-rising \Trise~$>$~5~d models by up to 1~--~2 orders of magnitude -- see Fig.~\ref{fig:MoriyaPolyGridMass}. This likely reflects the physics of CSM interaction: confined, dense CSM shells enable a rapid and efficient conversion of kinetic energy to radiation, producing brief, bright emission with fast rise times \citep[e.g.,][]{Moriya_2011, Chevalier_2012, Tinyanont_2022, Pearson_2022, Li_2023}. In contrast, longer rise times can result from either more extended CSM configurations where energy is released more gradually, or from different physical processes entirely, weakening the direct correlation between \Mcsm\ and light curve properties \citep[e.g.,][]{Moriya_2023b, Jacobson_2024b}.

The systematic under-prediction of \Mcsm\ for models with higher masses and longer rise times motivated our inclusion of post-peak magnitude as an additional predictor variable. For \Rcsm, the strongest correlations are consistently found in the subset of models with \Trise~$\leq$~5~d. 

%---- explore that the M23 model is uniform and not representative of IMF, possible consequences is that they no. high mass progenitors is largely over-predicted and might affect correlation

We restrict polynomial orders to three or less to avoid introducing unphysical complexity into the model. For \Mcsm, the reduction in residuals and increasing $R^2$ from first to third order indicates improved model performance, with third-order polynomials providing the best balance between model complexity and fit quality across both data regimes (\Trise~$\leq$~5~d and $>$~5~d), highlighting the importance of curvature terms in the relation. To better visualise the scatter in these correlations, we add 10\% random scatter to the plotted \Rcsm\ values in Fig.~\ref{fig:RadiusMoriyaPolyGrid}.

\Rcsm\ measurements become most reliable in regimes where \Mcsm\ is large enough to influence observables such as \Mabs\ and \Trise\ significantly (e.g., make \Mabs\ brighter or \Trise\ shorter). We determine the critical value -- below which the CSM does not notably influence the early light curve -- of \Mcsm\ by varying the physical parameters \MLR, \Rcsm, and $\beta$ independently (noting that \Mcsm\ is a function of these three variables within the model). Below this minimum \Mcsm, the CSM will become too diffuse to meaningfully influence the early light curve evolution, effectively transitioning to a regime where CSM interaction is negligible. This physical expectation is reinforced by the distinct bi-modality observed in M23's theoretical models, 

Fig.~\ref{fig:Moriya_absrise_massep}, which reveals these two distinct populations: one where CSM properties strongly correlate with observables (\Mabs\ and \Trise); and another showing no clear correlation, indicating negligible CSM influence. Our analysis reveals an observational transition at \Mcsm~$\approx 10^{-2.5}$~\Msol, below which we cannot detect significant changes in observed parameters (\Mabs\ and \Trise) -- see \cref{app:M23 CSM Lower Limit Appendix} for more details. Given this limitation, we restrict our subsequent analysis of \Rcsm\ and \MLR\ to events where the predicted \Mcsm\ exceeds this threshold. 

When we reanalysed the \Rcsm\ correlations with this \Mcsm\ threshold ($\geq$~$10^{-2.5}$~\Msol), we find substantially stronger correlations, especially for rapid-rise events (\Trise~$\leq$~5~d). For these rapid-rise events, our third-order polynomial fit achieves $R^2$~=~0.59 with an RMS scatter of 0.06~dex. Slower-rising events (\Trise~$>$~5~d) still show a weaker correlation with $R^2$~=~0.38 and larger scatter (RMS~=~0.10~dex). Based on these results, we adopt the third-order polynomial fits for \Rcsm\ in both time regimes.

\subsection{Mass-Loss Rate}
\label{sec:Measuring CSM Mass and Radius}

The conversion of \Mcsm\ estimates to \MLR, Eq. \ref{eqn:mlr}, requires the stellar wind velocity ($v_{wind}$). Although M23's models are parameterised using \MLR\ (which implicitly assumes a wind velocity), the resulting light curves depend solely on the CSM density profile at the time of explosion. Consequently, our \Mcsm\ measurements can be directly compared to their models, with the assumed wind velocity affecting only the conversion between \Mcsm\ and \MLR, not the underlying physics.

\begin{equation}
\centering
    \frac{\dot M}{\text{\Msol~yr}^{-1}} =  \left(\frac{M_{CSM}}{\text{\Msol}}\right) \left(\frac{v_{wind}}{10~\text{km~s}^{-1}}\right)  \left(\frac{R_{CSM}}{10^{14}~\text{cm}}\right)^{-1}
    \label{eqn:mlr}
\end{equation}

For Type II SNe, we assume a stellar wind velocity of 10~km~s$^{-1}$, consistent with previous literature \citep[e.g.,][]{Davies_2022, Moriya_2023}. We also calculate the time in which the mass was removed $t_{removal} \sim R_{CSM}$/$v_{wind}$. While M23 explored \Rcsm\ from $10^{14}$~--~$10^{15}$~cm, we cannot uniquely determine \MLR\  from our observations alone. For comparison with previous studies, we adopt a fiducial value of \Rcsm~=~5$\times10^{14}$~cm. %This allows us to estimate \MLR, though we no these values are sensitive to our assumptions about both \Rcsm and v_w

\subsection{Progenitor Property Volume Corrected Distributions}
\label{sec:Results}

\begin{figure}
    \begin{subfigure}[b]{0.47\textwidth}
         \includegraphics[width=1\textwidth]{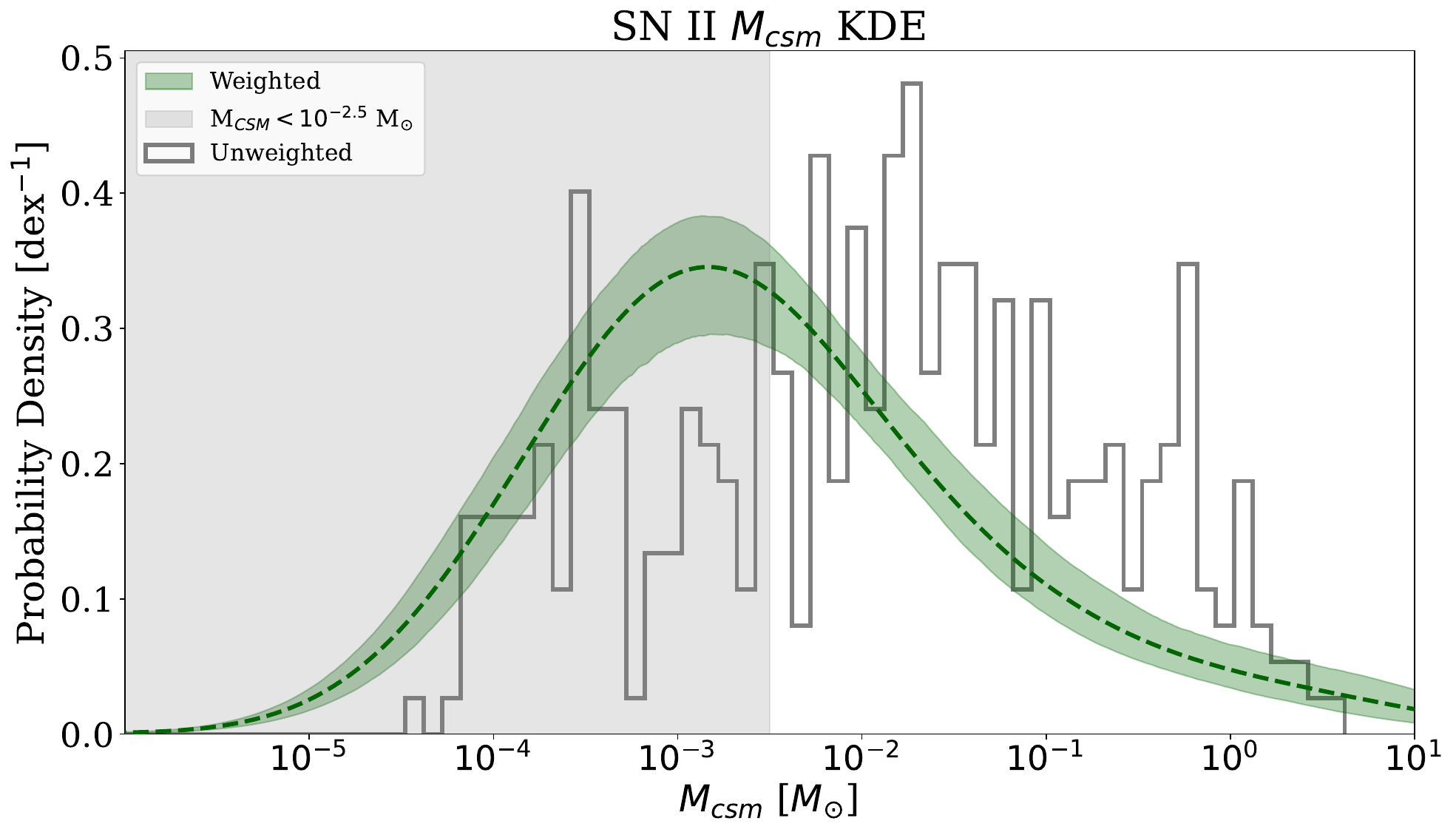}
         \caption{\Mcsm\ KDE Distribution.}
         \label{fig:II_csm_mass_kde}
\end{subfigure}
\begin{subfigure}[b]{0.47\textwidth}
         \includegraphics[width=1\textwidth]{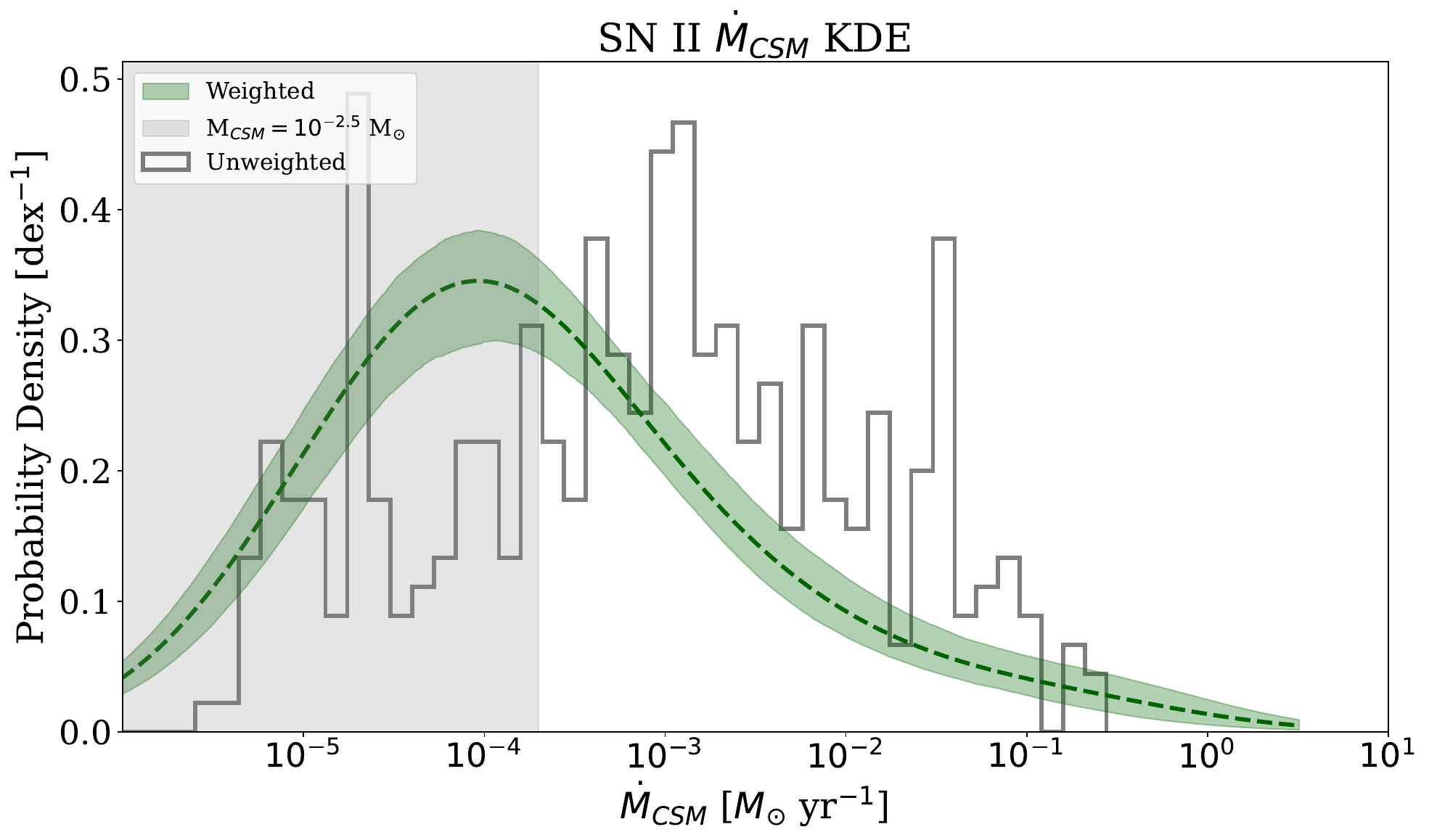}
         \caption{\MLR\  KDE Distribution.}
         \label{fig:II_mdot_kde}
\end{subfigure}
\begin{subfigure}[b]{0.47\textwidth}
         \includegraphics[width=1\textwidth]{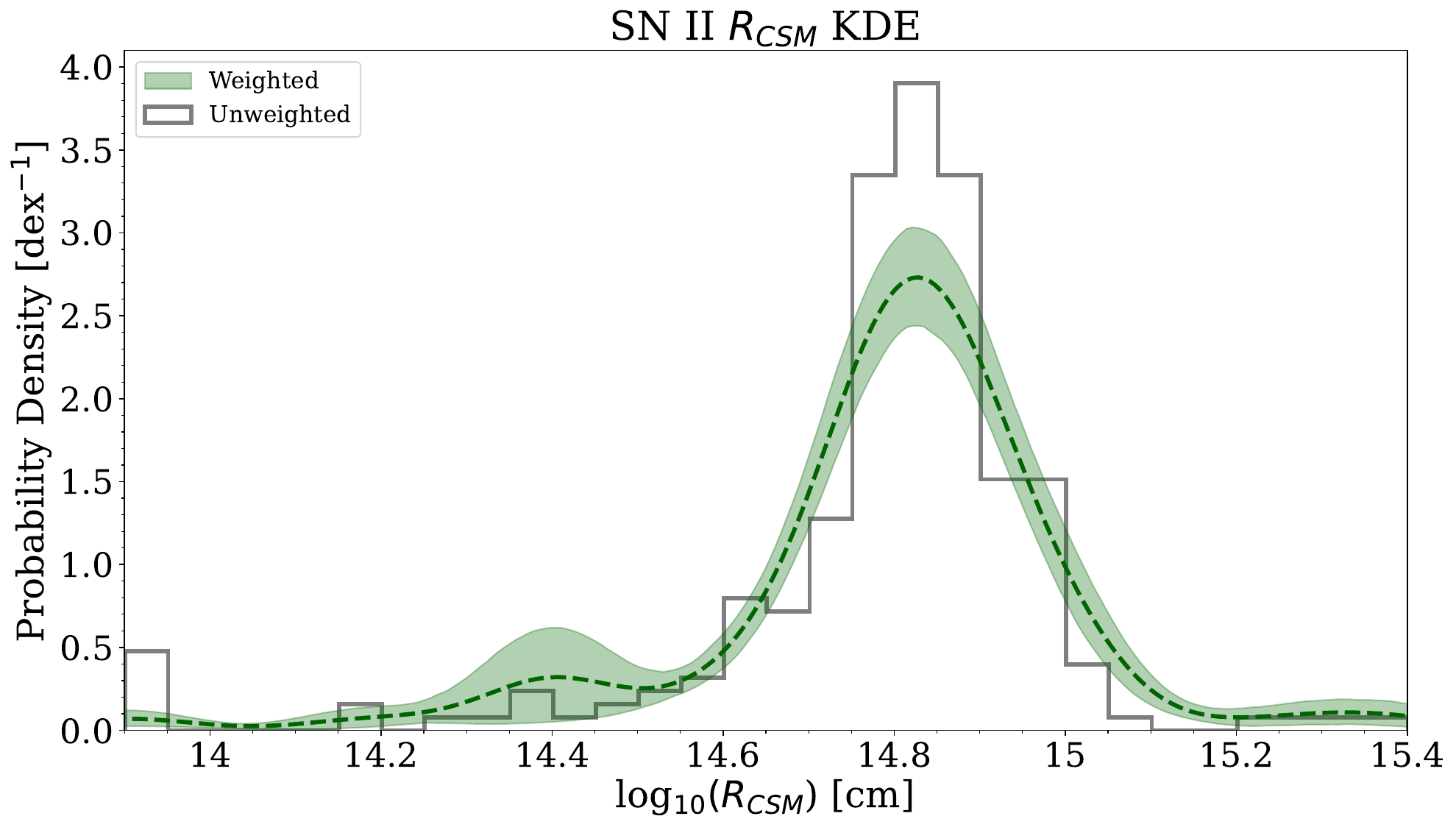}
         \caption{\Rcsm\ KDE Distribution.}
         \label{fig:II_csm_ext_kde}
\end{subfigure}
\caption{Type II KDE for \Mcsm, \MLR\ and \Rcsm\ along with their associated 80\% CI. The weighted distribution is in dark green (dashed line) and the unweighted normalised histogram is in black. The shaded region on the KDE for \Mcsm\ and \MLR\ show the region below 10$^{-2.5}$~\Msol\ (or corresponding to) where we find CSM does not impact observables.} 
\end{figure}

After weighting the distributions, we create weighted KDEs for empirically derived progenitor properties \Mcsm, \MLR\  and \Rcsm\ -- see  Figs. \ref{fig:II_csm_mass_kde}, \ref{fig:II_mdot_kde} and \ref{fig:II_csm_ext_kde} respectively. %A similar KDE distribution was made for \Mfe\ using Eq. \ref{eqn: iron_core_mass}, see Fig.~\ref{fig:II_iron_core_mass_kde}. 

\begin{figure}
\begin{subfigure}[b]{0.47\textwidth}
         \includegraphics[width=1\textwidth]{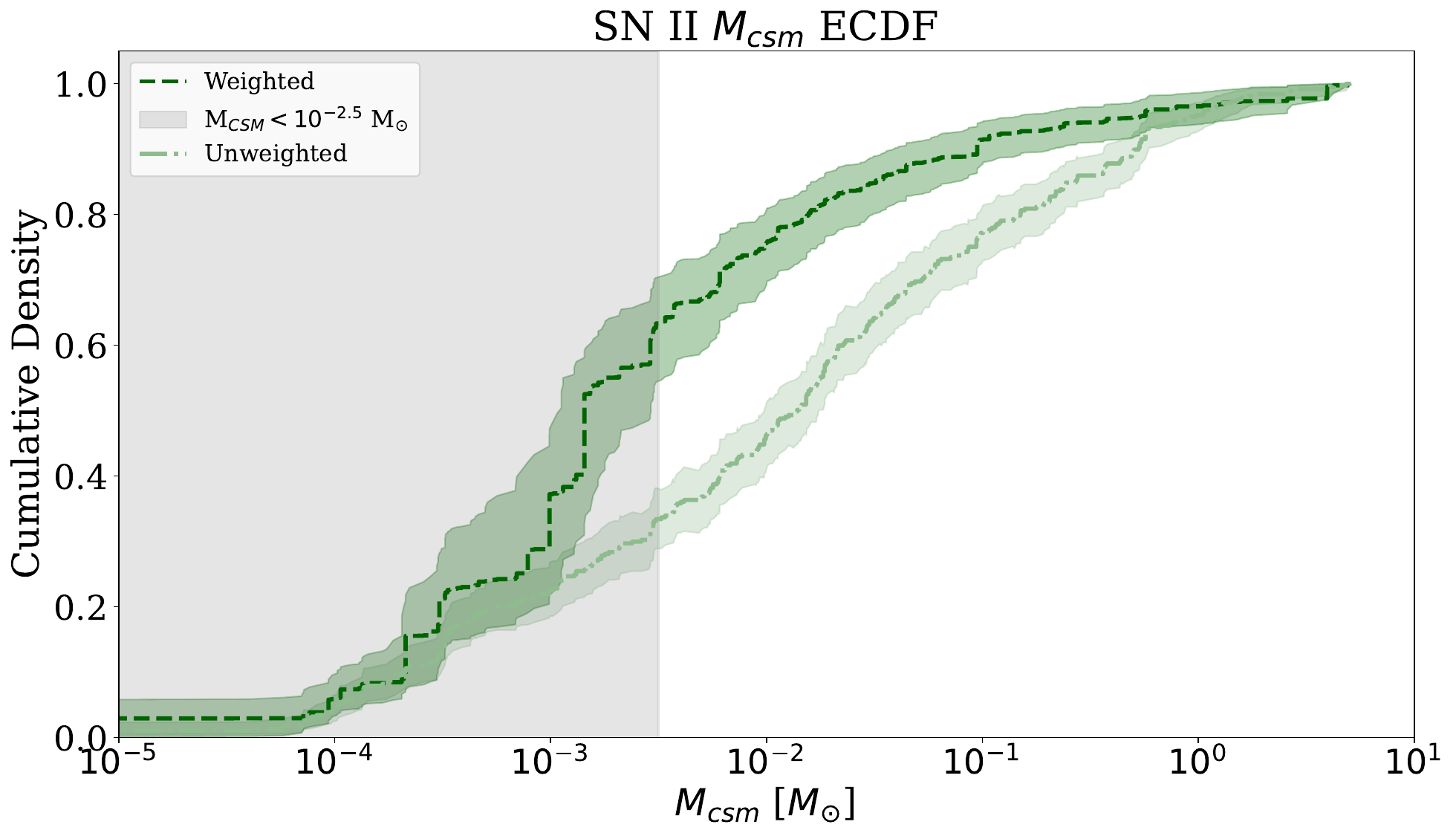}
         \caption{\Mcsm\ ECDF.}
         \label{fig:II_csm_mass_ecdf}
\end{subfigure}
\begin{subfigure}[b]{0.47\textwidth}
         \includegraphics[width=1\textwidth]{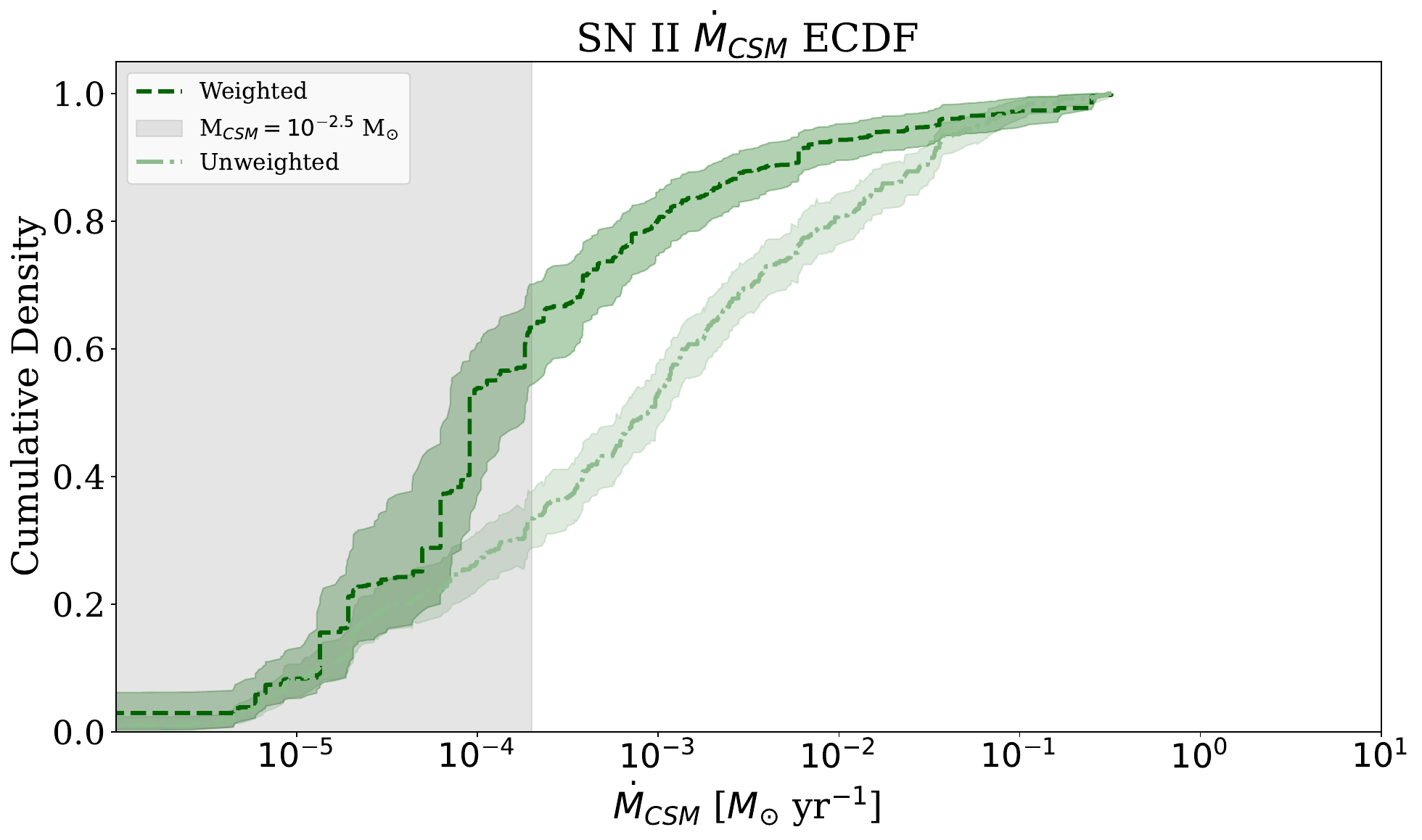}
         \caption{\MLR\  ECDF.}
         \label{fig:II_mdot_ecdf}
\end{subfigure}
\begin{subfigure}[b]{0.47\textwidth}
         \includegraphics[width=1\textwidth]{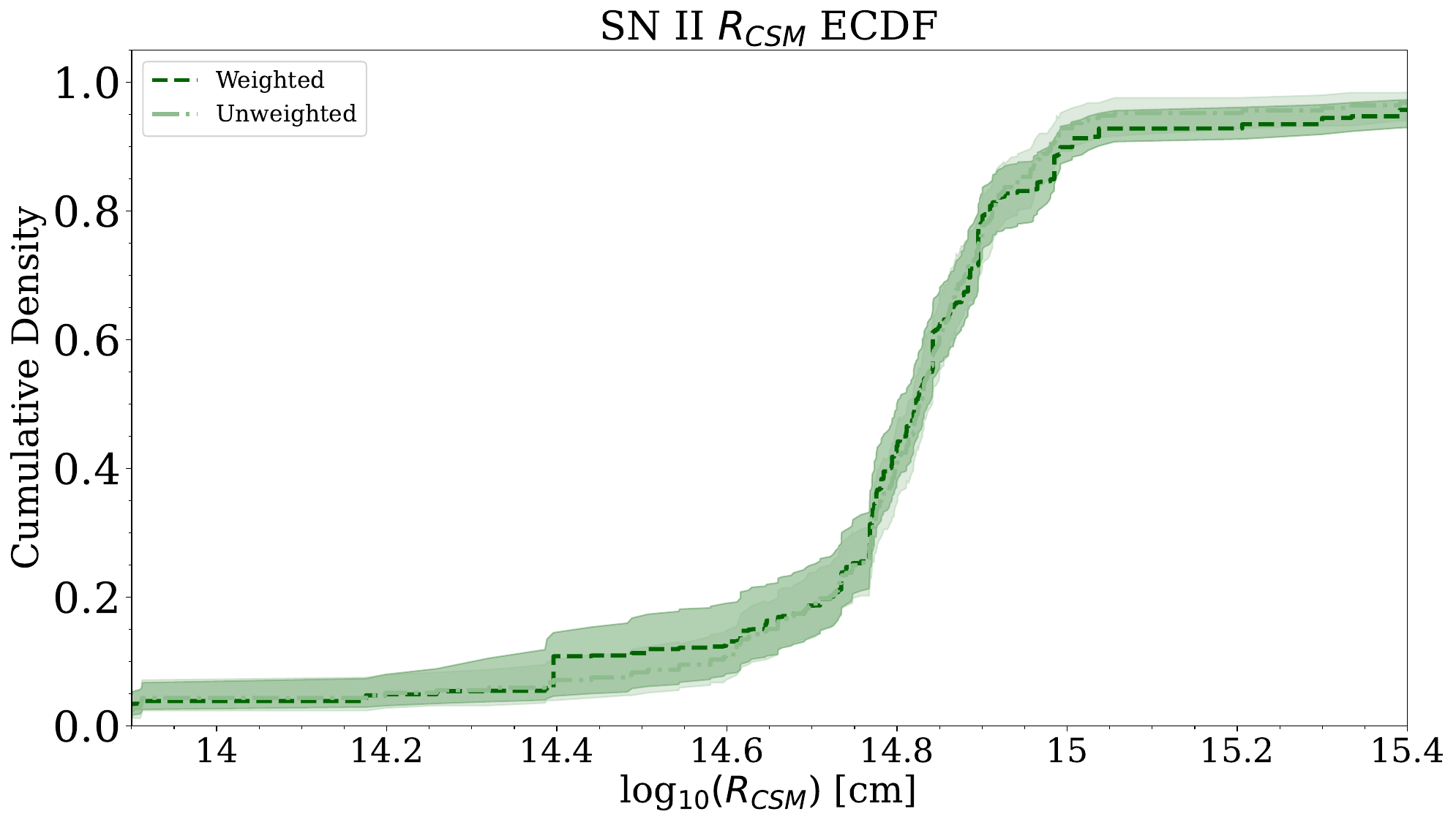}
         \caption{\Rcsm\ ECDF.}
         \label{fig:II_csm_ext_ecdf}
\end{subfigure}
\caption{Type II weighted (dark green) and unweighted (light green) ECDF for \Mcsm, \MLR\  and \Rcsm. We plot the 80\% CI for the weighted ECDF and the 95\% CI for the unweighted ECDF. For \Rcsm, we exclude data where the corresponding \Mcsm~$<$~$10^{-2.5}$~\Msol\ as we consider \Mcsm\ lower than this to have a negligible effect on the observable parameters. The shaded region on the ECDFs for \Mcsm\ and \MLR\ are the same as applied in Figs. \ref{fig:II_csm_mass_kde} and \ref{fig:II_mdot_kde}. }
\end{figure}

For comparison, recent studies of SN~2023ixf \citep[e.g.,][]{Hosseinzadeh_2023, Bostroem_2023, Hiramatsu_2023, Jencson_2023, Jacobson_2023, Li_2023, Zimmerman_2024} and SN~2024ggi \citep[e.g.,][]{Jacobson_2024a, Shrestha_2024, Chen_2024b, Chen_2024} prefer \MLR\  between $10^{-3} $~--~$10^{-2}$~\Msol~yr$^{-1}$ calculated by photometric and/or spectroscopic modelling of the event and its environment. For SN~2023ixf, we find \Mcsm~$\sim$~0.1~\Msol, \Rcsm~$\sim$~6$\times10^{14}$~cm and $\dot M$~$\sim$~1$\times10^{-2}$~\Msol~yr$^{-1}$ respectively. \citet{Singh_2024} and \citet{Moriya_2024} conduct a similar exercise, finding the best fitting M23 models to SN~2023ixf, and measure similar values for \MLR, $10^{-3}$~--$10^{-2}$~\Msol~yr$^{-1}$ and \Rcsm~$\sim$~5--10$\times10^{14}$~cm. For SN~2024ggi, we find \Mcsm~$\sim$~5$\times10^{-3}$~\Msol, \Rcsm~$\sim$~3$\times10^{14}$~cm and $\dot M$~$\sim$~2$\times10^{-4}$~\Msol~yr$^{-1}$ respectively. %Talk about the radii and how this is consistent with rise time of ggi being shorter-----

\begin{table*}
\begin{threeparttable}
\begin{tabular}{cccccccc}
\hline
Parameter & Units & Mean & 25th\%ile & 50th\%ile & 75th\%ile & Range & No. \\ \hline\hline
&&\multicolumn{4}{c}{\multirow{2}{*}{Weighted}} & &\\ 
&&&&&&&\\ \hline
\Mcsm & $\times10^{-3}$~\Msol  & $-$ & $-$ & $-$ & $ 9.55^{+3.48}_{-5.79} $ & [0.11,1.05$\times10^{3}$]\tnote{1} & 377 \\ 
\MLR  & $\times10^{-4}$~\Msol~yr$^{-1}$  &$-$ & $ -$ & $-$ & $ 6.03^{+2.20}_{-3.61} $ & [3.22,8.16$\times10{^2}$]\tnote{1} & 377 \\ 
\Rcsm & $10^{14}$~cm                     & $6.69 \pm 0.92$ & $ 5.58^{+0.28}_{-0.28} $ & $6.56 \pm 0.25 $ & $ 7.84^{+0.19}_{-0.11} $ & [1.98,14.11]& 253 \\ \hline
&&\multicolumn{4}{c}{\multirow{2}{*}{Unweighted}} & &\\ 
&&&&&&&\\\hline
\Mcsm & $\times10^{-3}$~\Msol  & $10.61^{+1.79}_{-1.54}$ & $ 1.34^{+1.91}_{-2.05} $ & $11.62^{+2.05}_{-1.74}$ & $ 93.87^{+36.22}_{-45.90} $ & - & - \\ 
\MLR  & $\times10^{-4}$~\Msol~yr$^{-1}$  &$6.70^{+1.15}_{-0.98}$ & $ 0.85^{+0.41}_{-0.49} $ & $7.54^{+1.40}_{-1.18}$ & $ 59.25^{+25.22}_{-30.85} $ & - & - \\ 
\Rcsm & $10^{14}$~cm                     & $6.55 \pm 0.44$ & $ 5.72^{+0.51}_{-0.18} $ & $6.63 \pm 0.12 $ & $ 7.90^{+0.33}_{-0.31} $ & - & - \\  \hline

\end{tabular}

\begin{tablenotes}\footnotesize %\tnote{1}
\item[1] The ranges reported are the 5th and 95th  percentiles to remove outliers beyond the limits of the original dataset. 
\end{tablenotes}
\end{threeparttable}

\caption{Mean and median of the volume corrected KDE for \Mcsm, \MLR\ and \Rcsm\ in the final sample. \Mcsm, \MLR\  and \Rcsm\ were inferred via linear relations involving GPR parameters. For \Rcsm, we exclude data where the corresponding \Mcsm\ ~$<$~$10^{-2.5}$~\Msol\ as we consider \Mcsm\ lower than this to have a negligible effect on the observable parameters and cannot constrain \Rcsm\ for lower \Mcsm. Values for \MLR\  use a fiducial value of 5$\times10^{14}$~cm for \Rcsm. Uncertainties reported here are the standard deviation on the bootstrapped values. The mean, 25th percentile and median values for \Mcsm\ are not reported here as they are below $10^{-2.5}$~\Msol\ and we cannot confidently constrain below this threshold. For the same reason, we do not report the mean, 25th percentile or median for \MLR\ as this is below the \MLR\ corresponding to \Mcsm~$= 10^{-2.5}$~\Msol.}%, the threshold below which we find \Mabs\ and \Trise\ are not influenced, thus}
\label{tab:final_csm_params_table}
\end{table*}

We report the key statistics for measured \Mcsm, \MLR\ and \Rcsm\ in Table~\ref{tab:final_csm_params_table} derived from their weighted respective KDEs. For the uncertainties on \Mcsm\ and \Rcsm\ quoted either in Table~\ref{tab:final_csm_params_table} or later, we use Eq. \ref{eqn:polyorder3} for \Mcsm\ and \Rcsm\ and resample each parameter 1,000 times within their uncertainties using a uniform distribution -- bounds set to be, for example, [\Mabs~$- \sigma_{\texttt{\Mabs}}$,\Mabs\ $+ \sigma_{\texttt{\Mabs}}$]. We then find the 1~$\sigma$ standard deviation. This method is repeated for \MLR\ using Eq. \ref{eqn:mlr}. 

Figs. \ref{fig:II_csm_mass_ecdf} and \ref{fig:II_mdot_ecdf} and \ref{fig:II_csm_ext_ecdf} are the ECDFs for \Mcsm, \MLR\  and \Rcsm\ which show the empirical distribution of parameters for the weighted and unweighted samples. For the weighted ECDFs, we show the 80\% CI via the same bootstrapping method as previously described. For the unweighted ECDF, we perform a similar bootstrap with replacement to select from the sample and find the 95\% CI empirically following a similar routine of finding the difference 2.5th and 97.5th  percentiles.

\section{Discussion}
\label{sec:Discussion}

\subsection{Luminosity and Rise Distributions}

The volume-corrected sample of 377 Type II SNe yields a mean peak absolute magnitude in rest-frame ZTF~$g$ band of \Mabs = $-16.59 \pm 0.29$ mag, with a median of \Mabs = $-16.71 \pm 0.25$ mag. To place these results in context of previous studies, we compile peak luminosities from recent surveys in Table~\ref{tab:previous_peaks}. While direct comparison is limited by filter differences -- earlier surveys typically used $UBVRI$ rather than Sloan or ZTF filters -- our mean and median weighted measurements in ZTF~$g$ can be broadly compared with Johnson-Cousins $B$ and $V$ band values, and show agreement within 1.5~$\sigma$ across all studies.

\begin{table*}
    \centering
    \begin{tabular}{cccccccc} \hline
        Reference                 & $M_{B,peak}$ [mag]    & $M_{V,peak}$ [mag] & No. & Extinction & Vol. & Class Sys. & Surveys \\ \hline \hline
        \citet{Anderson_2014}   & -                     & $-16.74 \pm 1.01$  & 68  & MW \& Host & No & Spec + Phot & CT; CTSN; SOIRS; CATS; CSP   \\
        \citet{Richardson_2014} & $-16.80 \pm 0.37 $    & -                  & 74  & MW \& Host & Yes  & Spec & ASC \\
        \citet{Galbany_2016}    & $-16.43 \pm 1.19$     & $-16.89 \pm 0.98 $ & 51  & MW         & No & Spec & C\&T; CTSN; SOIRS; CATS \\
        \citet{deJaeger_2019}   & $-16.57 \pm 1.14$     & $-16.74 \pm 0.92$  & 23  & MW         & No & Spec & LOSS \\ \hline
    \end{tabular}
    \caption{Previous measurements of the mean AB peak absolute magnitudes for Type II SNe. Uncertainties are those reported in the study. Extinction refers to whether the magnitudes were corrected for MW and host extinction, or for MW extinction only with no host correction. Vol. refers to whether there is a volume correction applied to the statistic. Classification refers to how the final samples were selected, all choose to spectroscopically (spec) classify, with \citet{Anderson_2014} performing some photometric (phot) typing to remove Type IIb, IIn and 1987A-like SNe. Survey refers to which surveys the samples belong to: CT -- Cerro Tololo SN program. CTSN -- Calán/Tololo SN. SOIRS --  Optical and Infrared Supernova Survey. CATS -- Carnegie Type II Supernova Program. CSP -- Carnegie Supernova Program. C\&T -- Calán/Tololo Supernova Program. LOSS -- Lick Observatory Supernova Search.    }
    \label{tab:previous_peaks}
\end{table*}

When comparing our volume corrected mean \Mabs\ to the entire M23 model grid (no weighting applied), we find the theoretical predictions are systematically brighter: mean \Mabs~=~$-17.88$~mag and median \Mabs~=~$-18.11$~mag. This luminosity difference reflects the the ability of our volume-complete sample to naturally capture the full diversity of Type II SNe, with the M23 grid focusing on systematically exploring parameter space rather than matching the observed luminosity distribution or the IMF.

We compared our sample with that of \citet{Das_2025}, which uses the ZTF Census of the Local Universe survey \citep[CLU;][]{De_2020}, and we are consist in the key overlapping measurement between the studies, with the luminosity distributions, Fig.~\ref{fig:II_abs_mag}, being in strong agreement.

We have limited overlap with \citet{PPessi_2024}, studying Type II SLSNe from ZTF, though we find $23^{+9}_{-7}$\% Type IIn SNe in our weighted sample have \Mabs\ brighter than $-19$~mag -- this is the most appropriate as we grouped Type IIn and SLSNe together -- with a large range in \Mabs\ $-17.01$ to $-22.20$~mag. Our unweighted median \Mabs\ for Type IIn,  $-19.32~\pm~0.13$~mag, agrees with median values in \citet{Hiramatsu_2024}. 

The observed correlation between \Mabs\ and \Trise\ for Type II SNe (excluding Types IIn and IIb) revealed in Section~\cref{sec:OverallDistributionAnalysis} is likely to be driven by a deficit of slow-rising, low-luminosity events. This weak correlation, also noted by \citet{Valenti_2016}, presents an intriguing discrepancy with theoretical predictions. While the M23 models (Fig.~\ref{fig:Moriya_absrise_massep}) predict this region of parameter space to be populated by explosions from low-mass progenitors, our BTS sample shows noticeably fewer such events than theoretically expected. Although we do detect some low-luminosity SNe with extended rise times (\Trise~$\gtrsim$~15~d), their relative scarcity compared to model predictions is significant. Section~\ref{sec:Long Risers} quantifies this population's occurrence rate relative to the broader Type II population. 

Type II SNe exhibit diverse rise times, with their distribution (Fig.~\ref{fig:rise distribution}) suggesting a continuous rather than bi-modal range of progenitor properties. The lack of bi-modality in the observed \Trise\ distributions is significant and likely reflects an overabundance of these events compared to their representation in the M23 grid. This continuous distribution is seen before and after the volume correction, Fig.~\ref{fig:II_rise}, suggesting this is not a result of lacking observations of a particular population. The prevalence of events with both rapid rise times and high luminosities provides compelling evidence for dense CSM being common among Type II SNe. These characteristics are consistent with SBO within CSM \citep[e.g.,][]{Moriya_2011, Chevalier_2012, Das_2017, Moriya_2018, Tinyanont_2022, Pearson_2022, Li_2023}, as models without significant CSM struggle to simultaneously produce such fast rise times and enhanced peak brightness.

To investigate whether our results depend on the extinction correction, we also examined the distributions of all parameters if no extinction correction was applied. \Mabs, \Trise\ and CSM parameters estimates changed by no more than 1~$\sigma$. Since the host extinction correction affects only a small fraction of our sample, our conclusions about the physical parameter distributions remain largely unaffected. The $1/V_{max}$ correction we apply to our observational sample addresses potential observational bias, but significant uncertainties remain in the faint end of the luminosity function -- explored in detail by \citet{Das_2025}. 

\subsection{Long Rising Type II SNe}
\label{sec:Long Risers}

Of note in Fig.~\ref{fig:rise distribution} are the events with long rise times, \Trise~$\gtrsim$~25~d. These events are slowly evolving, seemingly in the gap between Type IIn and the region of space the long simulated Type II light curves from \citet{Moriya_2023b} occupy. There are 5 events in our sample with rise times $\geq$~25~d that all have non-standard Type II light curves. Some of these SNe show resemblance to SN~1987A where the rise to peak is a slow hump or they have an early peak fainter than the main peak -- see ZTF18acbwaxk/SN~2018hna in Fig.~\ref{fig:LC_Diversity} for an example. To confirm the absence of narrow lines, and rule out misclassification of Type IIn SNe, we checked the classifying spectra and confirmed there were no narrow lines present in their spectral series -- most have multiple high-resolution spectra. 

When comparing the magnitude-limited BTS sample with the volume-limited ZTF CLU, only one event overlaps with the \citet{Sit_2023} study -- ZTF18acbwaxk/SN~2018hna (\Trise~$\sim$~35~d, \Mabs~$\sim -15.99$~mag uncorrected). This limited overlap stems from different selection criteria: BTS captures brighter events (peak magnitudes~$<$~18.5~mag) while CLU focuses on lower luminosity events in nearby galaxies. For Type II SNe with rise times $>$~25~d, we compute a rate relative to the CCSN rate found in \citet{Perley_2020a}.

Before applying a magnitude cut for completeness, we identify 5/481 Type II SNe with \Trise~$\geq$~25~d from our sample. Accounting for observational bias, we find these long-rising events constitute $2.16^{+1.93}_{-0.60}$\% of the Type II SN population, with a weighted mean \Mabs~=~$-$16.33~mag, compared to $-$16.59~mag for the overall sample. After implementing an 18.5~mag completeness cut, we retain 4 long-rising events in our sample of 377 SNe, corresponding to a bias-corrected fraction of $1.43^{+1.28}_{-0.15}$\% with no significant change in \Mabs. These low rates, though limited by small statistics, confirm these events are rare and align with previous findings: \citet{Sit_2023} report that SNe with \Trise~$>$~40~d comprise $1.4 \pm 0.3$\% of all CCSNe, while earlier studies found rates of 1.5~--~3\% \citep{Smartt_2009, Kleiser_2011, Pastorello_2012}.

\subsection{CSM Mass and Radial Extent} 

Our Type II SN sample reveals widespread evidence for substantial CSM present at the time of explosion. Accounting for observational biases through a volumetric weighting, $36^{+5}_{-7}$\% of events have \Mcsm~$\geq 10^{-2.5}$~\Msol, with the 80th percentile of the full sample being $1.56^{+1.12}_{-0.54}\times10^{-2}$~\Msol. In the unweighted sample, $67 \pm 6$\% of events show significant \Mcsm\ ($\geq$$10^{-2.5}$~\Msol). These \Mcsm\ estimates, and the corresponding \Rcsm, remain consistent when analysed without host-extinction corrections.

For events with massive CSM shells (\Mcsm~$\geq$$10^{-2.5}$~\Msol), our weighted sample shows a median \Rcsm\ of $\sim$~6$\times10^{14}$~cm, with nearly all events having inferred radii $\leq 10^{15}$~cm and a well-defined peak around this median (Figs. \ref{fig:II_csm_ext_kde} and \ref{fig:II_csm_ext_ecdf}). The upper limit of \Rcsm~$\leq 10^{15}$~cm in our sample reflects the parameter space explored by M23, with this particular methodology being insensitive to larger radii -- like those expected in Type IIn SNe. However, our analysis reveals a physically meaningful result: the rapid rise times observed in most events require both sufficient \Mcsm\ and relatively compact radii ($\sim$6$\times$10$^{14}$~cm median for \Mcsm~$>$10$^{-2.5}$~\Msol) to rapidly and efficiently convert kinetic energy to radiation, accelerating the early light curve evolution. 

The high \Mcsm\ ($\geq10^{-1}$\Msol) and \MLR\ ($\geq10^{-2}$~\Msol~yr$^{-1}$) we infer for a fraction of our sample (Figs. \ref{fig:II_csm_mass_kde} and \ref{fig:II_csm_mass_ecdf}) likely produce distinctive spectroscopic signatures from compact and dense CSM, such as flash features typically lasting $\leq$~1~week. While our study focuses on CSM shells with specific density profiles affecting early light curves \citep[$\rho\propto r^{-2}$;][]{Moriya_2023}, \citet{Dessart_2023} demonstrate that different CSM configurations can produce similar photometric evolution while predicting different spectroscopic features \citep[see][]{Moriya_2023b, Khatami_2023, Jacobson_2024b}. Spectroscopy is required to aid in breaking these degeneracies as it becoming clear that large amounts of CSM represent a common phenomenon rather than exceptional cases \citep[e.g.,][]{Forster_2018, Hossein_2022, Kozyreva_2022, Jacobson_2023, Hiramatsu_2023, Jencson_2023, Hosseinzadeh_2023, Irani_2023, Li_2023, Zimmerman_2024, Andrews_2024, Shrestha_2024, TPessi_2024, Xiang_2024, Chen_2024, Jacobson_2024a, Rehemtulla_2025}. 

For events with lower inferred \Mcsm\ ($< 10^{-2.5}$~\Msol), the impact of CSM on the early light curve is too weak to constrain \Rcsm. While many of these SNe are likely to possess non-negligible CSM, material at larger radii would likely not influence the early evolution, and would be too diffuse to influence the later light curve evolution in a clearly discernible way unless the mass is extremely high \citep[e.g.,][]{Irani_2023}. 

A limitation of the approach from this work is that for the fastest rising events ($\leq$~1~d), which likely require dense, compact CSM to achieve such rapid evolution, our measured rise times would represent upper limits, unresolved fast rises might require more substantial \Mcsm\ or closer and more compact CSM. While individual CSM parameters may have uncertainties due to model assumptions, observational constraints and scatter present in relationships, this frequency of substantial CSM in our volume corrected sample represents a robust statistical result, independent of the precise CSM parameterisation. Our analysis is further constrained by confidence in classifications, an issue we explore in greater detail in \cref{app:Systematic Misclassification}.

\subsection{Implications for Mass-Loss Mechanisms}
\label{sec:MLR Implications}

With a minimum \Mcsm, $10^{-2.5}$~\Msol, and observationally supported fiducial values for $v_{wind}$ and \Rcsm\ of 10~km~s$^{-1}$ and $5\times10^{14}$~cm, respectively, we find a characteristic \MLR\ of 2$\times10^{-4}$~\Msol~yr$^{-1}$. This characteristic value is higher, by $\sim$ two orders of magnitude, than values inferred from observations of local group RSGs e.g., $\dot M \sim 10^{-6}$~\Msol~yr$^{-1}$ \citep[e.g.,][]{Vink_2001, Smith_2014, Beasor_2018, Beasor_2020}. 

Our findings are instead closer to studies such as \citet{Morozova_2017, Moriya_2018, Bruch_2021,Bruch_2022,Irani_2023, Jacobson_2024b} that estimate \MLR\ to be $\sim 10^{-4}$~--~$10^{-1}$~\Msol~yr$^{-2}$ based on detailed analysis of spectroscopy and photometry of early SNe. Like \citet{Jacobson_2024b}, we find that Type II SNe exhibit a continuum of \MLR, representative of the heterogeneous morphology of light curves, with  $46^{+5}_{-13}$\% of the corrected population having \MLR~$\geq$~1$\times10^{-4}$~\Msol~yr$^{-1}$ in the last several decades before core-collapse.

Given a $v_{wind}$ of 10~km~s$^{-1}$ and the minimum and maximum values of \Rcsm, $10^{14}$~--~$10^{15}$~cm, we calculate the range of timescales in which the material is removed to be 3~--~32~yrs. The distinctly short timescales of mass-loss, compared to the lifespan of this evolutionary stage, further supports the need for a period of `enhanced' mass-loss. We argue the higher rates of mass-loss estimated from early photometry found here and by other studies \citep[e.g.,][]{Irani_2023, Jacobson_2024a, Silva_2024} are probing the end-of-life mass-loss rather than the typical \MLR\  of RSGs.

Any viable mass-loss mechanism must maintain sufficient CSM density at compact radii while preventing complete shell detachment. The mechanism must operate on timescales consistent with our inferred \MLR\ ($\geq$10$^{-4}$~\Msol~yr$^{-1}$) and produce velocities that allow the material to remain within \Rcsm~$\leq 10^{15}$~cm. Higher ejection velocities or more extended distributions would result in CSM densities too low to effectively interact with the SN ejecta and SBO during the early light curve evolution.

As an alternative to `enhanced' \MLR\ in the centuries before core-collapse, the dense chromosphere model of \citet{Fuller_2024} offers a compelling explanation for the apparent disparity between observed RSG mass-loss rates and those inferred from early SN evolution \citep[see][]{Fuller_2024}. A chromosphere exists in approximate hydrostatic equilibrium, with significantly higher densities above the stellar surface than predicted by constant or $\beta$-law wind velocity models, despite maintaining \MLR\ more consistent with local group RSG measurements  \citep[see Fig. 4 in][]{Fuller_2024}. This model is able to naturally produce the rapid photometric and spectroscopic evolution observed in fast-rising SNe while preserving realistic progenitor properties. While our \Mcsm\ estimates would remain largely unaffected, the inferred \MLR\ would decrease substantially due to the significantly different velocity structure in \citet{Fuller_2024} compared to those used in M23. Thus, the chromosphere model represents a promising alternative to `enhanced' \MLR, potentially resolving a tension between \MLR\ measured from local RSGs and early SN evolution, though additional development is required to fully understand the impact of chromospheres on the photometric and spectroscopic evolution across diverse SN populations.

\section{Conclusions}
\label{sec:Conclusions}

In this work, we have presented forced photometry and Gaussian process analysis of all spectroscopically classified H-rich SNe from the ZTF Bright Transient Survey, 1802 objects from May 1st, 2018 to December 31st, 2023, 981 of which pass various quality cuts outlined in BTS sample paper \citep{Perley_2020a}. We have modelled the light curves with GPR to return various empirical light curve parameters, with a focus on the rise times of 639 Type II SNe. Using various light curve parameters, we have created volume corrected ($V_{max}$ method) distributions from the BTS sample, allowing us to confidently report the following main conclusions for a highly complete sample of 377 Type II SNe (excluding Type IIn and IIb SNe, and after a magnitude cut at $\leq$~18.5~mag for completeness):

\begin{itemize}
\item We see large diversity in Type II light curve demographics but no clear separation in the luminosity-rise phase-space. The predicted bi-modality that appears when we measure the distributions of rise times, \Trise, from the simulated light curves of \citet{Moriya_2023} is not seen in our observed light curves. 

\item Based on the $1/V_{max}$ weighted sample of Type II SNe from this study, we find $36^{+5}_{-7}$\% of Type II SN progenitors have \Mcsm~$\geq$~$10^{-2.5}$~\Msol\ at the time of core-collapse. We find this is the minimum amount of \Mcsm\ needed to impact the observables like the rise time and peak magnitude, based on \citet{Moriya_2023} models. 

\item For an assumed progenitor wind velocity of 10~km~s$^{-1}$, a maximum CSM radius of  \Rcsm~$\approx 5\times10^{14}$~cm and \Mcsm~=~10$^{-2.5}$~\Msol, we estimate mass-loss rates of \MLR~$\sim2\times10^{-4}$\Msol~yr$^{-1}$ for events showing CSM-affected light curves. We constrain the period of this to occur within the last 3~--~32~years, consistent with recent findings from \citet{Bruch_2021}, \citet{Bruch_2022} and \citet{Jacobson_2024b} which suggest `enhanced' mass-loss is a common feature of RSG evolution in the final decades before core-collapse.

\end{itemize}

This supports findings in recent literature that CSM interactions contribute significantly to the early light curve and are prevalent in a large set Type II SNe. While not ubiquitous across Type II SNe, we show possessing large amounts of CSM in common amongst Type II SNe progenitors. For the $\sim$~36\% of Type II SNe (excluding Types IIn and IIb SNe) where CSM interactions dominate, we find that dense CSM both shortens the rise time to peak luminosity and enhances the early-time brightness. We have further highlighted the need to reconcile and address the disparity between light curve derived \MLR\ values and \MLR\ from RSG observations -- which are typically larger by $\approx 2$ orders of magnitudes than the rates inferred from local group RSG observations \citep[e.g.,][]{vanLoon_2005, Mauron_2011, Smith_2014, Beasor_2020, Stroh_2021, Strotjohann_2023}.

Mapping the true distribution of CSM properties and establishing robust connections between SNe and their progenitors requires deeper observations over longer baselines than currently available. The Vera Rubin Observatory, ZTF-III, and upcoming IR/UV missions will provide unprecedented multi-wavelength coverage with the depth and cadence needed to probe fainter CSM signatures and earlier epochs, essential for reconstructing progenitor mass-loss histories and understanding how they shape the observed diversity of Type II SNe.

\section{Acknowledgements}

Based on observations obtained with the Samuel Oschin Telescope 48-inch and the 60-inch Telescope at the Palomar Observatory as part of the Zwicky Transient Facility project. ZTF is supported by the National Science Foundation under Grants No. AST-1440341 and AST-2034437 and a collaboration including current partners Caltech, IPAC, the Oskar Klein Center at Stockholm University, the University of Maryland, University of California, Berkeley, the University of Wisconsin at Milwaukee, University of Warwick, Ruhr University, Cornell University, Northwestern University and Drexel University. Operations are conducted by COO, IPAC, and UW.

SED Machine is based upon work supported by the National Science Foundation under Grant No. 1106171. The ZTF forced-photometry service was funded under the Heising-Simons Foundation grant \#12540303 (PI: Graham). The Gordon and Betty Moore Foundation, through both the Data-Driven Investigator Program and a dedicated grant, provided critical funding for SkyPortal.

Numerical computations were in part carried out on PC cluster at the Center for Computational Astrophysics, National Astronomical Observatory of Japan. TJM is supported by the Grants-in-Aid for Scientific Research of the Japan Society for the Promotion of Science (JP24K00682, JP24H01824, JP21H04997, JP24H00002, JP24H00027, JP24K00668) and by the Australian Research Council (ARC) through the ARC's Discovery Projects funding scheme (project DP240101786).

A.~A.~M.~is partially supported by DoE award \#DE-SC0025599. 

W.J-G. is supported by NASA through the NASA Hubble Fellowship grant HSTHF2-51558.001-A awarded by the Space Telescope Science Institute, which is operated by the Association of Universities for Research in Astronomy, Inc., for NASA, under contract NAS5-26555.

This research has made use of the SVO Filter Profile Service ``Carlos Rodrigo", funded by MCIN/AEI/10.13039/501100011033/ through grant PID2023-146210NB-I00 .

This research has made use of the NASA/IPAC Extragalactic Database (NED), which is funded by the National Aeronautics and Space Administration and operated by the California Institute of Technology.

M.W.C acknowledges the support from the National Science Foundation with grant numbers  PHY-2308862 and PHY-2117997.

\section{Data Availability}
Alongside the upcoming publication, A.~A.~Miller et al.\ (2025, in prep.), there will be a large data release o the ZTF P48 light curves used in this work. 

The light curves and empirical properties measured for the SNe, as they relate to this work, can be found here: [\url{https://doi.org/10.5281/zenodo.15229515}].

Access to the public Bright Transient Survey sample explorer can be found here: [\url{https://sites.astro.caltech.edu/ztf/bts/explorer.php}].

\bibliographystyle{mnras}   % if natbib is available
\bibliography{references.bib} %bibtex file

%%%%%%%%%%%%%%%%%%%%%%%%%%%%%%%%%%%%%%%%%%%
%% Just a reminder that you may have to run bibtex
%% All of it up to \end{document} can be removed
%% if you don't like the warning.
%%%%%%%%%%%%%%%%%%%%%%%%%%%%%%%%%%%%%%%%%%%
\IfFileExists{\jobname.bbl}{}
 {\typeout{}
  \typeout{******************************************}
  \typeout{** Please run "bibtex \jobname" to optain}
  \typeout{** the bibliography and then re-run LaTeX}
  \typeout{** twice to fix the references!}
  \typeout{******************************************}
  \typeout{}
 }
 
\onecolumn

\appendix
\section{}

\subsection{Heavily Host-Extinguished}
\label{app: HostExtinct}

\begin{table}
    \centering
    \begin{tabular}{ccccccccc}
    \hline
    ZTF & TNS ID & Type & $z$ & \Mabs~[mag] & \Trise~[d] & \ColPeak~[mag] &  $A^{\mathrm{host}}_g$~[mag] \\ \hline\hline
ZTF18abdbysy & 2018cyg & II & 0.01127 & -14.40 $\pm$ 0.02 & 1.70 $\pm$ 0.18 & 0.71 $\pm$ 0.03 & 2.38 \\
ZTF18abvvmdf & 2018gts & II & 0.029597 & -16.70 $\pm$ 0.02 & 2.00 $\pm$ 0.14 & 0.59 $\pm$ 0.03 & 1.97 \\
ZTF19aamkmxv & 2019bxq & IIn & 0.014 & -16.66 $\pm$ 0.02 & 3.41 $\pm$ 0.16 & 0.77 $\pm$ 0.02 & 2.61 \\
ZTF19aamvape & 2019cjx & II & 0.03 & -17.69 $\pm$ 0.02 & 8.43 $\pm$ 0.39 & 0.30 $\pm$ 0.02 & 1.01 \\
ZTF19aayrosj & 2019hrb & II & 0.015064 & -15.86 $\pm$ 0.02 & 1.83 $\pm$ 0.14 & 0.27 $\pm$ 0.03 & 0.90 \\
ZTF19abgfuhh & 2019lgc & IIb & 0.0354 & -17.35 $\pm$ 0.02 & 3.99 $\pm$ 0.19 & 0.37 $\pm$ 0.03 & 1.26 \\
ZTF19abxtcio & 2019pof & IIb & 0.0155 & -15.79 $\pm$ 0.02 & 14.30 $\pm$ 0.68 & 0.46 $\pm$ 0.03 & 1.57 \\
ZTF20aaetrle & 2020sy & II & 0.02 & -16.99 $\pm$ 0.02 & 6.97 $\pm$ 0.68 & 0.47 $\pm$ 0.04 & 1.60 \\
ZTF20aaurfwa & 2020hem & IIn & 0.0935 & -20.37 $\pm$ 0.01 & 16.86 $\pm$ 0.51 & 0.32 $\pm$ 0.01 & 1.09 \\
ZTF20abfcrzj & 2020mob & IIb & 0.023244 & -16.86 $\pm$ 0.06 & 10.01 $\pm$ 0.68 & 0.26 $\pm$ 0.08 & 0.87 \\
ZTF20abpmqnr & 2020qmj & IIn & 0.022 & -18.53 $\pm$ 0.01 & 7.47 $\pm$ 0.18 & 0.42 $\pm$ 0.01 & 1.41 \\
ZTF20abwzqzo & 2020sbw & IIb & 0.023033 & -16.64 $\pm$ 0.07 & 7.97 $\pm$ 2.60 & 0.38 $\pm$ 0.10 & 1.27 \\
ZTF20aclkhnm & 2020xql & II & 0.036 & -17.07 $\pm$ 0.04 & 12.93 $\pm$ 1.61 & 0.56 $\pm$ 0.06 & 1.90 \\
ZTF20acnzkxb & 2020ykd & II & 0.02690421 & -16.95 $\pm$ 0.01 & 5.65 $\pm$ 0.28 & 0.40 $\pm$ 0.02 & 1.33 \\
ZTF20acpgokr & 2020yzi & II & 0.027 & -16.72 $\pm$ 0.02 & 2.08 $\pm$ 0.15 & 0.34 $\pm$ 0.03 & 1.15 \\
ZTF20acrzwvx & 2020aatb & II & 0.009954 & -16.41 $\pm$ 0.01 & 7.70 $\pm$ 0.52 & 0.45 $\pm$ 0.02 & 1.51 \\
ZTF20actqnhg & 2020aaxf & IIb & 0.014813 & -16.55 $\pm$ 0.02 & 5.01 $\pm$ 0.35 & 0.38 $\pm$ 0.03 & 1.27 \\
ZTF20acvevsn & 2020abqw & II & 0.01417 & -14.96 $\pm$ 0.02 & 1.73 $\pm$ 0.22 & 0.42 $\pm$ 0.04 & 1.42 \\
ZTF21aajgdeu & 2021cjd & II & 0.027929 & -16.71 $\pm$ 0.04 & 1.76 $\pm$ 0.19 & 0.36 $\pm$ 0.05 & 1.22 \\
ZTF21aakupth & 2021cvd & IIn & 0.023483 & -16.11 $\pm$ 0.02 & 4.33 $\pm$ 0.32 & 0.52 $\pm$ 0.03 & 1.76 \\
ZTF21aamwqim & 2021dru & II & 0.025878 & -16.51 $\pm$ 0.05 & 4.17 $\pm$ 0.70 & 0.46 $\pm$ 0.07 & 1.57 \\
ZTF21aavuqzr & 2021kat & IIn & 0.1013 & -20.24 $\pm$ 0.01 & 16.78 $\pm$ 0.50 & 0.35 $\pm$ 0.01 & 1.19 \\
ZTF21aaydxoo & 2021kwc & IIn & 0.021759 & -17.53 $\pm$ 0.01 & 4.85 $\pm$ 0.15 & 0.59 $\pm$ 0.01 & 1.99 \\
ZTF21aayfnjz & 2021kww & II & 0.023 & -17.52 $\pm$ 0.01 & 6.31 $\pm$ 0.13 & 0.31 $\pm$ 0.02 & 1.04 \\
ZTF21abfoyac & 2021pni & II & 0.033 & -18.27 $\pm$ 0.01 & 7.40 $\pm$ 0.25 & 0.38 $\pm$ 0.01 & 1.27 \\
ZTF21abujgmr & 2021wrr & IIn & 0.048 & -17.95 $\pm$ 0.01 & 6.86 $\pm$ 0.37 & 0.44 $\pm$ 0.01 & 1.47 \\
ZTF21abviabc & 2021wyn & II & 0.053467 & -18.20 $\pm$ 0.02 & 3.32 $\pm$ 0.39 & 0.75 $\pm$ 0.02 & 2.54 \\
ZTF21abyqrli & 2021ybc & IIb & 0.02925 & -17.29 $\pm$ 0.09 & 6.15 $\pm$ 1.26 & 0.29 $\pm$ 0.04 & 0.99 \\
ZTF22aagvxjc & 2022iep & IIn & 0.025 & -17.00 $\pm$ 0.01 & 16.40 $\pm$ 0.60 & 0.28 $\pm$ 0.01 & 0.95 \\
ZTF22aalorla & 2022lix & II & 0.06804 & -18.66 $\pm$ 0.01 & 5.52 $\pm$ 0.13 & 0.46 $\pm$ 0.01 & 1.54 \\
ZTF22aamjqvc & 2018elp & IIb & 0.030089 & -17.57 $\pm$ 0.01 & 5.56 $\pm$ 0.20 & 0.29 $\pm$ 0.02 & 0.99 \\
ZTF22aaotgrc & 2022ngb & IIb & 0.00965 & -16.18 $\pm$ 0.02 & 6.96 $\pm$ 0.13 & 0.66 $\pm$ 0.03 & 2.23 \\
ZTF22aapqaqe & 2022npv & II & 0.025177 & -17.32 $\pm$ 0.01 & 3.21 $\pm$ 0.12 & 0.46 $\pm$ 0.02 & 1.54 \\
ZTF22aawptbl & 2022pzh & II & 0.045 & -18.20 $\pm$ 0.02 & 5.79 $\pm$ 0.48 & 0.44 $\pm$ 0.02 & 1.49 \\
ZTF22ablvnwa & 2022xae & IIb & 0.045229 & -18.09 $\pm$ 0.04 & 4.64 $\pm$ 0.57 & 0.26 $\pm$ 0.05 & 0.87 \\
ZTF22abnejmu & 2022ycs & II & 0.01 & -15.50 $\pm$ 0.02 & 9.14 $\pm$ 1.46 & 0.48 $\pm$ 0.02 & 1.61 \\
ZTF22abssiet & 2022zmb & II & 0.01449 & -15.41 $\pm$ 0.02 & 1.73 $\pm$ 0.11 & 0.25 $\pm$ 0.03 & 0.85 \\
ZTF23aaawbsc & 2023aew & IIb & 0.025 & -18.55 $\pm$ 0.03 & 7.41 $\pm$ 0.22 & 0.30 $\pm$ 0.03 & 1.01 \\
ZTF23aaesmsf & 2023fsc & IIb & 0.02 & -17.81 $\pm$ 0.01 & 11.13 $\pm$ 0.41 & 0.34 $\pm$ 0.02 & 1.13 \\
ZTF23aazqmwp & 2023qec & II & 0.02079 & -17.50 $\pm$ 0.01 & 6.88 $\pm$ 0.21 & 0.26 $\pm$ 0.02 & 0.88 \\
ZTF23abjrolf & 2023uvh & II & 0.02676 & -16.69 $\pm$ 0.07 & 4.81 $\pm$ 1.15 & 0.46 $\pm$ 0.10 & 1.54 \\
    \hline
    \end{tabular}
    \caption{Properties of heavily dust-extinguished Type II SNe, identified by their red colours, \ColPeak$\geq$~0.25~mag, and moderate rise times, \Trise~$\leq$20~d, as shown in Fig.~\ref{fig:AllII_pcol_hist}. Table contains: ZTF object name; TNS name; spectroscopic classification; redshift; \Mabs\ in ZTF~$g$ at rest-frame and uncertainty; \Trise\ rise time [d] in ZTF~$g$ at rest-frame wavelength and uncertainty; $g-r$ colour at ZTF~$g$ peak time and uncertainty; host galaxy extinction in ZTF~$g$ band, method described in Section~\cref{sec:Extinction}.}
    \label{tab:dust_ext_sample}
\end{table}

Table \ref{tab:dust_ext_sample} contains significantly dust-extinguished Type II SNe we identify in our sample. These events are characterised by distinctly red colours, \ColPeak$\geq$~0.25~mag, at peak and moderate rise times ,\Trise~$\leq$20~d, placing them in a unique region of parameter space as illustrated in Fig.~\ref{fig:AllII_pcol_hist}. We correct for host extinction using \ColPeak\ and apply this correction to these SNe only, as described in Section \cref{sec:Extinction}.

\subsection{Rise Time Recovery}
\label{app:Rise Time Recovery Appendix}

%\subsection{Rise Time Recovery}
%\label{sec:Rise Time Recovery}
\begin{figure}
    \centering
    \includegraphics[width=1\textwidth]{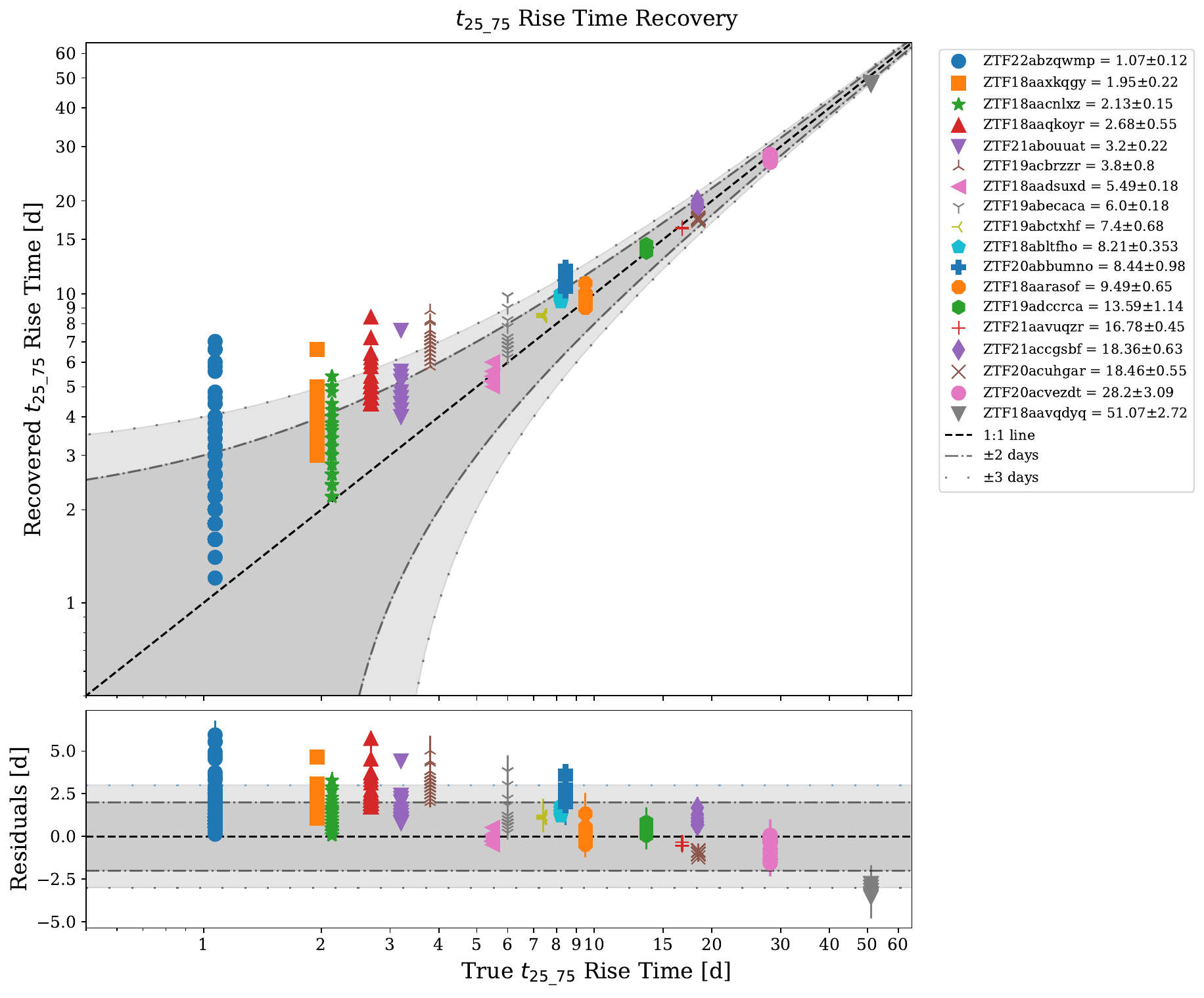}
    \caption{\Trise\ rise time recovery exploring the impact of resampling high cadenced light curve (see in the legend) to the sampling function of less well sampled light curves. The diagonal dashed line is the 1:1 line, the dot-dashed line encloses $\pm$~2~d and the dotted line encloses $\pm$~3~d. The bottom plot shows the residual between the `true' rise time of each event vs the measurements from resampling.}
    \label{fig:RiseResample}
\end{figure}

To assess the impact of the sampling function on our GPR measurements, we conducted a systematic resampling experiment using well-sampled light curves with well-constrained \Trise\ values. This involved taking thoroughly observed events (e.g., ZTF18aacnlxz/SN 2020aavr) and resampling their light curves (simulating alternative sampling functions) to match the observation cadence of more sparsely observed light curves in our sample. Figure \ref{fig:RiseResample} illustrates the resulting distribution of measured rise times across different intrinsic \Trise\ values.

Our analysis demonstrates that the GPR process reliably distinguishes between fast-rising (\Trise~$\leq$ 5~d) and slower-rising (\Trise~$>$~5~d) events. For the fastest risers (\Trise\ between 1--2~d), we observe substantial uncertainty with a spread of $\approx$~0.7~dex. This improves to $\approx$~0.4~dex for moderate risers (\Trise\ between 3--5~d) and further to $\approx$~0.2~dex for slower-rising events (\Trise~$>$~5~d). 

\subsection{Sample Redshift Distribution}
\label{app:Sample Redshift Distribution Appendix}
\begin{figure}
    \centering
    \includegraphics[width=0.8\textwidth]{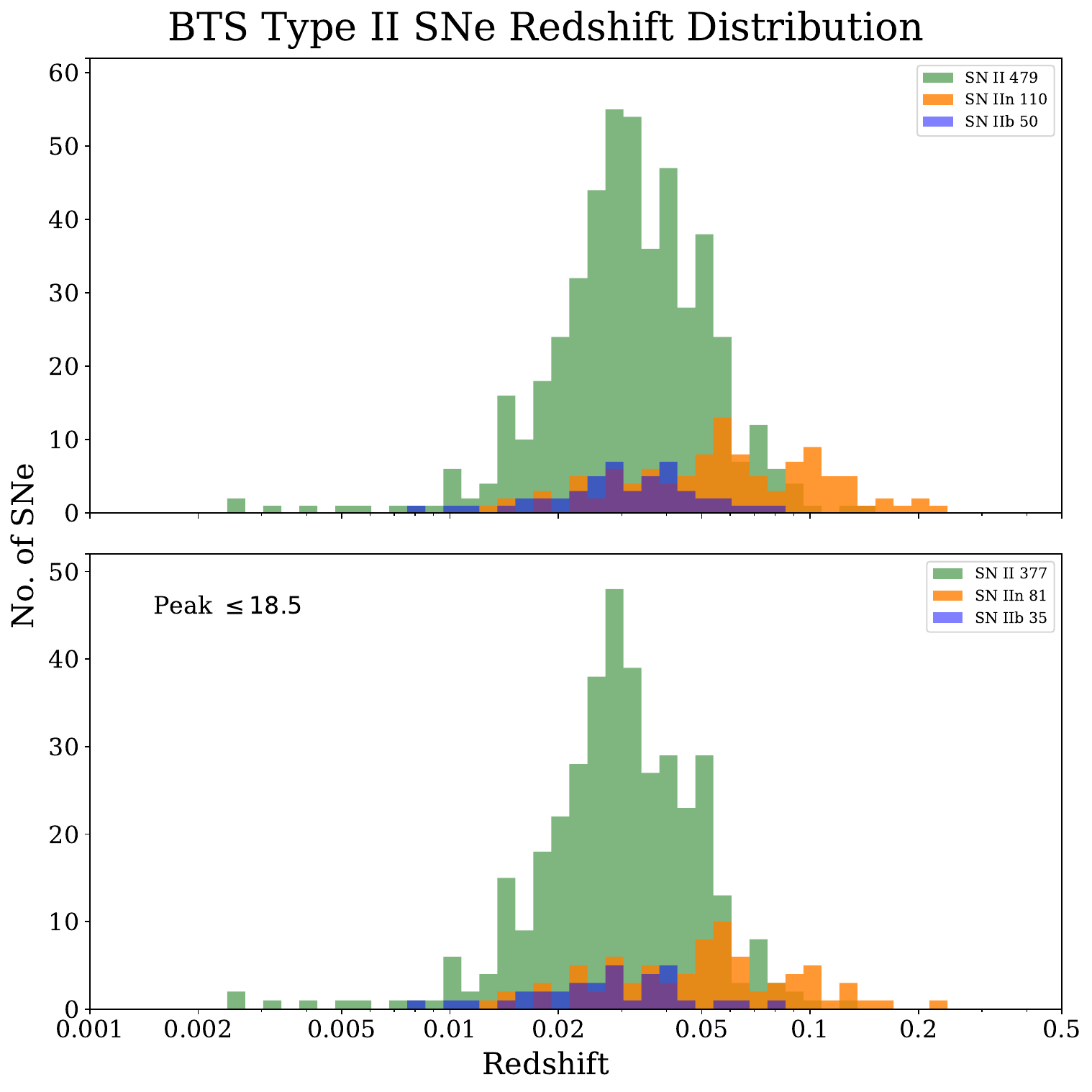}
    \caption{Distribution of redshift, $z$, across the for the whole sample (\textit{top}) and for the sample with a $m_{peak}$~$\leq$~18.5~mag (\textit{bottom}).}
    \label{fig:RedshiftDist}
\end{figure}

In Fig. ref{fig:RedshfitDist}, we show the distribution of redshift, $z$, across our SN sample. The upper panel shows the $z$ distribution for our full dataset, and the lower panel displays the highly complete ($\sim$95\%) sample limited to events with peak apparent magnitudes $m_{peak}$~$\leq$~18.5~mag. 

\subsection{Peak Colours}
\label{app:Peak Colours Appendix}
\begin{figure}   
\centering
    \begin{subfigure}[b]{0.47\textwidth}
        % \hspace*{-1.5cm} 
         \includegraphics[width=1\textwidth]{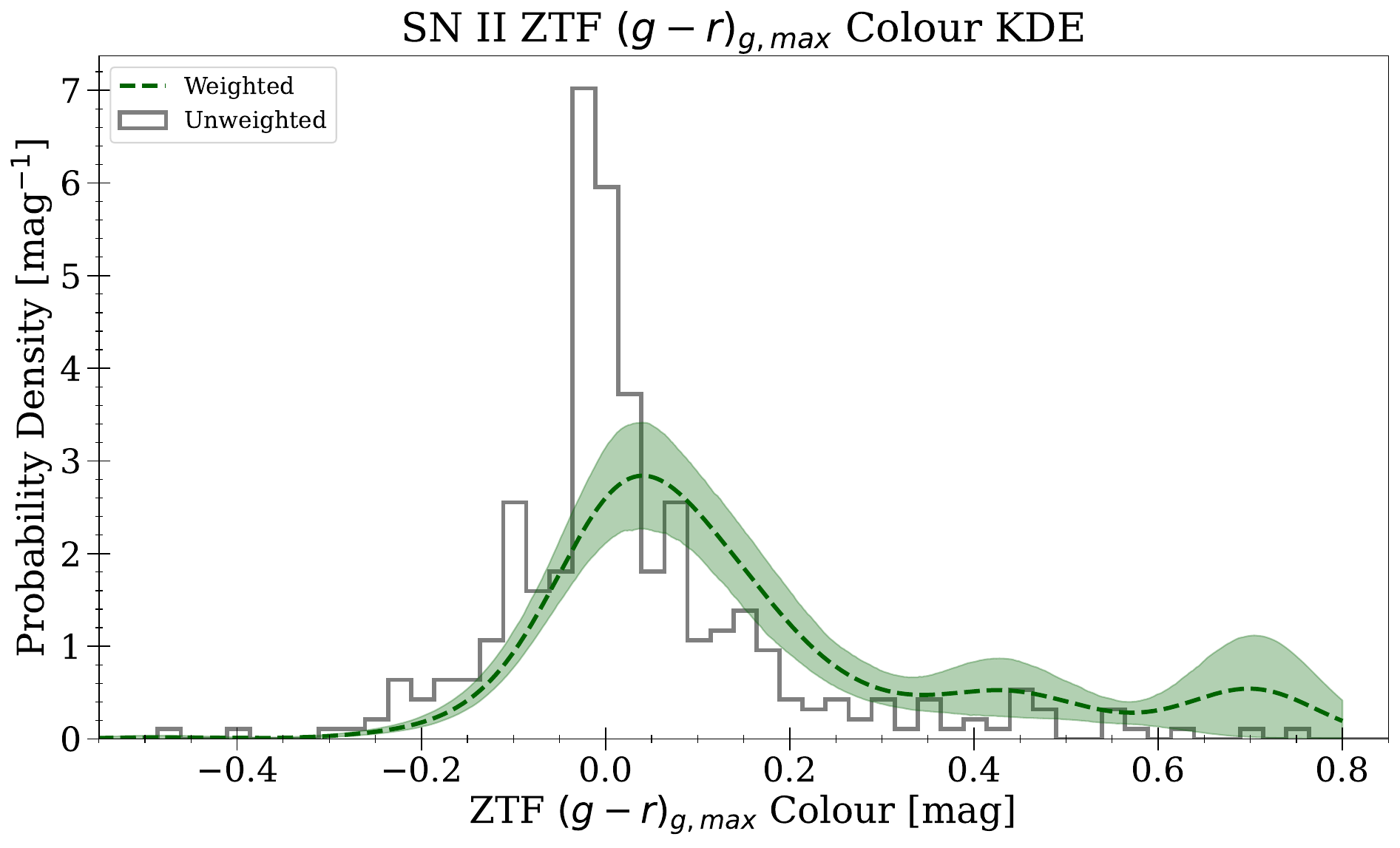}
         \caption{Type II Peak Colour, \ColPeak, KDE.}
         \label{fig:II_PCol_KDE}
    \end{subfigure}
    \begin{subfigure}[b]{0.47\textwidth}
        % \hspace*{-1.5cm} 
         \includegraphics[width=1\textwidth]{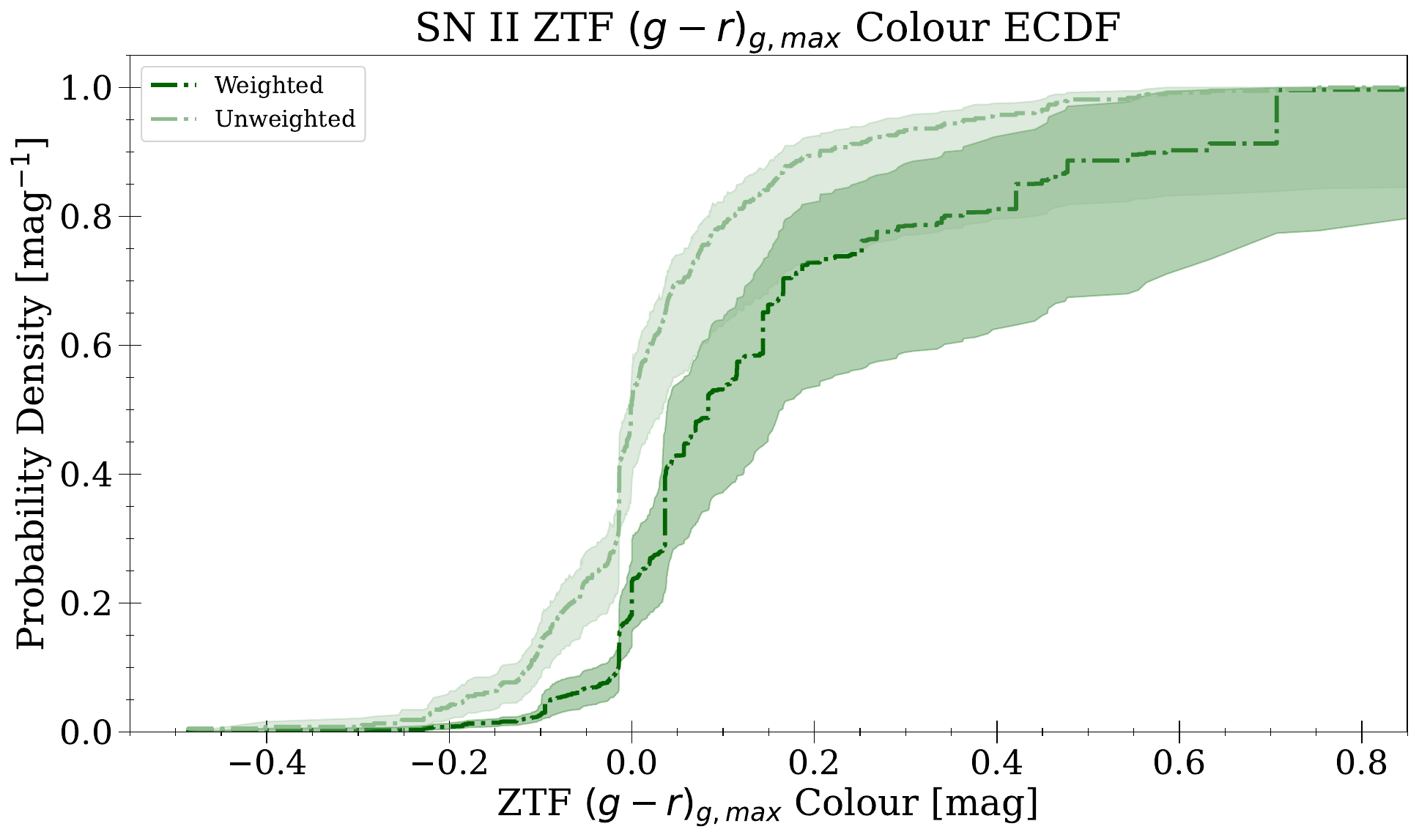}
        \caption{Type II Peak Colour, \ColPeak, ECDF.}
        \label{fig:II_PCol_ECDF}
    \end{subfigure}
    \end{figure}
    \begin{figure}\ContinuedFloat
    \centering
    \begin{subfigure}[b]{0.47\textwidth}
        % \hspace*{-1.5cm} 
         \includegraphics[width=1\textwidth]{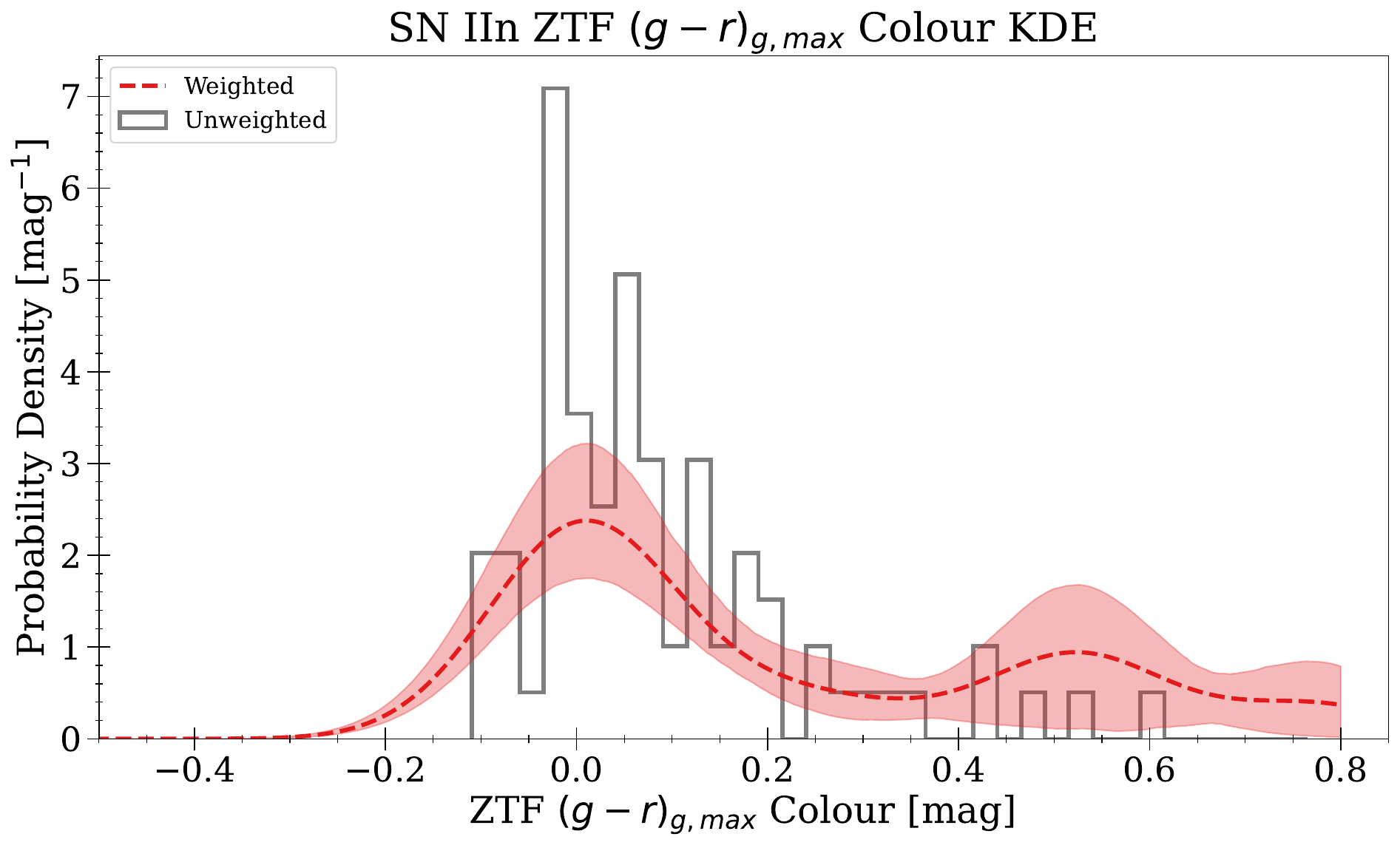}
         \caption{Type IIn Peak Colour, \ColPeak, KDE.}
         \label{fig:IIN_PCol_KDE}
    \end{subfigure}
    \begin{subfigure}[b]{0.47\textwidth}
        % \hspace*{-1.5cm} 
         \includegraphics[width=1\textwidth]{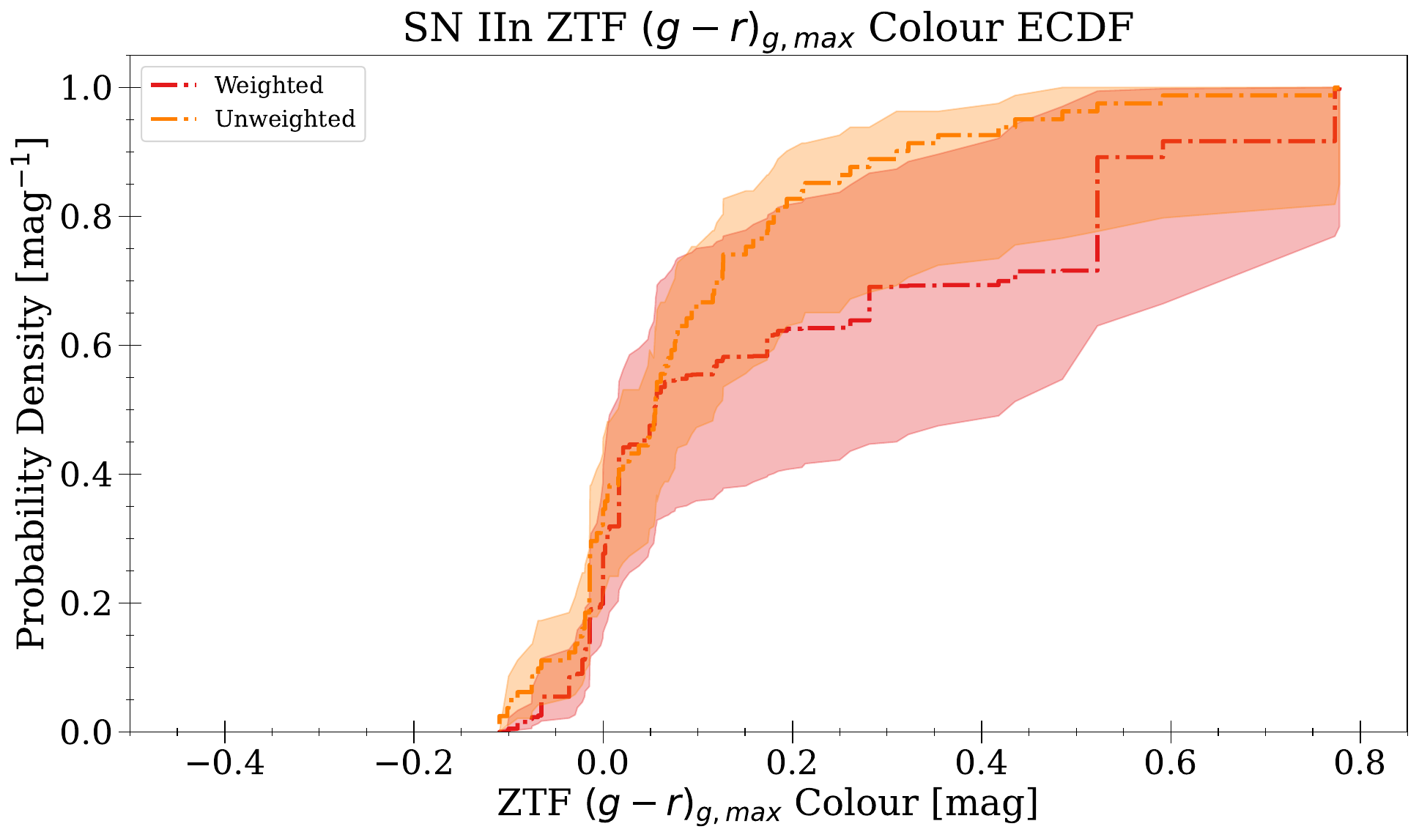}
        \caption{Type IIn Peak Colour, \ColPeak, ECDF.}
        \label{fig:IIN_PCol_ECDF}
    \end{subfigure}
        \end{figure}
    \begin{figure}\ContinuedFloat
    \centering
    \begin{subfigure}[b]{0.47\textwidth}
        % \hspace*{-1.5cm} 
         \includegraphics[width=1\textwidth]{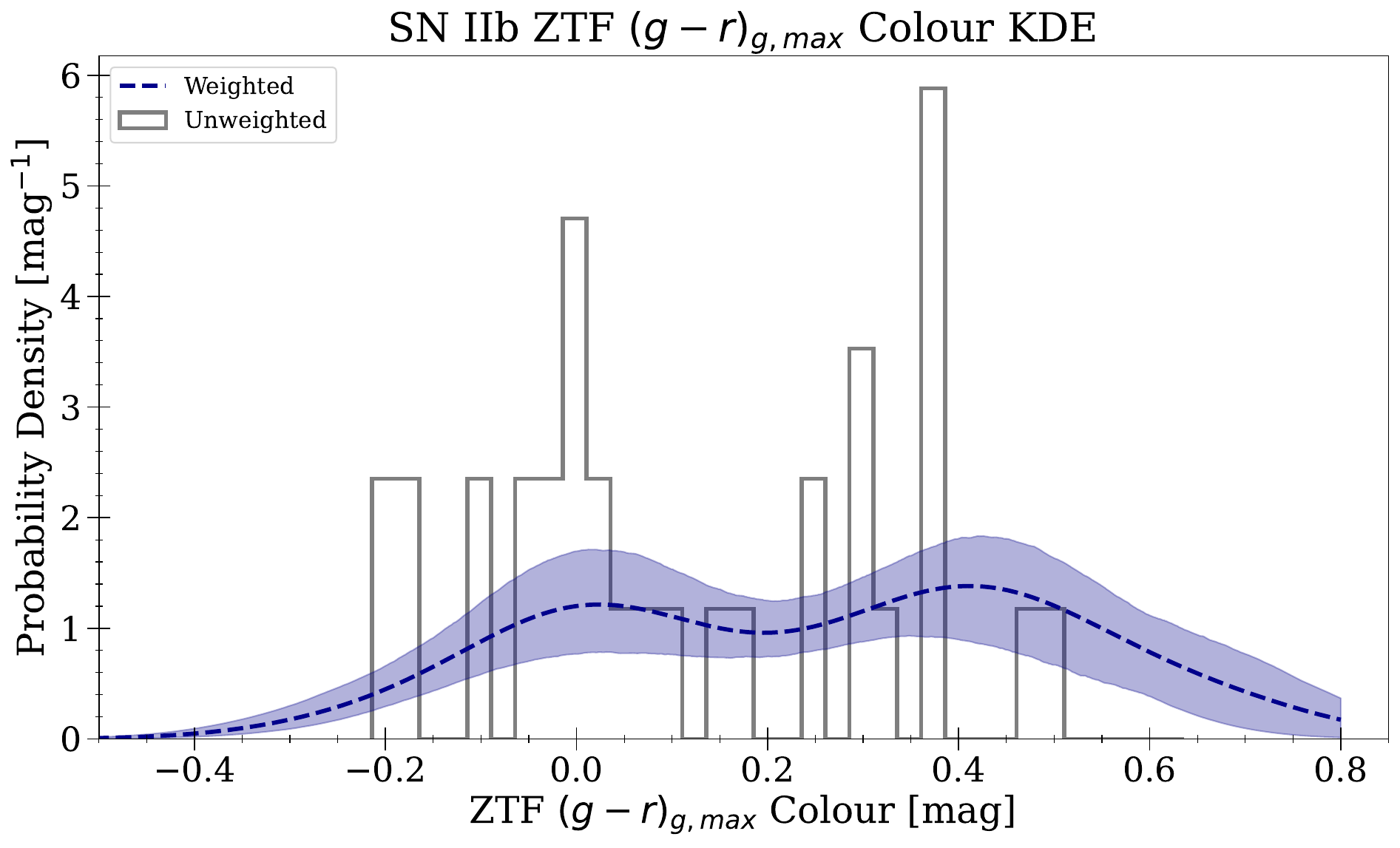}
         \caption{Type IIb Peak Colour, \ColPeak, KDE.}
         \label{fig:IIB_PCol_KDE}
    \end{subfigure}
    \begin{subfigure}[b]{0.47\textwidth}
        % \hspace*{-1.5cm} 
         \includegraphics[width=1\textwidth]{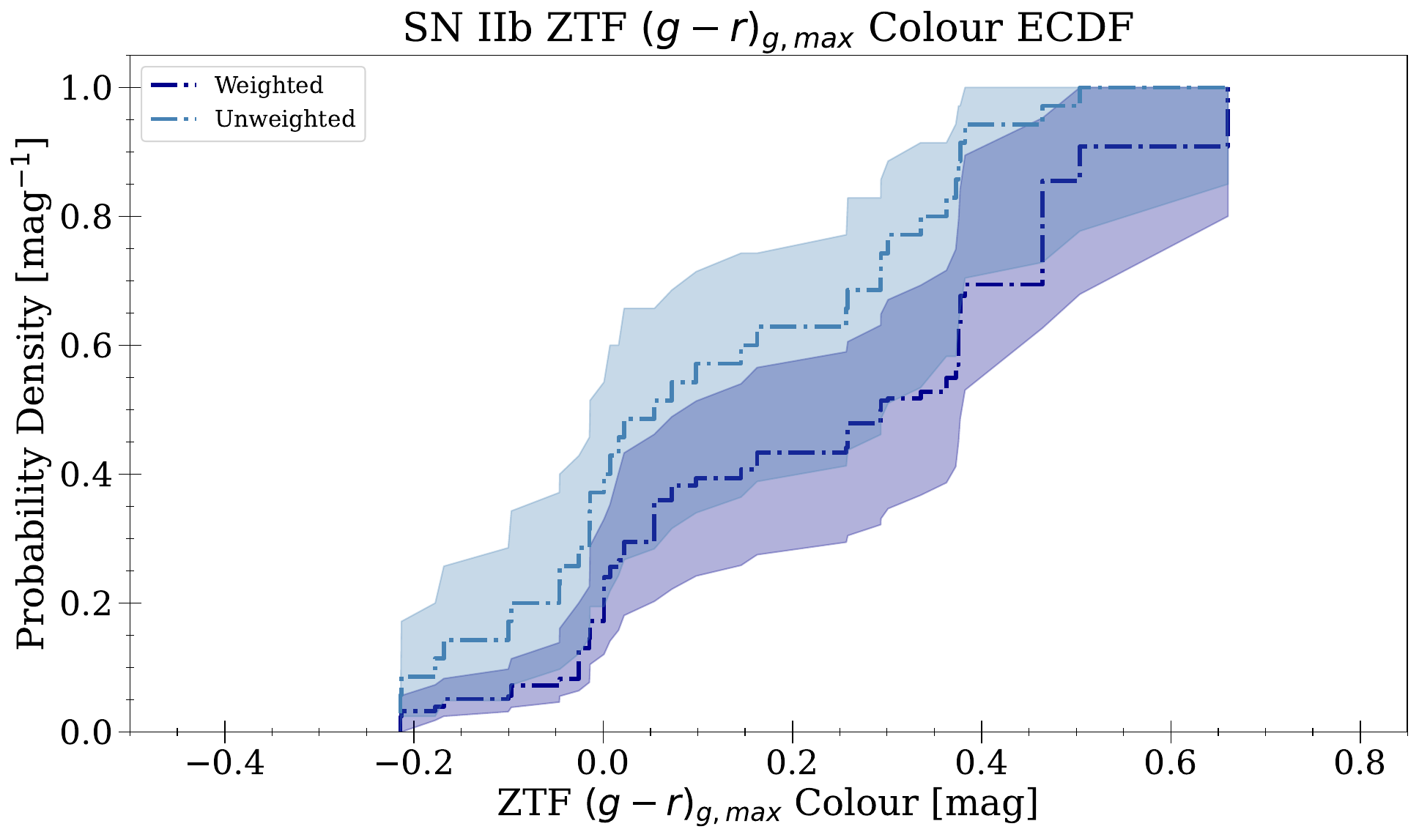}
        \caption{Type IIb Peak Colour, \ColPeak, ECDF.}
        \label{fig:IIB_PCol_ECDF}
    \end{subfigure}
\caption{KDE (\textit{left}) and ECDF (\textit{right}) for Type II (\textit{top}), Type IIn (\textit{middle}) and Type IIb (\textit{bottom}) showing the ZTF~$g-r$ colour at ZTF~$g$ peak, \ColPeak, for the purposes of correcting for host extinction using the colour at peak. A correction, detailed if Section~\cref{sec:Extinction} is applied to events with a $g-r\geq$0.25~mag and \Trise$< 20$~d.}
\label{fig:PeakColKDEECDF}
\end{figure}

In Figs. \ref{fig:II_PCol_KDE} -- \ref{fig:IIB_PCol_ECDF}, we present the KDE distributions (\emph{left}) and ECDFs (\emph{right}) of peak $g-r$ colours (\ColPeak) for Type II, Type IIn and Type IIb SNe. The panels display distributions for standard Type II (\emph{top}), Type IIn (\emph{middle}), and Type IIb (\emph{bottom}) SNe. We show various statistical quantities for each distribution in Table \ref{tab:risemagminmax}.

\newpage
\clearpage
\subsection{M23 Luminosity-Rise}
\label{app:MoriyaCMAP}

\begin{figure}
    \centering
    \includegraphics[width=0.9\textwidth]{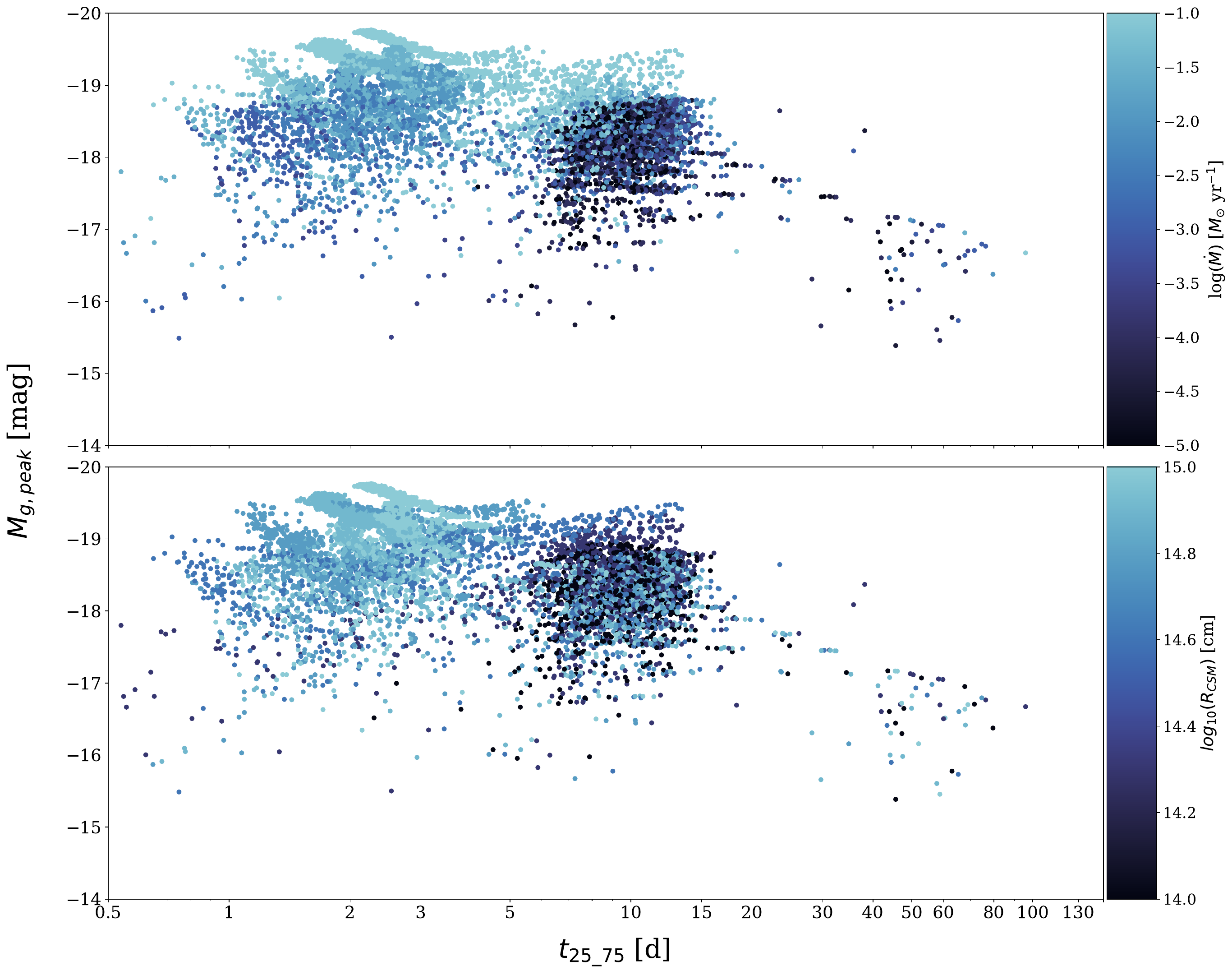}
    \caption{\Trise\ vs \Mabs\ distribution for the theoretical light curve grid from M23, with points drawn from a sample of 10,000 models, colour-coded by \MLR\ (\textit{top}) and $\log$(\Rcsm) (\textit{bottom}). The weighting is the same as applied in Fig.~\ref{fig:Moriya_absrise_massep}.}
    \label{fig:MoriyaCMAP}
\end{figure}

Fig.~\ref{fig:MoriyaCMAP} shows the distribution of \Trise\ vs. \Mabs\ for the theoretical light curve grid from M23. We present 10,000 model points using the same $V_{max} \times M_{ZAMS}^{-2.35}$ weighting scheme applied in Fig.~\ref{fig:Moriya_absrise_massep}, which accounts for both observational selection effects and the IMF \citep[e.g.,][]{Salpeter_1955}. The top panel colour-codes data by \MLR, while the bottom panel uses \Rcsm.

This visualisation reveals how CSM properties strongly influence the distribution of SNe in the \Trise-\Mabs\ plane. Fast risers (\Trise~$\leq$~5~d) typically have confined, dense CSM characterised by higher \MLR\ and smaller \Rcsm, producing moderately more luminous peaks. In contrast, slower risers (\Trise~$>$~5~d) typically exhibit less confined and less dense CSM with lower overall \Mcsm\ values and larger \Rcsm. Notably, even within the slower-rising population, the most luminous events still require substantial CSM masses, confirming that CSM mass remains a fundamental driver of peak luminosity across the distribution.

The clear separation between these populations emerges naturally from the underlying physics rather than from arbitrary parameter choices, suggesting fundamental differences in mass-loss mechanisms or progenitor structures. This bi-modality provides valuable context for interpreting the observed distribution of Type II SNe in our sample.

\newpage
\clearpage
\subsection{\Mfe\ Measurements}
\label{app:Mfe}

We extend the predictive capabilities of multi-output GPR extend to estimating the iron core mass, \Mfe, of the progenitor through Eq. \ref{eqn: iron_core_mass}, which exploits a tight correlation between \Mfe\ and the plateau luminosity at 50~d in simulated Type IIP light curves \citep[e.g.,][]{Barker_2022a, Barker_2022b}. The theoretical correlation indicates that more massive stellar cores lead to more energetic and luminous SNe, notably enhancing the bolometric luminosity during the plateau phase at approximately 50~d post-explosion \citep{Barker_2022a}. To quantify this relationship, we utilise the bolometric plateau luminosity at 50~d, $L_{bol,50 d}$. The plateau length is measured by analysing the gradient along the light curve and identifying significant changes in the slope. The ZTF~$g$ and $r$ band magnitudes are measured at 50~d after the plateau onset. A bolometric correction is then applied to convert these magnitudes into bolometric luminosity -- we adopt the methodology described by \citet{Lyman_2014}.

\begin{equation}
    \frac{M_{Fe,Core}}{\text{\Msol}} = 0.0978\times \left(\frac{L_{\text{bol,50 d}}}{10^{42}\ \text{erg s}^{-1} }\right)+ 1.29
    \label{eqn: iron_core_mass}
\end{equation}

\begin{figure}
\centering
\includegraphics[width=0.65\textwidth]{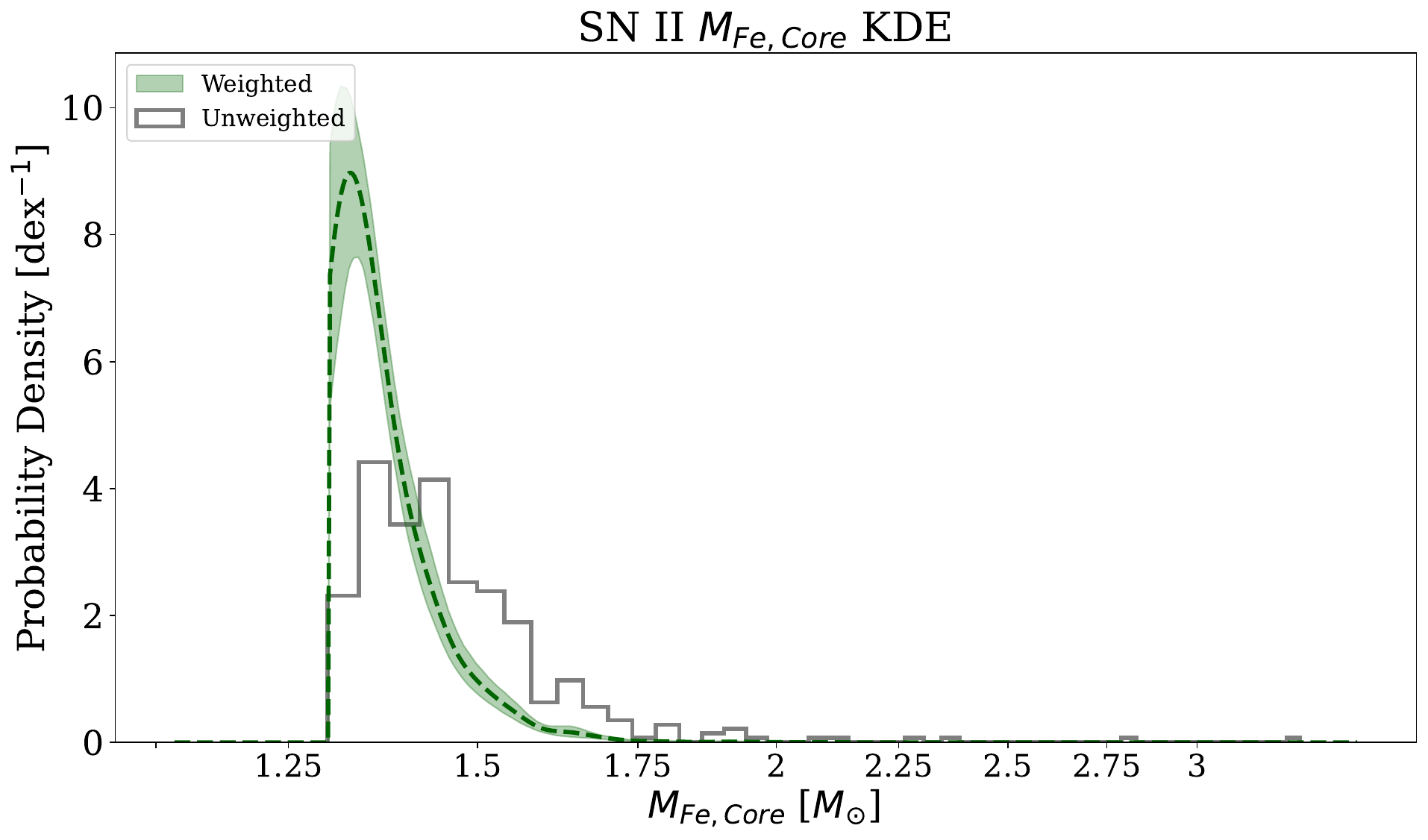}
\caption{Type II KDE for \Mfe\ along with the associated 80\% CI. The weighted distribution is in dark green (dashed line) and the unweighted normalised histogram is in black.} 
\label{fig:II_iron_core_mass_kde}
\end{figure}

The KDE distribution for \Mfe, Fig.~\ref{fig:II_iron_core_mass_kde}, shows a sharp cutoff at 1.3~\Msol, reflecting the lower limit of iron core masses in the models from which the correlation was derived \citep{Barker_2022a}. Since the KDE smoothing kernel could not properly handle this abrupt transition, we truncate the distribution at 1.3~\Msol\ and normalise the probability density to unity.

The weighted mean \Mfe\ of $1.36 \pm 0.01$~\Msol\ is consistent with the mean found in \citet{Barker_2022a} and \citet{Barker_2022b} of $1.4 \pm 0.05$~\Msol\ to within 1~$\sigma$ -- see Table~\ref{tab:final_mfe_table}. The distribution appears to be in agreement with the distribution created by \citet{Barker_2022a} and \citet{Barker_2022b}, as they find a range in \Mfe\ (1.3~--~1.5 $\pm$ 0.05~\Msol) after applying Eq. \ref{eqn: iron_core_mass} to CCSN samples from \cite{Anderson_2014} and  \citet{Guiterrez_2017b, Guiterrez_2017a}. 

While this correlation provides a useful estimate of the core mass, it assumes a direct relationship between core mass and explosion energy that, in reality, may be complicated by ejecta properties (mass and H-richness). Higher ejecta masses or more H-richness can extend and diminish the plateau luminosity independent of core mass \citep[e.g.,][]{Goldberg_2022}. The use of luminosity at 50~d may be particularly sensitive to hydrogen envelope mass variations, as it assumes complete H retention \citep[e.g.,][]{Goldberg_2022, Fang_2025}. A more robust approach might utilise the luminosity at half the plateau duration, which better accounts for diversity in envelope masses and better isolates the core mass contribution to the light curve evolution \citep{Fang_2025}.

%The simulations from which the \Mfe\ correlation is derived do not include extended envelopes or CSM shells, which our study shows are common features of Type II progenitors. While such CSM is likely swept up during the early light curve evolution and may have minimal direct impact at 50~d, its presence could affect the total energy budget and subsequent plateau luminosity, suggesting our derived core masses may be overestimates.

\begin{table}
\centering
\begin{threeparttable}
\begin{tabular}{cccccccc}
\hline
Parameter & Units & Mean & 25th\%ile & 50th\%ile & 75th\%ile & Range & No. \\ \hline\hline
&&\multicolumn{4}{c}{\multirow{2}{*}{Weighted}} & &\\ 
&&&&&&&\\ \hline
 \Mfe & \Msol & $1.36 \pm 0.01 $ & $ 1.31^{+0.02}_{-0.01} $ & $1.34 \pm 0.01 $ & $ 1.38^{+0.01}_{-0.04} $  & [1.30,3.31] & 354 \\ \hline

&&\multicolumn{4}{c}{\multirow{2}{*}{Unweighted}} & &\\ 
&&&&&&&\\\hline
 \Mfe & \Msol & $1.49 \pm 0.01 $ & $ 1.38^{+0.02}_{-0.01} $ & $1.45 \pm 0.01 $ & $ 1.53^{+0.01}_{-0.02} $  & -& -\\ \hline

\end{tabular}

\begin{tablenotes}\footnotesize %\tnote{1}
\item[1] The ranges reported are the 5th and 95th  percentiles to remove outliers beyond the limits of the original dataset. 
\end{tablenotes}
\end{threeparttable}

\caption{Mean and median of the volume corrected KDE for \Mfe\ in the final sample. Uncertainties reported here are the standard deviation on the bootstrapped values. }
\label{tab:final_mfe_table}
\end{table}

\subsection{M23 Relations}
\label{app:M23 Relations Appendix}
    
\begin{equation}
    \log_{10}(\text{U}) = C_0\ + C_1\times\text{V}\ + C_2\times\text{W} + C_3 \times\text{X}\ + C_4\times\text{Y}\ + C_5\times\text{Z}
    \label{eqn:polyorder1}
\end{equation}

\begin{multline}
    \log_{10}\left(\text{U}\right) = C_0 + C_1\times\text{V}^2 + C_2\times\left(\text{V}\times\text{W}\right)\ + C_3\times\left(\text{V}\times\text{X}\right)\ +  C_4\times\left(\text{V}\times\text{Y}\right)\ + C_5\times\left(\text{V}\times\text{Z}\right)\ + C_6\times\text{W}^2 + C_7\times\left(\text{W}\times\text{X}\right)\  + \\ C_8\times\left(\text{W}\times\text{Y}\right)\ + C_9\times\left(\text{W}\times\text{Z}\right)\ +  C_{10}\times\text{X}^2 + C_{11}\times\left(\text{X}\times\text{Y}\right)\ + C_{12}\times\left(\text{X}\times\text{Z}\right)\ + C_{13}\times\text{Y}^2 +  C_{14}\times\left(\text{Y}\times\text{Z}\right)\ + C_{15}\times\text{Z}^2\
    \label{eqn:polyorder2}
\end{multline}

\begin{multline}
    \log_{10}\left(\text{U}\right) = C_0 + C_1\times \text{V}^3 + C_2\times\left(\text{V}^2\times\text{W}\right)\ + C_3\times\left(\text{V}\times\text{W}^2\right)\ + C_4\times \text{W}^3 + C_5\times\left(\text{V}^2\times\text{X}\right)\ +  C_6\times\left(\text{V}\times\text{W}\times\text{X}\right)\ + \\ \ C_7\times\left(\text{W}^2\times\text{X}\right)\ + C_8\times\left(\text{V}\times\text{X}^2\right)\ + C_9\times\left(\text{W}\times\text{X}^2\right)\ + C_{10}\times \text{X}^3 + C_{11}\times\left(\text{V}^2\times\text{Y}\right)\ +  C_{12}\times\left(\text{V}\times\text{W}\times\text{Y}\right)\ + C_{13}\times\left(\text{W}^2\times\text{Y}\right)\ + \\ C_{14}\times\left(\text{V}\times\text{X}\times\text{Y}\right)\ + C_{15}\times\left(\text{W}\times\text{X}\times\text{Y}\right)\ + C_{16}\times\left(\text{X}^2\times\text{Y}\right)\ + C_{17}\times\left(\text{V}\times\text{Y}^2\right)\ + C_{18}\times\left(\text{W}\times\text{Y}^2\right)\ + C_{19}\times\left(\text{X}\times\text{Y}^2\right)\ + \\ C_{20}\times \text{Y}^3 + C_{21}\times\left(\text{V}^2\times\text{Z}\right)\ + C_{22}\times\left(\text{V}\times\text{W}\times\text{Z}\right)\ + C_{23}\times\left(\text{W}^2\times\text{Z}\right)\ + C_{24}\times\left(\text{V}\times\text{X}\times\text{Z}\right)\ + C_{25}\times\left(\text{W}\times\text{X}\times\text{Z}\right)\ + C_{26}\times\left(\text{X}^2\times\text{Z}\right)\ + \\C_{27}\times\left(\text{V}\times\text{Y}\times\text{Z}\right)\ + C_{28}\times\left(\text{W}\times\text{Y}\times\text{Z}\right)\ + C_{29}\times\left(\text{X}\times\text{Y}\times\text{Z}\right)\ + C_{30}\times\left(\text{Y}^2\times\text{Z}\right)\ + C_{31}\times\left(\text{V}\times\text{Z}^2\right)\ + C_{32}\times\left(\text{W}\times\text{Z}^2\right)\ + \\ C_{33}\times\left(\text{X}\times\text{Z}^2\right)\ + C_{34}\times\left(\text{Y}\times\text{Z}^2\right)\ + C_{35}\times \text{Z}^3 
    \label{eqn:polyorder3}
\end{multline}

\noindent For \Mcsm, U= \Mcsm, V = \Mabs, W = $\log_{10}(\text{\Ttwosix})$, X = $\log_{10}(\text{\Tsixnine})$, Y = \ColPeak\ \& Z = \Mten. \\
\noindent For \Rcsm, U = \Rcsm, V = \Mabs, W = $\log_{10}(\text{\Ttwofive})$, X = $\log_{10}(\text{\Tfiveeight})$, Y = \ColPeak\ \& Z = \Mfive.

\begin{table}
    \centering
    \begin{tabular}{c|c c c c c c c c c c c c c c c c c} \hline
     & $C_0$ & $C_1$ & $C_2$ & $C_3$ & $C_4$ & $C_5$ & $C_6$ & $C_7$ & $C_8$ & $C_9$ & $C_{10}$ & $C_{11}$ \\ \hline \hline
    \Mcsm\ $\leq$ 5~d & -4.51 & 2.28 & 2.07 & -0.24 & 1.07 & -2.88 & -0.19 & 0.0823 & 2.10 & -1.24 & -0.16 & -12.35   \\
    \Mcsm\ $>$ 5~d & -3.54 & 13.24 & -3.85 & 5.94 & 0.22 & -30.16 & -9.55 & -1.10 & 4.64 & 4.31 & -2.26 & 42.92  \\
    \Rcsm\ $\leq$ 5~d & 0.65 & 0.43 & -2.91 & 0.51 & 0.14 & 1.93 & 0.57 & 0.18 & -1.04 & 0.039 & -0.071 & -10.05  \\
    \Rcsm\ $>$ 5~d & -0.41 & 30.75 & 0.24 & -4.57 & 0.29 & 5.67 & 5.47 & -0.042 & -4.65 & -0.18 & 0.22 & 0.055 \\ \hline
    \end{tabular}
\hspace{0.15cm}
\begin{tabular}{c|c c c c c c c c c c c c c c c c c c } \hline
   &  $C_{12}$  & $C_{13}$ & $C_{14}$ & $C_{15}$ & $C_{16}$ & $C_{17}$ & $C_{18}$ & $C_{19}$ & $C_{20}$ & $C_{21}$ & $C_{22}$ & $C_{23}$  \\ \hline \hline
    \Mcsm\ $\leq$ 5~d& -16.12 & -2.31 & 5.66 & 2.76 & -0.13 & 8.06 & -2.24 & 2.19 & 0.84 & -6.88 & -3.95 & 0.20\\
    \Mcsm\ $>$ 5~d & 5.95 & 1.44 & 10.72 & -8.35 & 4.53 & -29.78 & 5.17 & 3.19 & -10.21 & -42.38 & 7.66 & -6.04 \\
    \Rcsm\ $\leq$ 5~d & 3.82 & -0.16 & 3.71 & -0.41 & 0.29 & 2.89 & 0.54 & -0.46 & -0.46 & -1.03 & 5.83 & -0.46  \\
    \Rcsm\ $>$ 5~d  & 6.36 & 0.0098 & -7.77 & 0.64 & -0.49 & -4.28 & -0.59 & 0.22 & -0.13 & -93.03 & -0.35 & 4.6\\ \hline
    \end{tabular}
\hspace{0.1cm}
\begin{tabular}{c|c c c c c c c c c c c c c} \hline
     & $C_{24}$ & $C_{25}$ & $C_{26}$ & $C_{27}$ & $C_{28}$ & $C_{29}$ & $C_{30}$ & $C_{31}$ & $C_{32}$ & $C_{33}$ & $C_{34}$ & $C_{35}$ \\ \hline \hline
    \Mcsm\ $\leq$ 5~d   & 5.85  & 0.085 & -2.11 & 25.25 & 16.32 & -5.72 & -8.03 & 6.91 & 1.87 & -2.97 & -12.91 & -2.31 \\
    \Mcsm\ $>$ 5~d & 61.57 & 9.91 & -5.03 & -86.78 & -6.18 & -10.02 & 29.54 & 45.11 & -3.81 & -31.43 & 43.9 & -15.97 \\
    \Rcsm\ $\leq$ 5~d & -3.99& -0.55 & 1.03 & 20.35 & -3.84 & -3.69 & -2.94 & 0.77 & -2.92 & 2.07 & -10.30 & -0.17 \\
    \Rcsm\ $>$ 5~d  & -11.87 & -5.48 & 4.68 & -0.54 & -6.38 & 7.74 & 4.31 & 93.82 & 0.11 & 6.21 & 0.49 & -31.53 \\ \hline
    \end{tabular}
    \caption{\Mcsm\ and \Rcsm\ coefficients.}
    \label{tab:M23_csm_coefficients}
\end{table}

\newpage

\subsection{M23 Radial Extent Predictions}
\label{app:M23 Radial Predictions Appendix}
\begin{figure}%\ContinuedFloat
\centering
    \includegraphics[width=0.9\textwidth]{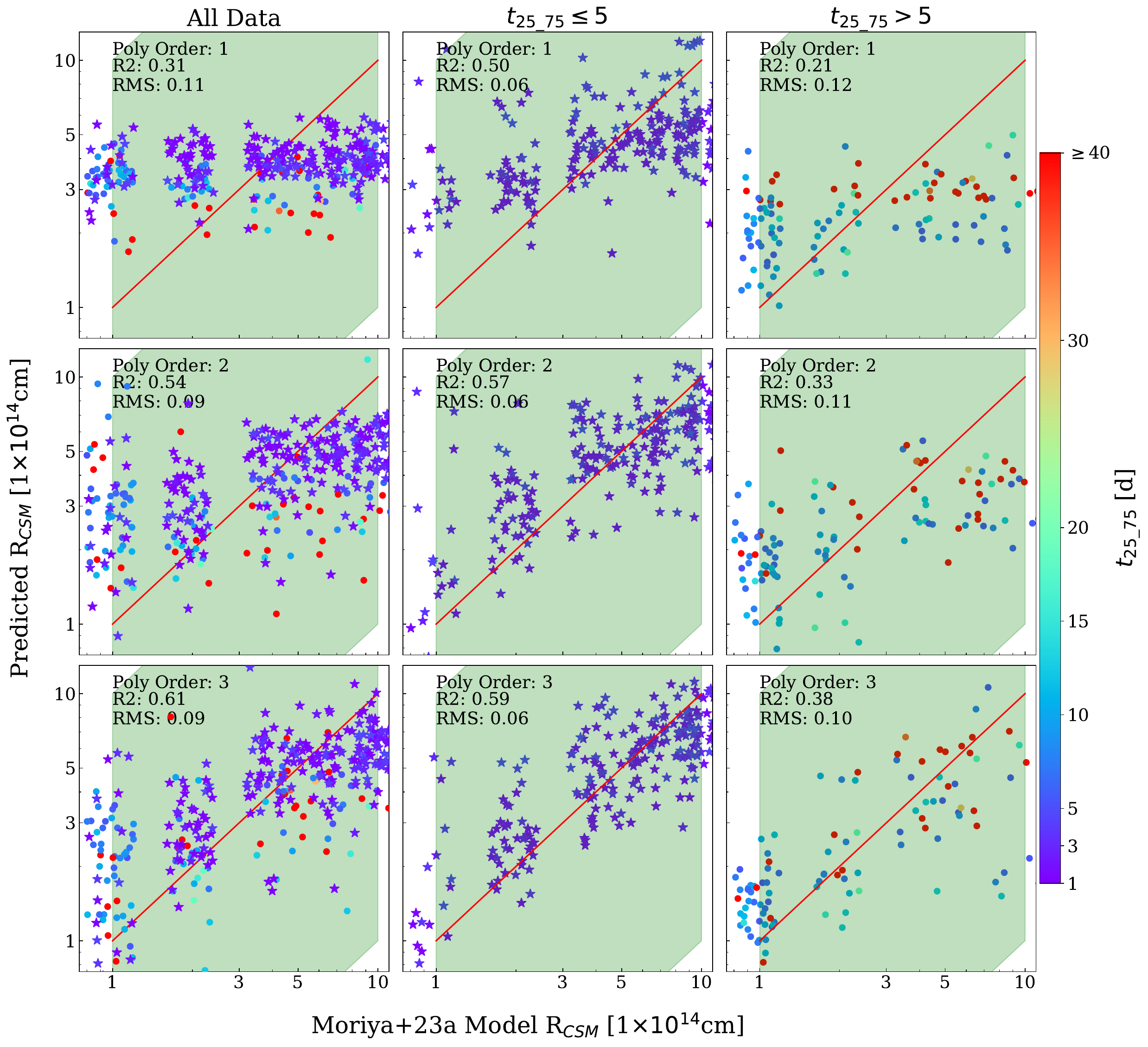}
     \caption{Multivariate analysis results of the predicted \Rcsm\ radial extent (y-axis) vs. the M23 \Rcsm\ radial extent (x-axis). The top, middle and bottom rows are polynomial orders 1, 2 and 3 respectively. The first column contains all the data and stars are those with \Trise~$\leq$~5~d with the 2nd and 3rd rows containing only data with \Trise~$\leq$~5~d and \Trise~$>$~5~d to show how the correlations predictive power decreases significantly for events with \Trise~$\geq$~5~d. The diagonal red line is the 1:1 line with the green shaded region showing 1 order of magnitude above and below. This is run only on M23 models where \Mcsm~$\geq$ 1$\times10^{-2.5}$~\Msol\ as we identify is Section~\cref{sec:CSMPolyReg} this to be the lower limit, above which \Mabs\ and \Trise\ were influenced. }
     \label{fig:RadiusMoriyaPolyGrid}
\end{figure}

Fig.~\ref{fig:RadiusMoriyaPolyGrid} shows the multivariate analysis comparing our polynomial regression-predicted CSM radial extent values (y-axis) against the corresponding M23 model values (x-axis), following an approach similar to Figure~\ref{fig:MoriyaPolyGridMass}. This systematic evaluation examines the performance of polynomial fits across different degrees (1st, 2nd, and 3rd order) and specific parameter regimes to determine the optimal method for characterising this relationship. Unlike our \Mcsm\ analysis, this investigation of \Rcsm\ is restricted to M23 models with \Mcsm~$\geq$ 1$\times10^{-2.5}$~\Msol, which Section~\cref{sec:CSMPolyReg} identifies as the threshold above which CSM significantly influences both \Mabs\ and \Trise. The results demonstrate that \Rcsm\ can only be reliably constrained for events with substantial \Mcsm.

\subsection{M23 \Mcsm\ Lower Limit}
\label{app:M23 CSM Lower Limit Appendix}

\begin{figure}
\centering
\includegraphics[width=0.9\textwidth]{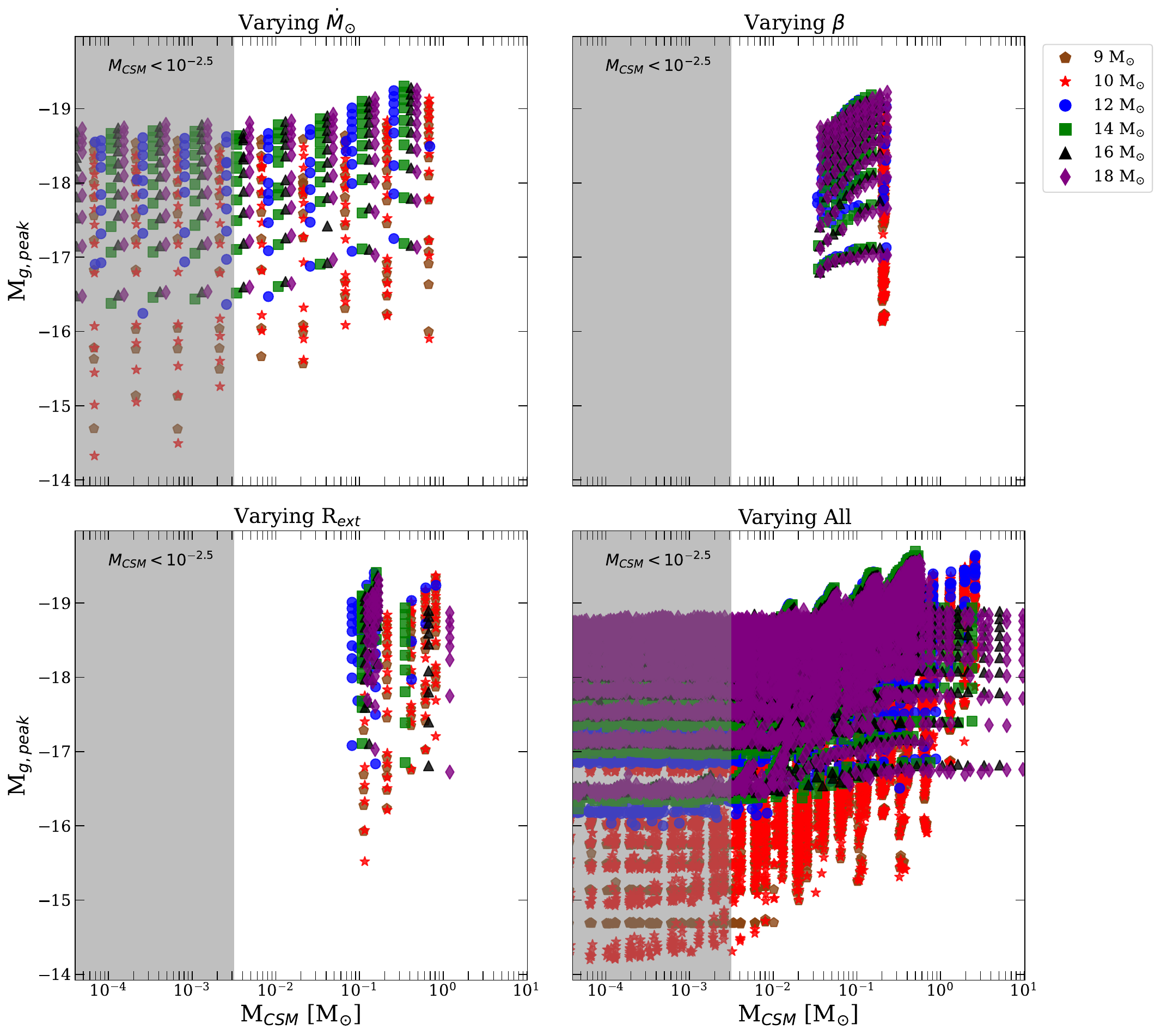}
\caption{Dependence of \Mabs\ on \Mcsm\ under different parameter variations, with fixed Ni mass and explosion energy. \emph{Top left} -- Varying \MLR\ with fixed $\beta$ and \Rcsm. \emph{Top right} -- Varying $\beta$ with fixed \MLR\ and \Rcsm. \emph{Bottom left} -- Varying \Rcsm\ with fixed \MLR\ and $\beta$. \emph{Bottom right} -- Combined variation of all CSM parameters (\MLR, $\beta$, \Rcsm). Each panel shows results for different progenitor masses (10 -- 18~\Msol). Below \Mcsm~$\approx 10^{-2.5}$~\Msol, CSM properties do not significantly influence the peak magnitude, indicating a transition to CSM-negligible evolution.}
\label{fig:CSM_mass_low_lim}
\end{figure}

Accurate measurements of \Rcsm\ becomes challenging when \Mcsm\ is insufficient to significantly influence observables such as \Mabs\ and \Trise. To establish a critical threshold below which CSM becomes virtually undetectable in early light curves, we systematically analysed how variations in key physical parameters -- \MLR, \Rcsm\ and $\beta$ -- affect observable properties.

We systematically varied these parameters while holding other key physical parameters constant (e.g., nickel mass). Our investigation revealed that when \Mcsm\ falls below approximately $10^{-2.5}$~\Msol, the CSM becomes too diffuse to meaningfully influence early light curve evolution. This threshold is evidenced by minimal variations in \Mabs\ below $10^{-2.5}$~\Msol\ and increasingly significant variations above this mass -- a pattern consistent across all progenitor masses. At this critical point, we observe a transition to a regime where CSM interaction becomes negligible in shaping the observable properties of the supernova. This theoretical expectation is strongly supported by the distinct bimodal distribution observed in the M23 models (Fig. \ref{fig:MoriyaCMAP}).

Given this fundamental limitation in detecting and characterising low-mass CSM environments, we restrict our subsequent analysis of \Rcsm\ to events where the predicted \Mcsm\ exceeds 10$^{-2.5}$~\Msol.

\subsection{Impact of Systematic Misclassifications}
\label{app:Systematic Misclassification}
%## Evaluation of Classification Robustness and Sample Integrity

Most SN classifications from the BTS rely on the low-resolution SEDM spectrograph (R $\sim$100). The limited spectral resolution and typically single-epoch observations near maximum light can make distinguishing certain SN subclasses challenging, particularly Type IIb from Type II and, to a lesser extent, Type IIn from Type II or host emission. Consequently, our Type II sample might contain some level of contamination from misclassified events, a consideration we quantitatively address here.

To quantify potential classification biases, we conducted Kolmogorov-Smirnov (KS) tests comparing Type II and Type IIb populations. KS tests of the unweighted \Mabs\ and \Trise\ distributions yielded $p$-values of 0.038 and 0.0030, respectively, indicating statistically significant differences between these populations. We identified an approximately 5\% shortfall of Type IIb SNe in our sample ($7.22^{+2.40}_{-1.84}$\% versus the expected $\sim$12.5\% from \citealt{Shivvers_2017}'s volume-complete sample). To assess the impact of possible misclassifications, we applied a conservative approach by removing the fastest-rising 5\% of Type II SNe -- those most likely to be misclassified Type IIbs and have the largest impact on our results -- and recalculated the \Mcsm\ KDE distribution. The fraction of Type II SNe with \Mcsm $\geq 10^{-2.5}$~\Msol\ remained consistent (38 -- 41\%) with our original finding ($\sim$36\%). This represents the most extreme scenario, confirming that potential misclassifications affect our results by less than 1~$\sigma$.

Similarly, our analysis yields a Type IIn rate of $4.34^{+1.49}_{-1.09}\%$ relative to Type II SNe, consistent with \citet{Shivvers_2017}.  We consider the possibility of misclassification between regular Type II SNe and Type IIn events to be minimal for several reasons: (1) Type IIn SNe typically exhibit higher luminosities and represent a small fraction of the overall population, resulting in negligible statistical impact after $V_{max}$ weighting; (2) BTS routinely conducts follow-up observations using higher-resolution spectrographs for suspected Type IIn events to refine classification; and (3) wee see that $<$2\% of our Type IIn sample exhibits photometric characteristics resembling typical Type II events (e.g., \Trise $\leq$~3~d and \Mabs $>-18$~mag, which represent the median values for our unweighted Type II sample; Table \ref{tab:risemagminmax}). Our robust classification methodology ensures complete Type IIn identification, and our focus on CSM around fast-rising SNe means the longer evolution timescales of Type IIn events minimally impact our conclusions.

We have also considered the potential impact of peculiar events resembling SN 1987A on our results. Such objects, characterised by moderate peak luminosities combined with unusually slow rise times (e.g., ZTF18acbwaxk), represent rare occurrences in the local universe. If several such events were misclassified or included within our sample, their statistical contribution would remain minimal given our $V_{max}$ weighting and large sample size.

\label{lastpage}
\end{document}